\documentclass[12pt]{article}
\usepackage{amsmath, amssymb}

\newtheorem{theorem}{Theorem}[section]

\newtheorem{lemma}[theorem]{Lemma}
\newtheorem{proposition}[theorem]{Proposition}
\newtheorem{definition}{Definition}[section]
\newtheorem{convention}[definition]{Convention}
\numberwithin{equation}{section}

\newcommand{\R}{\mathbb{R}}
\newcommand{\Complex}{\mathbb{C}}
\renewcommand{\Re}{\mathop{\mathrm{Re}}}
\renewcommand{\Im}{\mathop{\mathrm{Im}}}

\newcommand{\myspan}{\mathop{\rm span}}

\newcommand{\longto}{\longrightarrow}
\newcommand{\norm}[1]{\left\Vert #1 \right\Vert}
\newcommand{\abs}[1]{\left| #1 \right|}
\newcommand{\bka}[1]{\left \langle #1 \right \rangle}
\newcommand{\bke}[1]{\left( #1 \right)}
\newcommand{\bkt}[1]{\left[ #1 \right]}
\newcommand{\bket}[1]{\left\{ #1 \right\}}
\newcommand{\mb}[1]{\mathbf{#1}}

\newcommand{\myproof}{\textsc{Proof:}\ }
\newcommand{\myendproof}{\hspace*{\fill}
  {{\bf \small Q.E.D.}} \vspace{10pt}}

\newcommand{\Pc}{\, \mathbf{P}\! _\mathrm{c} \, \! }
\newcommand{\PcA}{\, \mathbf{P}\! _\mathrm{c} \! ^A \,}
\newcommand{\PcH}{\Pc}
\newcommand{\PcL}{\, \mathbf{P}\! _\mathrm{c} \! ^\L \,}
\newcommand{\Hc}{\, \mathbf{H} _\mathrm{c} \, \! }
\newcommand{\eigen}{\mathbf{E}}
\renewcommand{\L}{\mathcal{L}}

\newcommand{\epz}{\varepsilon_0}
\newcommand{\loc}{_{\mathrm{loc}}}
\newcommand{\al}{\alpha}

\newcommand{\la}{\lambda}

\newcommand{\wt}{\widetilde}
\newcommand{\conj}{\mathop{\,\mathbf{conj}\,}}  
\newcommand{\pd}{\partial}
\newcommand{\donothing}[1]{}
\newcommand{\wbar}[1]{\overline{\rule{0pt}{2.4mm} {#1}}}

\newcommand{\corr}{\longleftrightarrow}
\newcommand{\vect}[1]{\begin{bmatrix} #1 \end{bmatrix}}
\newcommand{\svect}[1]
{\bkt{\begin{smallmatrix} #1\end{smallmatrix}}}
\newcommand{\myfrac}[2]{\stackrel {\scriptstyle #1}{\scriptstyle #2}}
\newcommand{\les}{\lesssim}

\newcommand{\ev}{\omega}
\newcommand{\evector}{\mathbf{\Phi}}
\newcommand{\RE}{\mathop{\,\mathbf{RE}\,}}
\newcommand{\inomega}{\, \in \, \Omega}
\newcommand{\ssum}{{\textstyle \sum}}

\newcommand{\Dt}{\Delta t}
\newcommand{\NOT}[1]{\ \Big / \hspace*{-5mm} {#1}\ }

\begin{document}

\reversemarginpar \baselineskip 18pt

\title{Asymptotic Dynamics of Nonlinear Schr\"odinger Equations with
Many Bound States}

\author{\textsc{Tai-Peng Tsai}\\[3mm]
Institute for Advanced Study, Princeton, NJ 08540\\
E-mail: ttsai@ias.edu}

\date{}

\maketitle

\addtocounter{footnote}{1}

\addtocounter{footnote}{1}


\begin{abstract}

We consider a nonlinear Schr\"odinger equation with a bounded local
potential in $\R^3$. The linear Hamiltonian is assumed to have three or
more bound states with the eigenvalues satisfying some resonance
conditions. Suppose that the initial data is localized and small of order
$n$ in $H^1$, and that its ground state component is larger than
$n^{3-\epsilon}$ with $\epsilon>0$ small. We prove that the solution will
converge locally to a nonlinear ground state as the time tends to infinity.

\end{abstract}

\section{Introduction}

Consider the nonlinear Schr\"odinger equation
\begin{equation} \label{Sch}
i \partial _t \psi = (-\Delta + V) \psi + \la |\psi|^2 \psi,
\qquad \psi(t=0)= \psi_0,
\end{equation}
where $V=V(x)$ is a smooth localized real potential, $\la=\pm 1$
and $\psi=\psi(t,x):\R \times \R^3 \longto \Complex$ is a wave function.
For any solution $\psi(t)\in H^1(\R^3)$ the $L^2$-norm and the
Hamiltonian
\begin{equation} \label{1-2}
{\cal H}[\psi] = \int \frac 12 |\nabla \psi|^2 +
\frac 12 V |\psi|^2 + \frac 14 \la |\psi|^4 \, d x
\end{equation}
are constant for all $t$. The global well-posedness for small
solutions in $H^1(\R^3)$ can be proven using these conserved
quantities and a continuity argument, no matter what the sign of
$\la$ is.
We assume that the linear Hamiltonian $H_0:= -\Delta + V$ has
$N+1$ simple eigenvalues $e_0<e_1<\dots <e_N$ with normalized
eigenvectors $\phi_k$, $k=0,1,\ldots,N$, where $N\ge 2$. These
eigenvalues are
assumed to satisfy some resonance conditions to be specified later
on. 
%
%
%
The {\it nonlinear bound states} to the Schr\"odinger equation
\eqref{Sch} are solutions to the equation
\begin{equation}   \label{Q.eq}
    (-\Delta + V) Q + \la |Q|^2 Q = EQ,
\end{equation}
for some $E$. They are critical points to the Hamiltonian ${\cal
H}[\psi] $ defined in \eqref{1-2} subject to the constraint of
fixed $L^2$-norm. For any nonlinear bound state $Q=Q_E(x)$, the
function $\psi(t,x) = Q(x) e^{-i E t }$ is an exact solution to
the nonlinear Schr\"odinger equation \eqref{Sch}. We may obtain
$N+1$ families of such nonlinear bound states by standard
bifurcation theory, corresponding to the $N+1$ eigenvalues of the
linear Hamiltonian. For any $E$ sufficiently close to $e_0$ so
that $E-e_0$ and $\la$ have the same sign, there is a unique
positive solution $Q=Q_E$ to \eqref{Q.eq} which decays
exponentially  as $x \to \infty$. We call this family the {\it
nonlinear ground states} and we refer to it as $\bket{Q_{E}}_{E}$.
Similarly, for each $k>0$ there is a {\it nonlinear excited state}
family $\bket{Q_{k,E_k}}_{E_k}$ for $E_k$ near $e_k$.
These solutions are small and
$\norm{Q_{k,E_k}} \sim |E_k-e_k|^{1/2}$. See Lemma \ref{th:2-1}.

Our goal is to understand the long-time dynamics of the solutions
at the presence of nonlinear bound states. One first considers the
stability of nonlinear ground states. There are two different
concepts: {\it orbital stability} and {\it asymptotic stability}.
It is well-known that nonlinear ground states are orbitally stable
in the sense that the difference
\[
   \inf_{\Theta, E} \norm{ \psi(t) -Q_E \, e^{i \Theta} }_{L^2(\R^3)}
\]
remains small for all time $t$ if it is initially small. On the
other hand, one expects that the difference actually approaches
zero locally, as the majority of the difference is a dispersive
wave which escapes to infinity. Hence one expects that it is
asymptotically stable in the sense that
\begin{equation*}
\norm{ \psi(t) -Q_{E(t)} \, e^{i \Theta(t)} }_{L^2 \loc} \longto 0
\end{equation*}
as $t \to \infty$, for a suitable choice of $E(t)$
and $\Theta(t)$. Here $\norm{\cdot }_{L^2 \loc}$ denotes a local
$L^2$ norm, a precise choice
will be made later on in \eqref{Lploc.def}.
One also wants to determine the decay rate and
whether $E(t)$ has a limit.
A more difficult problem, which one studies next, is the asymptotic
dynamics of the solutions
when the initial data are away from nonlinear ground states.
As in the previous problem, one wants to identify their local behavior.

If $- \Delta + V$ has only one bound state, i.e., with no excited
states, the asymptotic stability is proved in \cite{SW1}. The
solution eventually settles down to some ground state
$Q_{E_\infty}$ with $E_\infty$ close to $E$ and the local
difference is bounded by $Ct^{-3/2}$, as the decay rate of the
free evolution $e^{i t \Delta}$. This result is extended in
\cite{PW} to all small initial data, not necessarily near ground
states. See also \cite{BP1,C1}.

Suppose $- \Delta + V$ has two bound states. We proved in
\cite{TY} that the evolution with initial data $\psi_0$ near some
$Q_E$ will eventually settle down to some ground state
$Q_{E_\infty}$ with $E_\infty$ close to $E$.  The local difference
is, however, only bounded by $C t^{-1/2}$ due to the persistence
of the excited state. The key mechanism here is the resonance
decay of the excited state due to resonance with the continuous
spectrum. See \cite{BP2,BS} for an one dimensional equation,
\cite{C2} its extension to higher dimensions, and \cite{SW2} for
real-valued nonlinear Klein-Gordon equations.

The problem becomes more delicate when the initial data are away
from ground states. Based on physical intuition, one expects that
all solutions near excited states decay to ground states unless
initially they are exactly excited states. However, we proved
\cite{TY3} that there exists a family of ``finite co-dimensional
manifolds'' in the space of initial data so that the dynamics
asymptotically converge to some excited states. Outside a small
neighborhood of these manifolds, the asymptotic profiles are given
by some ground states \cite{TY2}. We further showed \cite{TY4}
that there are exactly three asymptotic profiles: vacuum, excited
states or ground states. The last problem is also considered in
\cite{SW3}. Earlier works concerning related linear analysis were
obtained in \cite{G0,G, SS, SS2,W2}.

In this paper, we extend the results in \cite{TY, TY2} to the case
when $-\Delta+V$ has three or more bound states.

Our assumptions on the operator $-\Delta + V$ are as follows:

\noindent {\bf Assumption A0}: $- \Delta + V$ acting on
$L^2(\R^3)$ has $N+1$ simple eigenvalues $e_0<e_1< \ldots< e_N<0$,
$N\ge 2$, with normalized eigenvectors $\phi_0, \ldots, \phi_N$.

\noindent {\bf Assumption A1}: $V(x)$ is a real-valued function.
For $\la Q_E^2$ sufficiently small, the bottom of the continuous
spectrum to $-\Delta+ V+\la Q_E^2$, $0$, is not a generalized
eigenvalue, i.e., not an eigenvalue nor a resonance. Also, we
assume that $V$ satisfies the assumption in \cite{Y} so that the
$W^{k,p}$ estimates  $k\le 2$ for the wave operator
$W_{H_0}=\lim_{t\to \infty} e^{i t H_0}e^{it\Delta}$ hold for $k
\le 2$, i.e., there is  a small $s_1>0$ such that,
\[
|\nabla^\al V(x)| \le C \bka{x}^{-5- s_1}, \qquad \text{for }
|\al|\le 2 .
\]
Also, the functions $(x\cdot \nabla)^k V$, for $k=0,1,2,3$, are
$-\Delta$ bounded with a $-\Delta$ bound less than $1$:
\begin{equation*}
  \norm{(x\cdot \nabla)^k V\phi}_2 \le s_2 \norm{-\Delta\phi}_2 +
C\norm{\phi}_2,
  \qquad s_2 < 1 , \quad k=0,1,2,3  .
\end{equation*}

\noindent {\bf Assumption A2}: Resonance condition.
Denote $e_{k0} = e_k - e_0$.
We assume that
\begin{equation}
  e_0 < 2 \, e_1 .
\end{equation}
Hence $2e_{k0}>|e_0|$ for all $k>0$.
We further assume that, for some small $s_0>0$,
\begin{equation} \label{gamma0.def}
\gamma_0 \equiv \inf_{\myfrac{1\le k\le N}{|s|<s_0}} \lim_{r \to
0+} \Im \bke{ \phi_0\phi_k^2 \, , \, \frac {1}{-\Delta+V+e_0 - 2
e_k -s -r i} \PcH \phi_0\phi_k^2 } > 0.
\end{equation}

\noindent {\bf Assumption A3}: No-resonance condition. Let $j_{\text{max}}=3$.
For all $j=2,\ldots,j_{\text{max}}$ and for all $k_1,\ldots, k_j
,l_1,\ldots, l_j \in \bket{0, \ldots, N}$
with $\bket{k_1,\ldots, k_j} \not = \bket{l_1,\ldots, l_j}$ as
sets with multiplicities, (e.g., $\bket{0,0,1} \not  = \bket{0,1,1}$),
\begin{equation}
e_{k_1}+ \cdots + e_{k_j} \not = e_{l_1} + \cdots + e_{l_j} .
\end{equation}
%


Assumption A1 contains some standard conditions to assure that
most tools for linear Schr\"odinger operators apply. These
conditions are certainly not optimal. (For example, it is
sufficient to assume $0$ is not a resonance or eigenvalue of
$H_0$, which implies the same statement for $-\Delta+V+\la Q^2$.
See \cite{TY3}.)
The main assumption in A2 is the condition $e_0 < 2 e_1$. Since
the expression for $\gamma_0$ is quadratic, it is non-negative and
$\gamma_0 > 0$ holds generically. The condition $e_0 < 2 e_1$
states that the energies of the excited states are closer to the
continuum spectrum than to the ground state energy. It guarantees
that, for each $k>0$, $2 e_k-e_0>0$ becomes a resonance in the
continuum spectrum of $H_0$. ($H_0 + e_0 -2 e_k$ is not invertible
in $L^2$.) This resonance produces the main relaxation/growth
mechanism. If this condition fails, the resonance occurs in higher
order terms and a proof of relaxation will be much more
complicated. Also, the rate of decay will be different.
Assumption A3 is a new condition to avoid direct resonance between
the eigenvalues. It is trivial if $N=0,1$. It holds true
generically and is often seen in dynamical systems of ODE's. See
Example 2 at the end of this section for what may happen if this
assumption fails. If we relax the assumption $e_0 < 2 e_1$, we may
need to increase $j_{\text{max}}$.

Denote by $L^2_r$ the weighted $L^2$ spaces ($r$ may be positive
or negative),
\begin{equation} \label{L2r.def}
L^2_r(\R^3)= \bket{\phi \in L^2(\R^3): (1+x^2)^{r/2} \phi \in L^2
(\R^3)}.
\end{equation}
Fix $r_1>3$ large enough, to be determined by estimates
\eqref{eq:22-1B} and \eqref{eq:32-2}.
We denote by $L^p \loc$, $p=1,2$, the local $L^p$ spaces given by the norm
\begin{equation} \label{Lploc.def}
\norm{\phi}_{L^p \loc(\R^3)} \equiv
\bket{\int_{\R^3} (1+|x|)^{-2r_1} |\phi|^p d^3 x}^{1/p}.
\end{equation}

Now we are ready to state our main theorem.

\begin{theorem} \label{th:1-1}
Suppose the assumptions A0--A3 on $H_0=- \Delta+V$ hold and
let $\epsilon>0$ be any small constant.
There is a
small constant $n_0>0$ such that the following holds. Let
$\psi(t,x)$ be a solution of \eqref{Sch} with initial data $\psi_0$
decomposed as
\[
\psi_0 = x_0^0 \phi_0 +   x_1^0 \phi_1 + \cdots +  x_N^0 \phi_N + \xi_0
\]
with respect to $H_0$, where $x_k^0 \in \Complex$,
$k=0,1,\ldots,N$, and $\xi_0 \in \Hc(H_0)$. Suppose that
\begin{equation}  \label{psiz.asp}
\norm{\psi_0}_Y = n, \ 0 <  n \le  n_0 , \qquad |x_0^0| \ge
n^{3-\epsilon} , \qquad \norm{\xi_0}_Y  \le n/2,
\end{equation}
where $Y$ is the space for initial data,
\begin{equation} \label{Y.def}
Y \equiv  H^{1}\cap L^1 (\R^3).
\end{equation}
Then there exists an
$E_\infty$ with $\norm{Q_{E_\infty}}_{L^2}\sim n$ and a real
function $\Theta(t) = - E_\infty t + O(\log t)$  such that
\begin{equation}
\norm{\psi(t)-Q_{E_\infty} e^{ i \Theta(t)}}_{L^2 \loc} \le
C_2(1+t)^{-1/2} ,
\end{equation}
for some constant $C_2>0$ depending on $n$. Suppose, furthermore,
$\max_{k=1}^N|x_k^0|\ge n /100$, we also have a lower bound
\begin{equation}
\norm{\psi(t)-Q_{E_\infty} e^{ i \Theta(t)}}_{L^2 \loc} \ge
C_1(1+t)^{-1/2} ,
\end{equation}
for some constant $C_1>0$ depending on $n$.
\end{theorem}

The condition $ |x_0^0| \ge n^{3-\epsilon}$ is certainly not optimal
and can be greatly relaxed. It ensures that $\psi_0$ is away from
nonlinear excited states and that the dispersion component, $\xi_0$, is not
extremely large compared with $x_0^0$. This
condition, however, still allows the ground state component to be
much smaller than other components and exhibit the main phenomena.

Depending on the relative sizes of the bound states, there are
three regimes:
\begin{enumerate}
\item[I.] when an excited state is dominant;
\item[II.] when the ground state and the excited states are comparable;
\item[III.] when the ground state is dominant.
\end{enumerate}
Because the dominant terms are different in
different regimes, the natural linear
operators and the corresponding decompositions of the wave function
are different.
In regime II we can use $H_0$ as the linear operator and decompose
a function $\psi \in L^2$ as
\begin{equation} \label{1-6}
\psi = x_0 \phi_0 + \cdots + x_N \phi_N + \xi,
\end{equation}
with $x_k \in \Complex$ and $\xi \in \Hc(H_0)$. When the function
$\psi$ is close to a nonlinear ground state $Q_E$, i.e., in regime III,
it is natural to use $\L=\L_E$, the linearized operator around $Q_E$,
%
\begin{equation}
\L h = -i \bket{ (-\Delta + V -E + 2 \la Q^2)\,h + \la Q^2 \,\wbar
h\, },
\end{equation}
and decompose the wave function according to the spectral
decomposition with respect to $\L$. Notice that $\L$ is not
self-adjoint. Similarly, we can use linearized operators around
excited states in regime I.

The picture of the dynamics is as follows.
Suppose the ground state component is initially of
order $n^{3-\epsilon}$ and the solution is in regime II. The ground
state component will gradually grow while the excited states gradually
decay, until the solution enters regime III, i.e., when the ground
state component becomes much larger than other components.
This time interval is called the {\it transition regime}.
After entering regime III,  the solution will converge locally to
some nonlinear ground state as time tends to infinity,
with the excited state components vanishing and
the dispersion component escaping to infinity.
This time interval is called the {\it stabilization regime}.

As indicated above, we will use different operators and coordinate
systems for these two regimes. Besides technical problems
associated with changing coordinate systems, there is an intrinsic
difficulty related to the time reversibility of the Schr\"odinger
equations; It is not sufficient to control only the usual Sobolev
space norms at the time of coordinates changing. To overcome this
difficulty, we will use a concept introduced in \cite{TY2,TY4},
the {\it out-going estimates}, to capture the time-direction
sensitive information of the dispersive waves.

\donothing{ A technical complication occurs in dividing the two
regimes. Denote by $n$ and $z$ the sizes of the ground state and
excited states, respectively. In \cite{TY,TY2}, the analysis in
the stabilization regime can be extended to the region $z \le \epz
n$, for some constant $\epz \ll 1$. In our case, it is only valid
in the region
$z \le n^{1+\delta}$, for some $\delta >0$. To push $\delta \to
0+$, the constant $j_{\text{max}}$ in Assumption A3 has to go to
infinity. It will not solve the problem even if we assume
$j_{\text{max}}=\infty$, since our method requires a uniform gap
between any two numbers in the set $\bket{\sum_{j=1}^m (e_{k_j} -
e_{l_j}):m\le j_{\text{max}}}$. Because it takes a much longer
time in the transition regime to reach the region $z \le
n^{1+\delta}$ than the region $z \le \epz n$, finer estimates are
needed. }  

As it will be seen from examples below, there are energy transfers from
higher modes to lower modes. Although all excited states eventually vanish,
in an intermediate time an excited state may actually grow
because it gains more energy than it loses.
This complicates the analysis: one cannot prove the decay
of each individual excited state for all time. Instead, we have derived some
{\it monotonicity formulas} for their sum, see \eqref{4-52}.
Because the energy transfer between excited states is relatively small in the
stabilization regime, the phenomenon mentioned above is only apparent in
the transition regime.

We now give two examples illustrating some phenomena of
many bound states, in particular the relaxation/growth mechanism.
We first recall the concept of {\it phase factor}.
In our analysis it is essential to identify
the main oscillation factors of various terms. If we decompose the
solution $\psi(t)$ according to \eqref{1-6},  the
equation for a component $x_k(t)$ is
\begin{equation} \label{1-15}
i \dot x_k = e_k x_k + (\phi_k, \la \psi^2 \bar \psi).
\end{equation}
From the linear part we find that $x_k(t)$ has an oscillation
factor $e^{-ie_kt}$. We write $x_k(t)=e^{-ie_kt} u_k(t)$ and
say that its phase factor is $-e_k$. We will talk about phase
factors of polynomials in $x_k$ in a similar way. For example, the
phase factor of $x_lx_m \bar x_j$ is $-e_l-e_m+e_j$.

\bigskip

{\bf Example 1. Three bound states case in transition regime}

Suppose $H_0$ has three bound states and denote $x(t)=x_0(t)$,
$y(t)=x_1(t)$ and $z(t)=x_2(t)$.
The leading terms of $\xi$ are generated by bound states and are
cubic in $x_j$. Those with resonant coefficient functions, i.e.,
those with {\it negative phase factors}, give the relevant part of
$\xi$,
\begin{equation} \label{1-16}
\xi = (y^2 + z^2 + y z) \bar x + z^2 \bar y + \cdots
\end{equation}
Here we have ignored their coefficient functions, which are
non-local and complex because of the resonance with the continuous
spectrum. Denote $\chi = \ssum_{j=0}^N x_j \phi_j = x\phi_0 +
y\phi_1 + z \phi_2$. The main terms of \eqref{1-15} for $x_k$ are
\begin{equation} \label{1-17}
i \dot x_k = e_k x_k + (\phi_k, \, \la \chi^2 \bar \xi + 2 \la \chi \bar \chi
\xi) + \cdots,
\end{equation}
where
$\cdots$ denotes irrelevant terms.
Substituting \eqref{1-16} into \eqref{1-17},  we get
\begin{align} \dot x  &= \bkt{ \gamma_{22}^0 |z|^4
+ 4 \gamma_{12}^0|y|^2 |z|^2 + \gamma_{11}^0 |y|^4} \,x + \cdots
\nonumber
\\
\dot y &=\bkt{ \gamma_{22}^1 |z|^4 - 4 \gamma_{12}^0 |x|^2 |z|^2 -
2 \gamma_{11}^0 |x|^2|y|^2} \,y+ \cdots \label{model1}
\\
\dot z &=\bkt{ - 2 \gamma_{22}^0|x|^2 |z|^2 - 4 \gamma_{12}^0
|x|^2|y|^2 - 2 \gamma_{22}^1 |y|^2 |z|^2} \, z + \cdots \nonumber
\end{align}
Here $\cdots$ denotes irrelevant and error terms;
$\gamma_{11}^0$, $\gamma_{12}^0$, $\gamma_{22}^0$ and
$\gamma_{22}^1$ are non-negative constants to be defined later in \eqref{4-26}.
These constants
(except $\gamma_{22}^1$, depending on whether $e_1 - 2e_2 <0$) are
generically positive. They are the nonlinear analogue of the {\it Fermi
golden rule}, extensively studied in, e.g.,
\cite{Sigal,SW2,BP2,BS,C2,TY3}. The irrelevant terms have two
kinds. The first kind consists of terms with different phase
factors. They have few effect averaging over time and can be
removed using a {\it normal form} procedure, see Lemma
\ref{th:4-2}. The second kind consists of terms with same phase
factors but with purely imaginary coefficients. They only
contribute to the phase of $x_k$, not to the magnitude.
To illustrate further, let us assume $\gamma_{11}^0=\gamma_{12}^0
=\gamma_{22}^0=\gamma_{22}^1=1$ and denote $A=|x|^2$, $B=|y|^2$,
$C=|z|^2$. By \eqref{model1} we have
\begin{equation}\label{model2}
\begin{split}
\dot A /2&= 4 ABC + AB^2 + AC^2+ \cdots\\
\dot B /2&= -4 ABC -2 AB^2 + BC^2+ \cdots\\
\dot C /2&= -4 ABC -2 AC^2 -2 BC^2+ \cdots
\end{split}
\end{equation}
where $\cdots$ denotes irrelevant terms.
Although the above system is accurate only in the transition
regime, corresponding equations for other regimes are similar. From
\eqref{model2} we can read the energy transfers
from higher modes to lower modes. Only half of the energy decrease
in a higher mode goes to lower modes, while the other half goes to
dispersion (radiation). The lowest mode (ground state) is only
receiving energy while the highest mode ($x_N$) is only losing
energy. The intermediate modes receive energy from higher modes
and release energy to lower modes and radiation.

Richer phenomena occur when the constants $\gamma_{lm}^j$ have
different sizes. For example, suppose $\gamma_{22}^1 $ is much
larger than other constants and initially the energy is
concentrated in the third mode $z$. Then the second mode $y$ will
first grow exponentially, acquiring energy from $z$, and then
gradually decay.
Another phenomenon to be noticed is the interaction between three
modes, the $ABC$ terms in \eqref{model2}. They have the same coefficients
$\pm 4 \gamma_{12}^0$ in \eqref{model1}. Hence this interaction is more
apparent when $\gamma_{12}^0$ is much larger than other coefficients.
This phenomenon is
present only for many bound states case, $N\ge 2$.

\bigskip

{\bf Example 2. No-resonance assumption A3 violated}

Still assume $e_0 < 2e_1$ and three bound states.
The only possibility for assumption A3 to fail is
\begin{equation} \label{model3}
e_0 + 2 e_2 = 3 e_1.
\end{equation}
An example is $e_0 = -10$, $e_1=-4$ and $e_2=-1$. Note that
\begin{equation}
e_1 - 2 e_2 = e_0 - 2 e_1 < 0.
\end{equation}
Hence $\gamma_{22}^1$ is generically positive.
Because of \eqref{model3}, when we substitute \eqref{1-16} into
\eqref{1-17} we get new resonant terms in \eqref{model1}:
\begin{align}
\dot x  &=  [\cdots]x + y^2 \wbar {(z^2 \bar y)}
+ \cdots \nonumber
\\
\dot y &= [\cdots]y + z^2 \wbar {(y^2 \bar x)} + x \bar y (\bar y z^2) +
\cdots \label{model5}
\\
\dot z &= [\cdots]z + y \bar z (\bar x y^2) + \cdots \nonumber
\end{align}
The first group of terms on the right side are those in
\eqref{model1}. The cubic terms in the parentheses $()$ are from
\eqref{1-16}. For example, $y^2 \wbar {(z^2 \bar y)}$ has the
phase factor $-2e_1 - (-2e_2 + e_1) = -e_0$, the same as $x$, due
to \eqref{model3}. Although these new terms on the right side have
the same phase factors as the left side, their phases are not
exactly the same. Moreover, their coefficients are not quadratic
and it is unclear how to determine their signs. Hence it is
difficult to predict the dynamics of this system.

\donothing{ The rest of this paper is organized as follows. In
section 2 we collect results about nonlinear bound states and the
linearized operators. In section 3 we study the dynamics in the
stabilization regime. In section 4 we study the dynamics in the
transition regime.}

%
%

\section{Preliminaries}

We first fix the notation. Let $H^k$ denote the
Sobolev spaces $W^{k,2}(\R^3)$.
The weighted Sobolev space $L^2 _r(\R^3)$ is
defined in \eqref{L2r.def}. Denote by $\conj$ the conjugation operator.
The $L^2$ inner product $( \; , \; )$
is
\begin{equation} \label{L2.ip}
(f,g) = \int _{\R^3} \, \bar f \, g \; \, d^3 x .
\end{equation}
For a function $\phi \in L^2$, denote by $\phi^\perp$ the $L^2$-subspace
$\bket{g \in L^2: (\phi,g)=0}$.

In what follows we collect some facts about nonlinear bound states
and linear analysis. Their proofs can be found in
\cite{TY,TY2,TY3}. Although the proofs there are for two bound
states case, the proofs for the many bound states case are the
same.

\subsection{Nonlinear bound states}

Recall that nonlinear bound states of the equation \eqref{Sch} are
solutions of \eqref{Q.eq}. They are critical points of the energy
functional ${\mathcal H}[\psi]$ defined in \eqref{1-2}, subject to
the constraint of fixed $L^2$-norm. For each such solution $Q_E$,
the function $\psi(t,x) = Q_E(x) e^{-iEt}$ is an exact solution of
\eqref{Sch}. Since we have $N+1$ simple eigenvalues, we have $N+1$
families of corresponding nonlinear bound states. The existence
and basic properties of these  nonlinear bound states are
summarized in the following lemma. They are proven in
\cite{TY,TY2} using a contraction mapping argument.

\begin{lemma} \label{th:2-1}
Suppose $-\Delta+V$ satisfies assumptions A0--A1. Let $ n_0$ be
sufficiently small. For each eigenvalue $e_k$ with normalized
bound state $\phi_k$, $k=0,1,\ldots, N$, there is a family of
nonlinear bound states $\bket{Q_{k,E_k}}_{E_k}$ to \eqref{Sch} for
$E_k$ between $e_k$ and $e_k+ \la  n_0^2$ such that $Q_{k,E_k}$
are real, localized, smooth, and $ \la^{-1}(E_k-e_k)>0$,
\[
Q_{k,E_k} = n \,\phi_k +h, \qquad h \perp \phi_k,\quad h= O(n^3)
\text{ in } H^2,
\]
where $n = [(E_k-e_k)/(\la \int \phi_k^4 \, dx)]^{1/2}$.
Moreover, we have $\pd_{E_k} Q_{k,E_k} = O(n^{-2} ) \, Q_{k,E_k} +
O(n  )= O(n^{-1} )$, and $\pd_E^2 Q_{k,E} = O(n^{-3} )$.
For ground states we will drop the subscript and
write $Q_E$ and $R_{E}= \pd_{E} Q_{E}$. We have
$R_E = C n^{-2} Q_E + O(n)$. If we define
$c_1\equiv (Q_{E},R_{E})^{-1}$, we have $c_1 = O(1)$ and $\la c_1 >0$.
\end{lemma}

The following lemma summarizes the renormalization results near
ground states. They are proven in \cite{TY} using implicit
function theorem.
The $L^2$-subspace $M=M_E$ will be defined in Lemma \ref{th:L} (8).

\begin{lemma} \label{th:2-2}
Let $Y_1=Y$, defined in \eqref{Y.def}, or $Y_1 = L^2 \loc$,
defined in \eqref{Lploc.def}. There are small constants $n_0>0$ and
$\epz >0$ such that the following hold.
Suppose $\psi \in Y_1$ is close to
a nonlinear ground state  $Q_{E} \, e^{i \Theta} $ with
$\norm{\psi}_{Y_1} = n $,
$ \norm{\psi-Q_{E} \, e^{i \Theta}}_{Y_1}\le \tau n$,
 $0< n\le  n_0$,
$0< \tau\le \epz $.

(1) There are unique small $a,\theta \in \R$ and $h\in M_E$ such that
\begin{equation}\label{eq:2-6}
\psi = \bkt{Q_E + a R_E + h}\, e^{i(\Theta+\theta)}.
\end{equation}
Moreover, $\norm{Q_E}_{Y_1}\sim n$,  $a= O(\tau n^2)$, $h=O(\tau  n)$ and
$\theta=O(\tau )$.

(2) \textbf{(best approximation)}
There is a unique $E_*$ near  $E$  such that the component
along the $R_{E_*}$ direction as defined by  \eqref{eq:2-6} vanishes,
i.e., there are unique small  $\theta_* \in \R$ and $h_* \in M_{E_*}$
such that
\[
\psi = \bkt{Q_{E_*} +  h_*}\, e^{i  (\Theta_1+\theta_*)} .
\]
Moreover, $E-E_*=O(\tau  n^2)$,  $h_*=O(\tau n)$
and $\theta_*=O(\tau)$.
%

(3) Suppose $E'=E + \gamma$ with $|\gamma|\le  \tau^2 n^2 $.
By part (1) we can rewrite $\psi$ uniquely with respect to $E'$ as
\begin{equation*} 
 \psi = \bkt{Q_{E'} + a' R_{E'} + h'}\, e^{i(\Theta+\theta')},
\end{equation*}
where $ h' \in M_{E'}$; $h'$,  $a'$ and $\theta'$
are small. We have the estimates
\begin{equation}\label{shift}
E  + a  - E' - a' = O( \tau \gamma ), \quad
h  - h' = O( n^{-1} \tau \gamma ), \quad
\theta -\theta' = O( n^{-2} \tau \gamma ).
\end{equation}
Notice that $ n^{-2} \tau  \gamma \le C \epz ^3$ is small.
\end{lemma}

\subsection{Linear analysis}

We first recall some local decay estimates for $e^{-itH_0}$.
The decay estimate \eqref{eq:22-1A} is proved in  \cite{JSS,Y}
using estimates in \cite{JK,Ra}.
The estimate \eqref{eq:22-1B} is taken from \cite{SW2,TY}.
The estimate \eqref{eq:22-1B} holds
only if we take $r\to 0+$, not $r\to 0-$.

\begin{lemma}[decay estimates for $e^{-itH_0}$] \label{th:2-3}
Suppose that $H_0=-\Delta+V$ satisfies the Assumptions A0--A2.
For $q \in [2, \infty]$ and
$q'=q/(q-1)$,
\begin{equation}  \label{eq:22-1A}
\norm{  e^{-itH_0} \, \Pc  \phi }_{L^q} \le C \,|t|^ {-3
\bke{\frac 12 - \frac 1q}} \norm{\phi}_{L^{q'}}  .
\end{equation}
For sufficiently large $r_1$, for all $k,l,m \in \bket{0,\ldots,N}$,
we have
\begin{equation} \label{eq:22-1B}
\lim_{r\to 0+} \norm{ \bka{x}^{-r_1} \, \frac {e^{-it H_0}}
{(H_0  + e_k -  e_l - e_m - r i)} \Pc  \bka{x}^{-r_1} \phi
}_{L^2} \le C \bka{t}^{-3/2} \norm{\phi}_{L^2}.
\end{equation}

\end{lemma}

We now consider the linearized operators around nonlinear ground
states. Let $Q=Q_{E}$ be a nonlinear ground state with
$\norm{Q_{E}}_{L^2}$ small. If we consider
solutions $\psi(t,x)$ of \eqref{Sch} of the form
\[
\psi(t,x) = \bkt{ Q_E(x) + h(t,x)} \, e^{-i E t},
\]
with $h(t,x)$ small in a suitable sense, then $h(t,x)$ satisfies
\[
\pd_t h = \L h + \text{nonlinear terms,}
\]
where the linearized operator $\L =\L_E$ is defined by
\begin{equation} \label{L.def}
\L h = -i \bket{ (-\Delta + V -E + 2 \la Q^2)\,h + \la Q^2 \,\wbar
h\, }.
\end{equation}

The properties of $\L$ are best understood in the complexification of
$L^2(\R^3, \Complex)$.

\begin{definition} \label{CL2}

Identify $\Complex$ with $\R^2$ and $L^2=L^2(\R^3, \Complex)$ with
$L^2(\R^3, \R^2)$. Denote by $\Complex L^2=L^2(\R^3, \Complex^2)$
the complexification of $L^2(\R^3, \R^2)$. $\Complex L^2$ consists
of 2-dimensional vectors whose components are in $L^2$. We have
the natural embedding
\[
\mb{j}: f \in L^2 \longrightarrow  \vect{ \Re f \\ \Im f} \in
\Complex L^2.
\]
We equip $\Complex L^2$ with the natural inner product: For $f,g
\in \Complex L^2$, $f=\svect{f_1\\f_2}$, $g=\svect{g_1\\g_2}$, we
define
\begin{equation} 
(\!(f,g)\!) = \int_{\R^3} \bar f \cdot g \,d^3x = \int_{\R^3} (\bar
f_1 g_1 + \bar f_2 g_2)\,d^3x .
\end{equation}
Denote by $\RE$ the operator first taking the real part of
functions in $\Complex L^2$ and then pulling back to $L^2$:
\[
\RE: \Complex L^2 \to L^2, \quad \RE \vect{f\\g} = (\Re f)+i(\Re
g).
\]
We have $\RE \circ \, \mb{j}=\mb{id}_{L^2}$.
\end{definition}

The operator $\L $ can be naturally extended to an operator acting
on $\Complex L^2$ with the following matrix form:
\begin{equation} 
\begin{bmatrix}0 & L_-\\
 -L_+ &0\end{bmatrix}
,\qquad \text{where } \left\{
\begin{aligned}
L_- &= - \Delta + V -E + \la Q^2\\
L_+ &= - \Delta + V -E + 3\la Q^2.
\end{aligned} \right.
\end{equation}

Recall the Pauli matrices
\[
\sigma_1 = \begin{bmatrix}0&1\\1&0\end{bmatrix},\quad \sigma_2 =
\begin{bmatrix}0&-i\\i&0\end{bmatrix},\quad \sigma_3 =
\begin{bmatrix}1&0\\0&-1\end{bmatrix}.
\]
They are self-adjoint. We have $\RE \L = \L \RE$ and
\begin{equation}\sigma_1 \L =\L^*
\sigma_1, \qquad \sigma_3 \L = - \L\sigma_3 ,
\end{equation}
where $\L^*$ has the matrix form $\bkt{\begin{smallmatrix}0& -
L_+\\L_-&0\end{smallmatrix}}$.

We summarize the properties of $\L$ in the following lemma,
whose proof is the same as that in \cite{TY} and
\cite[Theorem 2.1]{TY3}. For
convenience of notation, we identify $L^2$ as a subspace of
$\Complex L^2$ and make no difference between $\psi \in L^2$ and
$\mb{j}(\psi)\in \Complex L^2$.

\begin{lemma} [spectral properties] \label{th:L}
Suppose the assumptions A0-A1 hold. Let $Q=Q_{E}$ be a nonlinear
ground state, $\norm{Q_E}_{L^2} = n$, $0<n\le n_0$. Let $\L=\L_E$
be defined as in \eqref{L.def}.

(1) The eigenvalues of $\L  $ are  $0$ and $\pm i \ev_k $,
$k=1,\ldots, N$, where $\ev_k = e_k - e_0 + O(n^2)$ are real and
positive. All eigenvalues are simple except $0$ which has
multiplicity two. The continuous spectrum of $\L$ is
\begin{equation}
\Sigma_c=\bket{ s i: \, s \in \R, |s| \ge |E|}.
\end{equation}
There is no embedded eigenvalue. The bottoms of the continuous
spectrum, $\pm i E$, are not eigenvalue nor resonance.

(2) The $0$-eigenspace is spanned by $\svect{0\\Q}$ and
$\svect{R\\0}$. Note $\L \svect{0\\Q} = 0$ and $\svect{R\\0}$ is a
generalized $0$-eigenvector with $\L \svect{R\\0}=-\svect{0\\Q}$.
We denote
\[
S = \myspan_\R \bket{\vect{0\\Q},\vect{R\\0}} \subset L^2 ,\quad
(\text{or } S=\myspan_\R \bket{i Q, R}).
\]

(3) For each eigenvalue $i\ev_k$, $k=1,\ldots,N$, there is an
eigenvector $\evector_k$ of the form $\evector_k = \svect{u_k \\
-i v_k}$, where $u_k$ and $v_k$ are real-valued $L^2$-functions
satisfying
\[
L_+ u_k = \ev_k v_k, \qquad L_- v_k = \ev_k u_k, \qquad
(u_k,v_k)=1.
\]
Moreover, they are perturbations of $\phi_k$: $u_k,v_k=\phi_k +
O(n^2)$. $\wbar \evector_k = \svect{u_k \\ i v_k}$ is an
eigenvector of $-i\ev_k$. We denote the combined eigenspaces of
$\pm \ev _k$ as
\[
\Complex \eigen_k = \myspan_\Complex \bket{\evector_k,\wbar
\evector_k} \subset \Complex L^2,\qquad \eigen_k = \myspan_\R
\bket{\vect{u_k\\0},\vect{0\\v_k}} \subset L^2.
\]

(4) The continuous spectrum subspace, $\Hc(\L)$, is equal to
\begin{equation*} 
\Hc(\L) = \bket{ \psi \in L^2: (\!(\sigma_1 \psi, f)\!)=0 , \
\forall f \in S \oplus \eigen_1 \oplus \cdots \oplus \eigen_N}.
\end{equation*}

(5) The space $L^2(\R^3,\Complex)$, as a real vector space, can be
decomposed as the direct sum of $N+2$ $\L$-invariant subspaces:
\begin{equation} \label{L2.dec}
L^2(\R^3, \Complex) = S \oplus \eigen_1 \oplus \cdots \oplus
\eigen_N \oplus \Hc(\L).
\end{equation}
For any $f$ and $g$ belonging to two different invariant
subspaces, we have the orthogonality relation
\begin{equation}\label{ortho}
(\!(\sigma_1 f,\, g)\!)=0.
\end{equation}

(6) For any function $\zeta_k \in \eigen_k$, $k=1,\ldots,N$, there
is a unique $\al_k \in \Complex$ so that
\[
\zeta_k = \RE \al_k \evector_k.
\]
Since $\L \RE = \RE \L$, we have
\[
\L  \zeta_k =  \RE i \ev_k \al_k \evector _k ,\qquad e^{t\L }
\zeta_k = \RE e^{ti \ev_k  } \al_k \evector_k .
\]

(7) By the orthogonality relation \eqref{ortho}, any $\psi \in
L^2$ can be decomposed with respect to \eqref{L2.dec} as
\begin{equation} 
\psi = aR +biQ + \sum_{k=1}^N \RE \al_k
\evector_k + \eta,
\end{equation}
with $\eta \in \Hc(\L )$, $\al_k = c_k + i d_k$,
\begin{equation}  
\begin{gathered}
a=(Q,R)^{-1}(Q,\Re \psi),\\
b=(Q,R)^{-1}(R,\Im \psi),
\end{gathered}
\qquad
\begin{gathered}
c_k=(u_k,v_k)^{-1}(v_k,\Re \psi),\\
d_k=(u_k,v_k)^{-1}(u_k,\Im \psi).
\end{gathered}
\end{equation}
Note $(Q,R)^{-1}=c_1=O(1)$ and $(u_k,v_k)^{-1}=1$.

(8) Let $M=\eigen_1 \oplus \cdots \oplus \eigen_N \oplus \Hc(\L)$.
We have $L^2(\R^3, \Complex) = S \oplus M$ and $M = \vect{Q^\perp
\\ R^\perp}$. For $m=0,1,2$, there is a constant $C>1$ such that, for all
$\phi \in M \cap H^2$ and all $t\in \R$, we have
\begin{equation}
C^{-1}\norm{\phi}_{H^m} \le \norm{e^{t\L}\phi}_{H^m} \le C \norm{\phi}_{H^m} .
\end{equation}

\end{lemma}

As in \cite{TY}, in order to prove various estimates and make
explicit computations, we will introduce an $L^2$-subspace ${\bf
X}$ and two operators $A:{\bf X}\to {\bf X} $ and $U:M\to {\bf X}$
so that
\begin{equation} \label{22-11}
    \L \, \big|_M \;  = \; U^{-1} (-i) A U  .
\end{equation}
Explicitly, let ${\bf X}$ be the $L^2$-subspace orthogonal to $Q$:
\begin{equation} \label{X.def}
{\bf X}=\Pi(L^2) = \bket{\phi \in L^2(\R^3): \phi \perp Q} ;
\qquad  {\bf X} \corr \vect{Q^\perp\\Q^\perp},
\end{equation}
where $\Pi$ is the orthogonal projection which eliminates
$Q$-direction: $\Pi h = h - \frac {(Q, h)}{(Q,Q)}\, Q$. Let $P_M$
be the projection (not orthogonal) from $L^2$ onto $M$ according
to the decomposition $L^2(\R^3) = S \oplus M $. $P_M$ has the
matrix form $\begin{bmatrix} P_1  & 0 \\ 0 & P_2 \end{bmatrix}$,
where the projections $P_1$ and $P_2$ are given by ($c_1=(Q,R)^{-1}$)
\begin{equation}\label{eq:22-14}
\begin{split}
P_1&: L^2 \longto Q^\perp, \quad P_1= \mb{id} - c_1 |R\rangle
\,\langle Q|  ,
\\
P_2&: L^2 \longto R^\perp, \quad P_2= \mb{id} -  c_1|Q\rangle
\,\langle R|  .
\end{split}
\end{equation}
Clearly $P_1 R=0$ and $P_2 Q=0$. One can check easily that the maps
\begin{equation}
\label{eq:22-15} R_{MX}\equiv
\begin{bmatrix} I & 0  \\
0 &\Pi \end{bmatrix} :M \longto {\bf X} \; ,  \qquad R_{XM}\equiv
\begin{bmatrix} I & 0  \\
0 & P_2 \end{bmatrix} :{\bf X} \longto M,
\end{equation}
are inverse to each other.
We now define $H= L_-$ and
\begin{equation} \label{A.def}
 A\equiv\bkt{(H^2 + H^{1/2} \Pi 2 \la Q^2 \Pi H^{1/2}) }^{1/2}
 = \bkt{ H^{1/2} L_+ H^{1/2} }^{1/2} .
\end{equation}
$A$ is a self-adjoint operator acting in $L^2(\R^3)$, with $Q$ as
a $0$-eigenvector. We shall often view $A$ as an operator
restricted to its invariant subspace ${\bf X}$. Define
\begin{equation} \label{U0.def}
U_0: {\bf X} \longto {\bf X}, \qquad
U_0\equiv \begin{bmatrix}  A^{1/2} H^{-1/2}  & 0 \\
0 &  A^{-1/2} H^{1/2}  \end{bmatrix}  ,
\end{equation}
and let
\begin{equation} \label{U.def}
U\equiv U_0 R_{MX} :M \longto {\bf X}  ,  \qquad U^{-1} \equiv
R_{XM} U_0^{-1}:{\bf X} \longto M  .
\end{equation}
Notice that $H^{-1/2}$ is defined  only in $Q^\perp$. We summarize
the properties of $A$ and $U$ in the following lemma.

\begin{lemma} [Similarity relation] \label{th:AU}
(1) Let ${\bf X}$, $A$, $U$ and $U^{-1}$ be defined as in
\eqref{X.def}, \eqref{A.def} and \eqref{U.def}, respectively. Then
\eqref{22-11} holds in $M$. If we denote by $P^A_{k}$,
$k=1,\dots,N$, and $\PcA$ the orthogonal projections onto the
eigenspaces and continuous spectrum subspace of $A$, we have
\begin{equation}
U P^\L_{k } = P^A_{k } U P_M, \qquad U\PcL = \PcA U P_M .
\end{equation}

(2) The operators $U:M \to {\bf X}$ and $U^{-1}:{\bf X} \to M$ are
bounded in $W^{k,p}$ and $L^2_r$ norms for $k=0,1,2$, $1\le p <
\infty$, and $|r|\le r_1$. Here $r_1>0$ is determined by
\eqref{eq:32-2} later. The operator $U-1$ is local and bounded by
$n^2$, and hence so is $[U,i]=[U-1,i]$, in the sense that
\begin{equation} \label{Ulocal.est}
\norm{[U,i]\phi}_{L^{1}\cap L^{5/4}} \le C n^2 \norm{\phi}_{L^5}.
\end{equation}
We have
\begin{equation}\label{Upm.dec}
U=U_+ + U_- \conj, \qquad U^{-1}=U_+^* - U_-^* \conj,
\end{equation}
where $\conj$ is the conjugation operator with the Pauli matrix
$\sigma_3$ as its matrix form, and the duals of $U_+$ and $U_-$ are
respect to the $L^2$ inner product \eqref{L2.ip}.
The operators $U_+$ and $U_-$ are
given by
\begin{align}
\label{Upm.def}
U_\pm &= \frac 12(\Pi A^{1/2} H^{-1/2}P_1 \pm \Pi A^{-1/2} H^{1/2} \Pi),
\\
\label{Upmstar.def} (U_\pm)^* &= \frac 12( P_2  H^{-1/2}A^{1/2}
\Pi \pm \Pi H^{1/2} A^{-1/2}\Pi).
\end{align}
In particular, $U_+=1 + O(n^2)$, $U_- = O(n^2)$. They are not
self-adjoint but they commute with $i$ and $\conj$.

(3) For $m=0,1,2$, there is a constant $C>1$ so that
\begin{equation}  \label{eq:32-0}
C^{-1} \norm{ \phi }_{H^m} \le \norm{e^{-itA} \phi }_{H^m} \le C
\norm{ \phi }_{H^m},
\end{equation}
for all $\phi \in {\bf X} \cap H^m$ and all $t \in \R$. For $q \in
[2, \infty]$ and $q'=q/(q-1)$,
\begin{equation}  \label{eq:32-1}
\norm{  e^{-itA} \, \PcA \Pi \phi }_{L^q} \le C \,|t|^ {-3
\bke{\frac 12 - \frac 1q}} \norm{\phi}_{L^{q'}}  .
\end{equation}
For sufficiently large $r_1>0$ and for all $k,l \in \bket{1,\ldots,N}$,
we have
\begin{equation} \label{eq:32-2}
\norm{ \bka{x}^{-r_1} \, e^{-itA} \, \frac 1{(A - 0i -\ev_k -\ev_l
) } \PcA \Pi \bka{x}^{-r_1} \phi }_{L^2} \le C \bka{t}^{-3/2}
\norm{\phi}_{L^2},
\end{equation}
where 
$0i$ means $r i$ with $\lim_{r \to 0+}$ outside of the norm. Finally,
\begin{multline}
\bke{ \phi_0\phi_k^2 \, , \, \Im \frac {1}{A -0i - 2\ev_k} \PcA
\Pi \phi_0\phi_k^2 } \\= \bke{ \phi_0\phi_k^2 \, , \, \Im \frac
{1}{H_0-E -0i - 2 \ev_k} \PcH \phi_0\phi_k^2 } + O(n^2) > 0  .
\end{multline}
\end{lemma}

Estimate \eqref{eq:32-1} for $A=- \Delta + V$ was proven in
\cite{JSS,Y} using estimates from \cite{JK,Ra}. Estimate
\eqref{eq:32-2} for $A=(-\Delta+V+m^2)^{1/2}$ was proven in
\cite{SW2}. Lemma \ref{th:AU} is a summary of \cite[Lemmas
2.5--2.9]{TY}. We omit the proof.

%
%

\section{Stabilization regime}

In this section we study the dynamics of the solution when it is
close to nonlinear ground states. We want to show that the solution
$\psi(t)$
converges to some nonlinear ground state locally as the time tends to
infinity. In this time regime, the natural decomposition of
$\psi(t)$ is
\begin{equation} \label{psi.dec2}
\psi(t) = \bkt{Q _E + a(t) R_E + \zeta(t) + \eta(t) } e^{-i Et + i
\theta(t)}
\end{equation}
with respect to a fixed $E$. Here $a(t),\theta(t) \in \R$,
$\zeta(t)=\zeta_1(t) +\cdots + \zeta_N(t)$, $\zeta_k \in \eigen_k (\L_E)$,
$k=1,\ldots, N$, and
$\eta(t) \in \Hc (\L_E)$; see Lemma \ref{th:L}.
Define
\begin{equation}\label{gamma0p.def}
\gamma_0^+\equiv \max_{\myfrac{1\le k,l \le N}{|s|<s_0}}
\lim_{r \to 0+} \Im \bke{ \phi_0\phi_k \phi_l, \frac
1{-\Delta+V+e_0-e_k-e_l-s-r i}\Pc \phi_0\phi_k \phi_l },
\end{equation}
and (recall $c_1=(Q,R)^{-1}$ and $\gamma_0$ is defined in \eqref{gamma0.def})
\begin{equation}\label{D.def}
D= 6 N |c_1| \gamma_0^+ /\gamma_0 = O(1).
\end{equation}
We will prove the following theorem.

\begin{theorem}\label{th:S-1}
Assume the assumptions A0--A3. There are small constant
$n_0,\epz>0$ such that the following holds. Suppose that the initial
data $\psi_0$ with $\norm{\psi_0}_{H^1} \ll 1$
is close to a nonlinear ground state $Q_{E_0}e^{i \Theta_0}$ in
$L^2\loc$-norm with $\norm{Q_{E_0}}_{L^2} = n \le n_{0}$, and that in
the decomposition \eqref{psi.dec2} of $\psi_0$ with $E=E_0$
one has
\begin{equation} \label{S-2}
\bke{\ssum_{k=1}^N \norm{\zeta_{k,E_0}}_{L^2}^2}^{1/2}
\le \tfrac 12 \rho_0, \qquad
|a_{E_0}| \le  D \rho_0 ^2, \qquad
\rho_0 \le \epz n.
\end{equation}
Suppose, furthermore, for all $E$
close to $E_0$ with $|E -E_0 |\le 3D \rho_0^2$, the dispersive part
$ {\eta_E} (0)$ in the decomposition \eqref{psi.dec2} satisfies
\begin{equation} \label{out:thS1}
\begin{split}
\norm{e^{s\L} \eta_E(0) }_{L^5} &\le n^{4/5} \rho(s) ^{8/5},
\\
\norm{e^{s\L} \eta_E(0) }_{L^2 \loc} &\le \Lambda(s) \equiv
(1+s)^{-1/2} \rho^2(s),
\end{split}
\end{equation}
for all $s \ge 0$,
where 
\begin{equation} \label{tls.def}
\rho(s) = \bkt{\rho_0^{-2}+  N^{-1} \gamma_0 n^2 s}^{-1/2}.
\end{equation}
Then there is a frequency $E_\infty$ with $|E_\infty -E_0|\le 3D \rho_0^2$
 and a function $\Theta(t)= - E_\infty t + O(\log(t))$ for
$t\in [0,\infty)$ such that, for  some constant $C_3>1$ independent of $n$,
\begin{equation}
 \norm{\psi(t) - Q_{E_\infty}e^{i \Theta(t)} }_{L^2 \loc}
\le C_3 \rho(t), \qquad (t\ge 0).
\end{equation}
Suppose, furthermore, that
$\bke{\ssum_{k=1}^N \norm{\zeta_{k,E_0}}_{L^2}^2}^{1/2} \ge \tfrac 14 \rho_0$.
Then we also have a lower bound
\begin{equation} \label{lowerbound}
C_3^{-1} \rho(t) \le \norm{\psi(t) -
Q_{E_\infty}e^{i \Theta(t)} }_{L^2 \loc}, \qquad (t\ge 0).
\end{equation}

\end{theorem}

We call Eq.~\eqref{out:thS1} the {\it out-going estimates} of
$\eta(0)$. The theorem holds true if they are replaced
by the following stronger but simpler assumption that
\begin{equation} \label{S-6}
\psi_0 \in Y \equiv H^1 \cap L^1(\R^3), \qquad
\norm{\eta(0)}_{Y}\le \rho_0^2,
\end{equation}
since \eqref{S-6} implies \eqref{out:thS1} by Lemma \ref{th:2-3}.
Eq.~\eqref{S-6} means that the data $\psi_0$ is localized. In
contrast, Eq.~\eqref{out:thS1} only requires the data to be
``out-going'' in some sense. This will be useful when we prove
Theorem \ref{th:1-1} using Theorem \ref{th:S-1} in section 4.

Our strategy of proof is as follows. For each $T>0$, we choose
$Q_{E(T)}$ to be the best approximation of $\psi(T)$ given by
Lemma \ref{th:2-2}. We will prove estimates for the components of
$\psi(t)$ in the decomposition \eqref{psi.dec2} with respect to
$E=E(T)$ for $t \in [0, T]$. We will then use a continuity
argument to show that $E(T)$ can always be chosen and we have
uniform estimates as $T \to \infty$. It then follows that $E(T)$
converges to some $E_\infty$ close to $E(0)$ as $T \to \infty$.

\subsection{Equations}

Let $Q=Q_E$ be a fixed nonlinear ground state with frequency $E$
near $e_0$, and $R=R_E=\partial_E Q_E$. We write the solution
$\psi(t,x)$ of \eqref{Sch} in the form
\begin{equation} \label{3-7}
\psi(t,x) = [Q_E(x)+a(t)R_E(x)+h(t,x)] \, e^{-iEt +i \theta(t)} ,
\end{equation}
where $a(t),\theta(t)\in \R$ and $h(t,.) \in M_E$. $M_E$ is
defined in Lemma \ref{th:L} (8). Substituting the ansatz
\eqref{3-7} into \eqref{Sch} and using $\L iQ =0$ and $\L R = -
i Q$, we get
\begin{equation} \label{S-8}
\pd_t h = \L h + i^{-1} (F +\dot \theta (Q+aR+h)) - a i Q- \dot a R.
\end{equation}
Here
\[
F =\la Q( 2|h_a|^2 + h_a ^2 ) + \la |h_a |^2 h_a, \qquad  h_a = aR
+ h.
\]

We want to choose $a(t)$ and $\theta(t)$ so that $h(t) \in M$,
that is, $h(0) \in M $ and $ i^{-1} (F +\dot \theta (Q+aR+h))- ai
Q -\dot a R \in M$. Since $M=\svect{ Q^\perp \\ R^\perp }$, $a(t)$
and $\theta(t)$ satisfy
\begin{align*}
 \bke{ Q, \Im (F +\dot \theta h) - \dot a R  } &=0 ,
\\
\bke{ R, \Re (F +\dot \theta (Q+aR+h)) -a Q} &=0 .
\end{align*}
Denote $c_1=(Q,R)^{-1}$. We have
\begin{align*}
\dot a &= (c_1 Q, \Im (F +\dot \theta h) ) ,
\\
\dot \theta &= -\bkt{a+ \bke{ c_1R,\, \Re F } }  \cdot  \bkt{1+
a(c_1R,R)+(c_1R,\Re h)}^{-1}.
\end{align*}
Eq.~\eqref{S-8} for $h$ becomes
\begin{equation} \label{S-9}
\pd_t h = \L h + P_M F_{all} ,\qquad F_{all} =  i^{-1} (F +\dot
\theta (a R +h)) .
\end{equation}

We decompose $h(t)$ with respect to the spectral decomposition
\eqref{L2.dec},
\[
h(t)=\zeta(t)+\eta(t) , \qquad \zeta=\zeta_1+ \cdots + \zeta_N,
\]
where $\zeta_k\in \eigen_{k}(\L)$, $ k=1 ,\ldots , N$, and $\eta
\in \Hc(\L)$. For each $\zeta_k(t)$ we associate a function
$z_k(t) \in \Complex$ by writing
$\zeta_k= \Re (z_k)\, u_k + \Im (z_k)\, iv_k
$. In other words, $\zeta_k (t) = \RE z_k(t) \evector _k$. If we
define $ u_k^\pm= \tfrac 12 ( u_k \pm v_k )$, we can write
\begin{equation}\label{zetak.dec}
\zeta_k= \Re (z_k)\, u_k + \Im (z_k)\, iv_k =z_k u_k^+ + \bar z_k
u_k^- .
\end{equation}
Projecting \eqref{S-9} to $\eigen_{k}(\L)$ we get
\begin{align*}
\dot z_k &= -i \ev _k \, z_k + (v_k, \Re F_{all}) + i(u_k, \Im
F_{all}) = -i \ev _k \, z_k + (u_k^+, F_{all}) -(u_k^-, \wbar
F_{all}).
\end{align*}
From the linear part, we identify the {\it phase factor} of $z_k$
as $-\ev_k$. Hence we define $p_k(t)=e^{i \ev _k t}z_k(t)$, which
has the same magnitude as $z_k$ but with no strong oscillation.
$p_k(t)$ satisfy
\begin{align*}
e^{-i \ev _k t}\, \dot p_k(t) &=(u_k^+, F_{all}) -(u_k^-, \wbar
F_{all})
\\
&=i^{-1}\bket{  (u_k^+, F)+(u_k^-, \wbar F)+ \bkt{(u_k^+,
h)+(u_k^-, \wbar h)+ (u_k,R) a} \dot \theta } .
\end{align*}
Also,
projecting \eqref{S-9} to $\Hc(\L)$ we get
$\pd_t \eta = \L \, \eta + \PcL F_{all}$.

Summarizing, for
\begin{align*}
\psi (t) &= (Q + a(t) R + h(t)) \, e^{-iEt+i \theta(t)},
\\
h  &= \zeta + \eta , \quad \zeta = \zeta_1+\cdots +\zeta_N , \quad
\zeta_k= z_k u_k^+ + \bar z_k u_k^- , \quad z_k=e^{-i \ev _k t}\,
p_k,
\end{align*}
we have
\begin{equation} \label{eq:all}
\begin{split}
&\dot a = (c_1 Q, \Im (F +\dot \theta h) ) ,
\\
&i e^{-i \ev _k t}\, \dot p_k =   (u_k^+, F)+(u_k^-, \wbar F)+
\bkt{(u_k^+, h)+(u_k^-, \wbar h)+ (u_k,R) a} \dot \theta ,
\\
&\partial _t \eta = \L \eta + \PcL i^{-1} (F +\dot \theta (a R +
h)) ,
\end{split}
\end{equation}
where $c_1=(Q,R)^{-1}$,
\begin{align}
\label{F.def} F &=
\la Q( 2|h_a|^2 + h_a ^2 ) + \la |h_a |^2 h_a, \qquad  h_a = aR+ h,
%
\\
\dot \theta &= F_\theta \equiv -\bkt{a+ \bke{ c_1R,\, \Re F } }
\cdot \bkt{1+ a(c_1R,R)+(c_1R,\Re h)}^{-1} . \label{Ftheta.def}
\end{align}
This is a system of equations involving $a$, $z_k$ and $\eta$
only. Note that $\theta$ enters
\eqref{eq:all} only via $\dot \theta=F_\theta$.
It will appear in the form $e^{i \theta}$  when we integrate $\eta$.
Hence we do not need estimates of $\theta$ for the proof.

For convenience, we will use the following convention.
\begin{convention} \label{convention}
For $k=1,2,\cdots , N$, denote
\begin{equation}  
 \ev_{-k} = - \ev _k , \qquad  z_{-k} = \bar z_k, \qquad
 p_{-k} = \bar p_k .
\end{equation}
We have $z_{k}(t) = e ^{-i \ev_k t} p_k(t)$ for both $k>0$ and
$k<0$. We also denote
\begin{equation}
\Omega = \bket{ \pm 1, \ldots, \pm N}.
\end{equation}
\end{convention}

\subsection{Decompositions of $F$, $a$ and $\eta$}

Most quantities in our system of equations are strongly oscillatory.
It is necessary to identify
their oscillatory parts before we can estimate. In this subsection
we identify the leading oscillatory terms of $a$ and $\eta$, and decompose
$F$ according to order.
We will treat $z_k$ and $a$ again in \S 3.3.
Note that
\begin{equation}\label{n.order}
Q=O(n), \quad R=O(n^{-1}), \quad c_1 = O(1), \quad
u_k^+ = \phi_k + O(n^2), \quad u_k^- = O(n^2).
\end{equation}
In fact, since $\norm{Q_{E_0}} _{L^2} = n$ and $|E-E_0| \le 3D \epz^2 n^2$,
we have  $\norm{Q_{E}} _{L^2} = [1+O(\epz^2)]n$ and
$Q=(1+O(\epz^2)) n \phi_0+O(n^3)$.
We will prove that
\begin{equation}\label{order}
|z_k(t)|\le  C t^{-1/2}, \quad |a(t)|+ \norm{\eta(t)}_{L^2\loc}\le C
t^{-1}, \qquad \text{as } t\to \infty.
\end{equation}
Hence the main term in $h_a= aR + \zeta+\eta$ is $\zeta$.
Therefore, the main part of $F$, defined in \eqref{F.def}, is
\begin{equation} \label{F2.def}
F_1 = \la Q( 2|\zeta|^2+ \zeta^2 ).
\end{equation}

\subsubsection{Decomposition of $a$}

We now identify the main oscillatory terms of $a(t)$. Recall from
\eqref{eq:all} that $ \dot a= (c_1Q, \Im F+\dot \theta h)$,
$c_1=(Q,R)^{-1}$. We shall impose the boundary condition of $a$ at
$t=T$, which is in fact a condition imposed on the choice of
$E(T)$. Hence we use the following equivalent integral equation:
\[
a(t) = a(T)+ \int_T^ t (c_1Q, \Im F+\dot \theta h)(s) \, ds .
\]
The main term of $\Im (F+\dot \theta h)$ is
$\Im F_1 = \Im \la Q\zeta^2 $.
Thus the main oscillatory terms of $a(t)$ are from the integral
$\int_T^t A^{(2)} \, ds$ with
 \[
A^{(2)} \equiv (c_1Q, \Im \la Q \zeta^2) =
(c_1\la Q^2, \Im \sum_{k,l=1}^N \zeta_k \zeta_l) .
\]
 Since
\begin{align*}
\Im  \zeta_k \zeta_l &= \Im (z_k u_k^+ + \bar z_k u_k^-)(z_l u_l^+
+ \bar z_l u_l^-) \\ &=\Im (z_k z_l)(u_k^+u_l^+ -u_k^-u_l^-) + \Im
(z_k \bar z_l)(u_k^+u_l^- -u_k^-u_l^+),
\end{align*}
 we have
\begin{equation*}
A^{(2)} = \sum_{k,l=1}^N \bket{ a_{kl,1}\Im (z_k z_l)  +
a_{kl,2}\Im (z_k \bar z_l) },
\end{equation*}
 where $a_{kl,1}=(c_1 \la Q^2,(u_k^+u_l^+ -u_k^-u_l^-))$ and
$a_{kl,2}=(c_1 \la Q^2,u_k^+u_l^- -u_k^-u_l^+)$ are real constants
bounded by $n^2$.
We can integrate by parts $A^{(2)}$ to get:
\begin{align} \nonumber
\int_T^t A^{(2)} \, ds &= \Im\sum_{k,l=1}^N \int_T^t a_{kl,1}(z_k
z_l)  + a_{kl,2}(z_k \bar z_l) \, ds
\\ \nonumber
&=\Im  \sum_{k,l=1}^N\int_T^t e^{-i(\ev _k+\ev _l)s} a_{kl,1}(p_k
p_l) + e^{-i(\ev _k-\ev _l)s} a_{kl,2}(p_k \bar p_l) \, ds
\\
&= \Im \sum_{k,l=1}^N \bkt{i a_{kl,3}z_k z_l+  i a_{kl,4}z_k \bar
z_l}_T^t - \int_T^t A_{2,rmd} (s) \, ds , \label{S-19}
\end{align}
where
\begin{align}
a_{kl,3}&=(\ev _k+\ev _l)^{-1}\, a_{kl,1} , \quad
a_{kl,4}=\delta_k^l \,(\ev _k-\ev _l)^{-1}\, a_{kl,2},
\\
\label{A2rmd.def}
A_{2,rmd} & = \Im \sum_{k,l=1}^N \bket{e^{-i(\ev _k+\ev _l)s} i
a_{kl,3}\frac d{ds}(p_k p_l) + e^{-i(\ev _k-\ev _l)s} i
a_{kl,4}\frac d{ds}(p_k \bar p_l) }.
\end{align}
Here we put $\delta_k^l$ in the definition of $a_{kl,4}$ to impose
$a_{kk,4}=0$. We have the factor $\delta_k^l$ in $a_{kl,4}$ since
$\Im z_k \bar z_l=0$ if $k=l$. Also note that $a_{kl,3}$ and
$a_{kl,4}$ are real constants
bounded by $n^2$.
%
%
The first part in \eqref{S-19} can be rewritten as
$a^{(2)}(t)-a^{(2)}(T)$, where
\begin{equation}
a^{(2)}=\Im \sum_{k,l=1}^N i a_{kl,3}z_k z_l+  i a_{kl,4}z_k \bar
z_l = \sum_{k,l\inomega} a_{kl} z_k z_l .
\end{equation}
Here we have used Convention \ref{convention}. The constants
$a_{kl} = \tfrac 12 a_{kl,3}$ if $k$ and $l$ have the same sign;
$a_{kl} = \tfrac 12 a_{kl,4}$ otherwise. In particular, $a_{kl}$
are real constants bounded by $n^2$. $a^{(2)}(t)$  contains the
main oscillatory part of $a$. We denote the rest of $a(t)$ by
$b(t)$,
\begin{equation} \label{a.dec}
   a(t) = a^{(2)}(t) + b(t) .
\end{equation}
Thus $b(t) = a(T)-a^{(2)}(T) + \int_T^ t \dot b (s) \, ds$ with
\begin{equation}  \label{eq:b1}
\dot b =  (c_1Q, \Im [F-F_1+\dot \theta h]) - A_{2,rmd}.
\end{equation}
%
%
As we will see later that $\dot b(t)$ is smaller than $\dot
a^{(2)}(t)$. However, $b(t)$ is the main part of $a(t)$: We have
$b(t)\les \rho^2(t)$ while $a^{(2)}(t)\les n^2\rho^2(t)$.

\subsubsection{Decompositions of $F$}

Recall \eqref{F.def},
\begin{align*}
 F &=\la Q( 2|h_a|^2 + h_a ^2 ) + \la |h_a |^2 h_a, \qquad  h_a = aR+ h.
\end{align*}
In view of \eqref{order}, we decompose $h_a = \zeta + bR + (\eta+a^{(2)}R)$
and decompose $F$ as
\begin{equation} 
F = F_1 +F_2 + F_3 + F_4+ F_5,
\end{equation}
where
\begin{equation}\label{F.dec}
\begin{split}
F_1 &= \la Q( 2|\zeta|^2+ \zeta^2 ) ,
\\
F_2 &=  2 \la QR b(2\zeta+\bar \zeta)+ 3 \la  QR^2 b^2
+ \la (\zeta+bR)^2 (\bar \zeta+bR)  ,
\\
F_3 & =   2 \la QR a^{(2)} (2\zeta+\bar \zeta),
\\
F_4 &= 2\la Q[(\zeta +\bar \zeta) \eta^{(2)}+\zeta \bar
\eta^{(2)}] ,
\\
F_5 &= 2\la Q[(\zeta +\bar \zeta) \eta^{(3)}+  \zeta \bar
\eta^{(3)}]
\\
&\quad  + \la Q \bkt{ 2 |\eta_a|^2 + \eta_a^2} + 2 \la  QR b(
2\eta_a + \bar \eta_a)
\qquad \qquad (\eta_a = \eta + a^{(2)}R)
 \\
&\quad + \la (aR+ h)^2 (aR+\bar h)- \la (\zeta+bR)^2 (\bar \zeta+bR) .
\end{split}
\end{equation}
Here $F_1$ consists of terms of order $nz^2$; $F_2$, $F_3$ and $F_4$
consist of
terms no smaller than $n^2 z^3$; and $F_5$ higher order terms.

\subsubsection{Decomposition of $\eta$}

We now identify the main terms in $\eta$. We first recall from
\eqref{eq:all} that
\[
\pd _t \eta = \L \eta + \PcL i^{-1} [F +\dot \theta (a R + \zeta+
\eta)] .
\]
Using Lemma \ref{th:AU} that $\L = U^{-1}(-i)A U$ on $\Hc(\L)$ and
$U \PcL = \PcA U$, we have
\begin{align*}
\pd _t U\eta &= -iA U\eta + \PcA U i^{-1} [F +\dot \theta (a R +
\zeta+ \eta)]
\\
&= -iA U\eta - i \dot \theta  U\eta + \PcA U i^{-1} [F +\dot
\theta (a R + \zeta)] - \PcA [U,i] \dot \theta  \eta .
\end{align*}
Here we have used the commutator $[U,i]$ to interchange $U$ and
$i$ so as to produce the term $i \dot \theta  U\eta$. This term is
a global linear term in $U\eta$ and cannot be treated as error
(however $[U,i] U^{-1} \dot \theta  \eta$ is a local error
term). We can eliminate it by introducing
\begin{equation} \label{eta1.def}
\tilde \eta(t) \equiv e^{i\theta(t)}U\eta(t), \qquad
\theta(t) = \int_0^t F_\theta(s) \, ds .
\end{equation}
We have  $\tilde \eta(0) = U\eta(0)$ and
\begin{equation*}
\partial _t \tilde \eta  = -iA \tilde \eta
+e^{i \theta} \PcA U i^{-1} [F +\dot \theta (a R + \zeta)]
 - e^{i \theta} \PcA [U,i]\dot
\theta  \eta .
\end{equation*}
Hence $\tilde \eta(t)$ satisfies the integral equation (using
$e^{-iA t} \wt \eta(0) = U e^{t\L}\eta(0)$)
\begin{align}        \label{eq:eta1}
\tilde \eta(t) &=  U e^{t\L}\eta(0) + \int_0^t e^{-iA (t-s)} \PcA
F_\eta (s) \, d s ,
\\
F_\eta &\equiv e^{i \theta}  U i^{-1} [F +\dot \theta (a R +
\zeta)]  -e^{i \theta} [U,i] \dot \theta  \eta .
\end{align}
Since $U$ and $U^{-1}$ are bounded in Sobolev spaces by Lemma
\ref{th:AU}, and
\begin{equation} \label{eq:eta}
\eta(t) = U^{-1} e^{-i \theta(t)} \tilde \eta (t) ,
\end{equation}
for the purpose of estimation we can treat $\eta$ and $\tilde
\eta$ as the same.

To identify the main term of $\tilde \eta$, we decompose $F_\eta$
as follows,
\begin{align}
F_\eta&= F_{\eta,2} + F_{\eta,3} ,\nonumber
\\
F_{\eta,2} &= e^{i \theta}  U i^{-1} F_1 , \label{Feta}
\\
F_{\eta,3} &=e^{i \theta}  U i^{-1} [(F- F_1 )+\dot \theta (a
R + \zeta)] -e^{i \theta} [U,i] \dot \theta  \eta. \nonumber
\end{align}
The leading part of $\tilde \eta$ is from $F_{\eta,2}$.
Recall $F_1 = \la Q (\zeta^2 + 2\zeta \bar \zeta)$, and
 $U= U_+ + \conj U_-$ with $ U_+$ and $ U_-$ commuting with
$i$ and $\conj$, see \eqref{Upm.dec}. Hence
\begin{align*}
F_{\eta,2} &= e^{i \theta}  U i^{-1} \la Q (\zeta^2 + 2\zeta \bar \zeta)
\\
& = e^{i \theta} i^{-1} (U_+ - \conj U_-)\la Q (\zeta^2 + 2\zeta \bar \zeta)
\\
& = e^{i \theta} i^{-1} \bket{U_+\la Q (\zeta^2 + 2\zeta \bar \zeta)
 -  U_-\la Q (\bar \zeta^2 + 2\zeta \bar \zeta)}.
\end{align*}
Substituting $\zeta=\sum_{l=1}^N \zeta_l$ and
$\zeta_l = z_l u_l^+ + \bar z_l u_l^-$, we have
\begin{align*}
F_{\eta,2}
& = e^{i \theta} i^{-1} \sum_{k,l=1}^N
\bket{U_+\la Q (\zeta_k \zeta_l + \zeta_k \bar \zeta_l + \bar \zeta_k \zeta_l)
 -  U_-\la Q (\bar \zeta_k \bar \zeta_l + \zeta_k \bar \zeta_l
 + \bar \zeta_k \zeta_l)}
\\
&= e^{i \theta} i^{-1} \sum_{k,l \inomega} z_k z_l \Phi_{kl}.
\end{align*}
In the last line, Convention \ref{convention} is used. In particular,
$z_k = \bar z_{|k|}$ if $k<0$.
The functions $\Phi_{kl}$ are defined as follows.
For $k<0$, denote $u_k^+ = u_{|k|}^+$ and  $u_k^- = u_{|k|}^-$. We define
\begin{equation}\label{Phikl.def}
\begin{split}
 k,l>0: & \quad \Phi_{kl} =
      U_+ \la Q(u_k^+ u_l^+ + u_k^+ u_l^- +  u_k^- u_l^+)
\\ & \qquad \qquad
    - U_- \la Q(u_k^- u_l^- + u_k^+ u_l^- +  u_k^- u_l^+),
\\
 k>0,l<0: & \quad \Phi_{kl} =
      U_+ \la Q(u_k^+ u_l^- + u_k^+ u_l^+ +  u_k^- u_l^-)
\\ & \qquad \qquad
    - U_- \la Q(u_k^- u_l^+ + u_k^+ u_l^+ +  u_k^- u_l^-),
\\
 k<0,l>0: & \quad \Phi_{kl} = \Phi_{lk},
\\
 k,l<0: & \quad \Phi_{kl} =
      U_+ \la Q(u_k^- u_l^- + u_k^+ u_l^- +  u_k^- u_l^+)
\\ & \qquad \qquad
    - U_- \la Q(u_k^+ u_l^+ + u_k^+ u_l^- +  u_k^- u_l^+).
\end{split}
\end{equation}
Since $u_k^-=O(n^2)$ and $U_-=O(n^2)$, we have $\Phi_{kl} = O(n^3)$
if $k,l<0$, $\Phi_{kl} = O(n)$ otherwise.

We now integrate $F_{\eta,2}$. Using $z_k(s) = e^{-i\ev_ks}p_k(s)$
(see Convention \ref{convention}) and integrating by parts we get
\begin{align*}
\int_0^t e^{-iA(t-s)} &\PcA F_{\eta,2}\, d s =\sum
 \int_0^t
e^{-iA(t-s)} e^{i \theta} i^{-1} z_k z_l \PcA \Phi_{kl}  \, ds
\\
&=\sum
i^{-1} \int_0^t e^{-iAt} e^{is(A-0i- \ev _k - \ev _l)}\, (e^{i
\theta} p_k p_l) \PcA \Phi_{kl} \, ds
\\
& =\sum e^{i \theta} \, z_k z_l \wt{\eta}_{kl}
-\sum e^{-iAt}(e^{i \theta}
\,z_k z_l)(0) \wt{\eta}_{kl}
\\
&\quad -\sum
\int_0^t  e^{-iAt} e^{is(A-0i- \ev _k - \ev _l)}\frac d
{ds} (e^{i \theta} p_k p_l(s)) \wt{\eta}_{kl} \, ds ,
\end{align*}
where the summation is over $k,l \inomega$ and
\begin{equation}  \label{eta.kl}
\wt{\eta}_{kl}=\lim_{r \to 0+}\frac {-1}{A-\ev _k-\ev _l
-r i} \,  \PcA \Phi_{kl} , \qquad (k,l \inomega).
\end{equation}
Here we add $-r i$ since $A-\ev _k-\ev _l$ may not be invertible.
We take $r \to 0+$, not  $r \to 0-$, to get the
decay of the second and the last terms above, see Lemma \ref{th:AU}
(3).
Recall $\ev_k \sim e_k -e_0>0$ for $k>0$ and $\ev_k = - \ev_{|k|}$
if $k<0$. Since $A$ has a spectral gap $\sim |e_0|$,
 $A-\ev _k-\ev _l$ is not invertible only if
$k>0$ and $l>0$, by Assumption A2.
Hence $\wt{\eta}_{kl}\in L^2$ unless both $k$ and
$l$ are positive.

We denote the main term above by $\tilde \eta^{(2)}$,
\begin{equation} \label{tdeta2.def}
\tilde\eta^{(2)}= \sum_{k,l\inomega} e^{i \theta} \,
z_k z_l \wt{\eta}_{kl},
\end{equation}
and denote the rest of $\tilde \eta$ by $\tilde \eta^{(3)}$.
We decompose $\tilde \eta^{(3)} =\tilde \eta^{(3)}_1+ \cdots
+\tilde\eta^{(3)}_4$ where
\begin{align}
\nonumber\tilde \eta^{(3)}_1  &=U e^{t\L } \eta(0),
\\
\label{eta.all} \tilde \eta^{(3)}_2 &=-\sum_{k,l\inomega} e^{-iAt}
(e^{i \theta} z_k z_l)(0) \wt{\eta}_{kl} = - e^{-iAt}
\tilde\eta^{(2)}(0),
\\
\nonumber \tilde \eta^{(3)}_3 &=-\sum_{k,l\inomega}
\int_0^t  e^{-iAt}  e^{is(A-0i- \ev _k - \ev _l)} \frac d {d s}
(e^{i \theta} p_k p_l(s)) \wt{\eta}_{kl} \, d s,
\\
\nonumber \tilde\eta^{(3)}_4 &= \int_0^t  e^{-iA(t-s)} \PcA
F_{\eta,3} \, d s.
\end{align}
We have
\[
\tilde \eta =\tilde \eta^{(2)}+\tilde \eta^{(3)},\qquad
\tilde \eta^{(3)} =\tilde \eta^{(3)}_1+ \cdots +\tilde\eta^{(3)}_4.
\]
Correspondingly, we define
\begin{equation} \label{eta23.def}
\begin{split}
\eta^{(2)}&= U^{-1}e^{-i\theta}\tilde\eta^{(2)} =
U^{-1}  \sum_{k,l\inomega} z_k z_l \wt{\eta}_{kl},
\\
\eta^{(3)}&= U^{-1} e^{-i\theta}\tilde\eta^{(3)}, \qquad
\eta^{(3)}_j= U^{-1} e^{-i\theta}\tilde\eta^{(3)}_j, \quad (j=1,\dots,4).
\end{split}
\end{equation}

\begin{lemma} \label{th:S-2}
Suppose, for a fixed time, for some $\rho \ll n\le n_0$,
\begin{equation} \label{S-39}
\begin{split}
&\norm{Q}=n , \qquad 
\norm{\eta}_{L^2 \cap L^5} \ll 1, \qquad \norm{\eta}_{L^2 \loc} \le n,
\\
&\max_{k=1,\ldots,N}|z_k| \le \rho, \qquad
|a| \le C \rho^2.
\end{split}
\end{equation}
%
%
Denote
\begin{equation}\label{S:X.def}
\begin{split}
\wt X &= \rho^2  \norm{\eta}_{L^2\loc}
+ n \norm{\eta}_{L^2 \loc} ^2 + \norm{\eta^3}_{L^1 \loc},
\\
X &=  n \rho  \norm{\eta}_{L^2\loc}
+ n \norm{\eta}_{L^2 \loc} ^2 + \norm{\eta^3}_{L^1 \loc}.
\end{split}
\end{equation}
We have
\begin{equation}\label{S-42}
\begin{split}
&\norm{F_5}_{L^1 \loc} \les n \rho^4 + n \rho
\norm{\eta^{(3)}}_{L^2\loc} + \wt X,
\\
&\norm{F_3+F_4+F_5}_{L^1 \loc }  \les n^2 \rho^3 + X,
\\
&\norm{F-F_1}_{L^1 \loc} \les \rho^3 + X,
\qquad
\norm{F}_{L^1 \loc} \les n \rho^2  + X,
\\
&\norm{F_\theta}_{L^1 \loc} \les \rho^2 + n^{-1}X,
\\
&\max_k |\dot p_k| \les n \rho^2  + X,
\qquad
|\dot b| \les n \rho^3 + n X,
\\
&\norm{F_{\eta,3}}_{L^{5/4}\cap L^1} \les \rho^3 + n\rho \norm{\eta}_{L^2\loc}
+ n \norm{\eta}_{L^2\loc \cap L^5}^2 + \norm{\eta^3}_{L^{5/4}\cap L^1}
+ n^2 \rho^2 \norm{\eta}_{L^5} , 
\\
&\norm{F_{\eta}}_{L^{5/4}} \les n\rho^2 + n\rho
\norm{\eta}_{L^2\loc} + n \norm{\eta}_{L^2\loc\cap L^5}^2 +
\norm{\eta^3}_{L^{5/4}}+ n^2 \rho^2 \norm{\eta}_{L^5} .
\end{split}
\end{equation}

\end{lemma}

\myproof By \eqref{S-39} we have $|a^{(2)}| \les n^2\rho^2$, $|b|
\les \rho^2$ and $\norm{\eta^{(2)}}_{L^2\loc} \le n \rho^2$. Also
recall \eqref{n.order}. Hence for $F_5$ defined in \eqref{F.dec}
we have
\begin{align*}
\norm{F_5}_{L^1 \loc } &\les n \rho  \norm{\eta^{(3)}}_{L^2\loc}
+ n \bke{\norm{\eta}_{L^2\loc} + n \rho^2}^2
+ \rho^2 \bke{\norm{\eta}_{L^2\loc} + n \rho^2}
+  \norm{\eta^3}_{L^1\loc}
\\
&\les n \rho^4 + n \rho  \norm{\eta^{(3)}}_{L^2\loc}
+ \rho^2  \norm{\eta}_{L^2\loc}
+ n \norm{\eta}_{L^2 \loc} ^2 + \norm{\eta^3}_{L^1\loc}
\\
&=  n \rho^4 + n \rho  \norm{\eta^{(3)}}_{L^2\loc} + \wt X,
\end{align*}
where  $\wt X$ is defined in \eqref{S:X.def}.
Combining $\eta^{(2)} + \eta^{(3)} = \eta$ in the definition
\eqref{F.dec} of $F_4$ and $F_5$, we can use the same  argument for
$\norm{F_5}_{L^1 \loc }$ to show
\begin{align*}
\norm{F_4+F_5}_{L^1 \loc } &
\les n \rho^4 + n \rho  \norm{\eta}_{L^2\loc}+  \wt X \les n \rho^4 + X.
\end{align*}
By their explicit form in \eqref{F.dec}, we have
$\norm{F_3}_{L^1 \loc } \les n^2 \rho^3$,
 $\norm{F_2}_{L^1 \loc } \sim
|b z| + b^2/n + (z + b/n)^3 \les \rho^3$, and $\norm{F_1}_{L^1 \loc }
\les n \rho^2$. Hence
\[
\norm{F_3+F_4+F_5}_{L^1 \loc }  \les n^2 \rho^3 + X, \quad
\norm{F-F_1}_{L^1 \loc }  \les  \rho^3 + X, \quad
\norm{F}_{L^1 \loc }  \les n \rho^2 + X.
\]
From definition \eqref{Ftheta.def} of $F_\theta$ we have
\[
|F_\theta| \les |a| + n^{-1}\norm{F}_{L^1 \loc}
\les \rho^2 + n^{-1}(n\rho^2 + X) \les \rho^2 + n^{-1}X.
\]
By \eqref{eq:all}, we have
\begin{align*}
|\dot p_k| &\les \norm{F}_{L^1 \loc} + (\rho +\norm{\eta}_{L^2\loc})
|F_\theta|
\\
&\les n \rho^2 + X + (\rho +\norm{\eta}_{L^2\loc}) (\rho^2 + n^{-1}X)
\les n \rho^2 + X .
\end{align*}
By \eqref{eq:b1},
\[
|\dot b| \le C n \norm{F-F_1}_{L^1 \loc} + |(c_1Q,\Im h)|\cdot |F_\theta|
+ |A_{2,rmd}|.
\]
Note that $|(c_1Q,\Im h)| \les n^3 \norm{h}_{L^1 \loc}$ since $\Im
h\perp R$ and $Q=C n^2 R + O(n^3)$ by Lemma \ref{th:2-1}. Also
$|A_{2,rmd}| \les n^2 \rho \max_k|\dot p_k|$ by definition
\eqref{A2rmd.def} of $A_{2,rmd}$. Thus
\begin{align*}
|\dot b| &\les n(\rho^3 +X)
+ n^3(\rho +\norm{\eta}_{L^2\loc}) (\rho^2 + n^{-1}X)
+ n^2 \rho (n \rho^2 + X)
\les n \rho^3 + n X .
\end{align*}

Let $r=1$ or $r=5/4$. The estimate for $\norm{F-F_1}_{L^r}$ is the same as
that for $\norm{F-F_1}_{L^1\loc}$ except for the non-local term
$\la \eta^2 \bar \eta$ and
$\norm{n \eta^2}_{L^r} \les n \norm{ \eta}_{L^2\loc \cap L^{5}}^2$. Thus
\[
\norm{F-F_1}_{L^r} \les
\rho^3 + n \rho \norm{\eta}_{L^2\loc} + n \norm{\eta}_{L^2\loc \cap L^{5}}^2
+  \norm{\eta^3}_{L^r} .
\]
From the definition \eqref{Feta} of $F_{\eta,3}$, the boundedness of
$U$, and the estimate \eqref{Ulocal.est} for $[U,i]$, we have
\begin{align*}
\norm{F_{\eta,3}}_{L^r}
&\les \norm{F-F_1}_{L^r} + |F_ \theta|(n^{-1}|a|+\max|z_k|)
+  n^2 |F_ \theta|\norm{\eta}_{L^5}
\\
&\les \rho^3 + n\rho \norm{\eta}_{L^2\loc} + n \norm{\eta}_{L^2\loc\cap L^{5}}^2
+ \norm{\eta^3}_{L^r}
\\
&\qquad + (\rho^2 + n^{-1}X)(\rho + n^2 \norm{\eta}_{L^5})
\\
&\les \rho^3 + n\rho \norm{\eta}_{L^2\loc} + n
\norm{\eta}_{L^2\loc\cap L^{5}}^2 + \norm{\eta^3}_{L^r} + n^2 \rho^2
\norm{\eta}_{L^5} .
\end{align*}
Here we also used $\norm{\eta^3}_{L^1 \loc}\les  \norm{\eta^3}_{L^r}$.
Since $\norm{F_{\eta}}_{L^{5/4}} \les n \rho^2 + \norm{F_{\eta,3}}_{L^{5/4}}$,
the estimate for  $\norm{F_{\eta}}_{L^{5/4}}$ in \eqref{S-42} follows.
\myendproof

\subsection{Normal forms for equations of bound states}

Recall that $z_k(t)=e^{-i\ev _kt}p_k(t)$ and $a(t)=a^{(2)}(t)+
b(t)$. Since many terms in the equations of $p_k$ and $b$ are
oscillatory and only contribute to the phases, we now derive the
normal forms for the equations of $p_k$ and $b$, where those terms
are removed.
Throughout this subsection, Convention 
\ref{convention} for $\ev_k$, $z_k$ and $p_k$ with $k<0$ is understood.

\begin{lemma} [Normal Form] \label{th:S-NF}
Suppose, for some $\rho(t) \ll n\le n_0$,
\begin{equation}
\begin{split}
&\norm{Q}=n, \qquad
\norm{\eta(t)}_{L^2 \cap L^5} \ll 1, \qquad
\norm{\eta(t)}_{L^2 \loc} \le n,
\\
&\max_{k=1,\ldots,N}|z_k(t)| \le \rho(t), \qquad
|a(t)| \le C \rho^2(t).
\end{split}
\end{equation}
There are perturbations $q_k(t)$ of $p_k(t)$, $k=1,\ldots,N$, and
$\beta(t)$ of $b(t)$, to be defined in \eqref{qk.def} and \eqref{beta.def},
satisfying
\begin{equation} \label{q-p.est}
|q_k - p_k|\le C n \rho^2 , \qquad |b - \beta| \le  C n \rho^3 + n^2 \rho
\norm{\eta}_{L^2\loc},
\end{equation}
such that
\begin{equation}\label{qk.eq}
\dot q_k =\sum_{l= 1,\ldots,N} D_{kl} |q_l|^2 q_k + i J_k q_k + g_k,
\end{equation}
\begin{equation}\label{beta.eq}
\dot \beta =\sum_{1\le k \le l \le N} B_{kl}|z_k|^2|z_l|^2  + g_b,
\end{equation}
where $J_k(t)$ are real functions bounded by $C\rho^2$ to be defined
in \eqref{fk.def}. The
constants $D_{kl}$ are complex and bounded by $n^2$, with their
real parts given by \eqref{ReDkl.def} satisfying
\begin{equation} \label{3-45}
- 5 \gamma_0^+ n^2 \le \Re D_{kl}\le  C n^4,\quad
 \Re D_{kk} \le - \gamma_0  n^2.
\end{equation}
The order one constants $\gamma_0$ and $\gamma_0^+$ are defined
in \eqref{gamma0.def} and \eqref{gamma0p.def}.
The constants $B_{kl}$ are real and $B_{kl} = - \frac {1}2 \,c_1
\Re D_{kl}+O( n^4)$. Moreover,
\begin{equation}\label{S:gk.est}
|g_k|\les n \rho^4 + n^4 \rho^3 
+ n \rho \norm{\eta^{(3)} }_{L^2\loc} + \wt X,
\end{equation}
\begin{equation}\label{S:gb.est}
|g_b| \les n \rho^5 + n^4 \rho^4 +  n \rho^2
\norm{ \eta^{(3)}}_{L^2\loc} + n^3 \rho^2 \norm{\eta}_{L^2\loc} +
n^2 \norm{\eta}_{L^2\loc}^2 + n \norm{\eta^3}_{L^1\loc},
\end{equation}
where $\wt X =  \rho^2  \norm{\eta}_{L^2\loc}
+ n \norm{\eta}_{L^2 \loc} ^2 + \norm{\eta^3}_{L^1 \loc}$
is defined in \eqref{S:X.def}.

\end{lemma}

\myproof We first observe a consequence of Assumption A3: Suppose
$k_0, k_1,\ldots, k_j \in \bket{1,\ldots,N}$ with $j \le
j_{\text{max}}=3$ satisfy
\begin{equation} \label{A3p}
\ev _{k_0} \pm \ev _{k_1} \pm \cdots  \pm \ev _{k_j} =0.
\end{equation}
Then $j$ is odd and, after a relabeling, the two sets (with
multiplicities) $\bket{k_0,\ldots, k_{\frac{j-1}2}}$ and
$\bket{k_{\frac{j+1}2}, \ldots,k_{j}}$ are the same. The reason is
as follows. Recall $\ev_k = e_k - e_0 + O(n^2)>0$, and hence the
expression in \eqref{A3p} is approximately equal to
\begin{equation} \label{A3p2}
(e_{k_0} - e_0) \pm (e_{k_1} - e_0) \pm \cdots \pm (e_{k_j} - e_0) .
\end{equation}
For the expression in \eqref{A3p} to be zero, one of the $\ev_{k_m}$
must be with a negative sign. Hence two $e_0$'s in \eqref{A3p2} cancel
and \eqref{A3p2} is the sum of $j$ $e_k$'s minus the sum of another $j$
$e_k$'s. Since $j\le j_{\text{max}}$,
Assumption A3 then implies that they can be divided to two equal sets.
Since there are $(j-1)$ $e_0$'s and all $k_m$'s are positive, $j$ must be odd,
these $e_0$'s cancel each other, and the other $e_k$'s form two equal sets.

{\bf Part 1. Excited states.} Recall from \eqref{eq:all}
the equations for $p_k$:
\begin{equation} \label{pk.eq1}
\dot p_k = i^{-1}e^{i\ev _kt}\,
\bkt{(u_k^+, F)+(u_k^-, \wbar F)+
\bkt{(u_k^+, h)+(u_k^-, \wbar h)+ (u_k,R) a} F_ \theta },
\end{equation}
where $h=\zeta_1 + \dots + \zeta_N + \eta$. The nonlinear terms
$F$ and $F_\theta$ are given in \eqref{F.def}--\eqref{Ftheta.def}
and $F=F_1+\cdots +F_5$ is decomposed in \eqref{F.dec}. We shall
derive the normal form
of \eqref{pk.eq1} using integration by parts for those terms
no smaller than $n^2 \rho^3$.

\medskip

\noindent{\bf Step 1} \quad Integration of terms of order $nz^2$.

Recall that the main part of $F$ is
$F_1= \la Q (2 \zeta \bar \zeta + \zeta ^2)$.
The main terms on the right side of \eqref{pk.eq1} are among the first
two groups of terms involving $F$,
\begin{equation}\label{pk.eq2}
i^{-1}e^{i\ev _kt}\bket{ (u_k^+, F_1)+(u_k^-, \wbar F_1)} .
\end{equation}
We denote the rest by $R_k$,
\begin{equation} \label{Rk.def}
\begin{split}
R_k = i^{-1}e^{i\ev _kt}\, &\Big[
 (u_k^+, F-F_1)+(u_k^-, \wbar F-\wbar F_1)
\\
&\quad + \bkt{(u_k^+, h)+(u_k^-, \wbar h)+ (u_k,R) a} F_ \theta
\Big],
\end{split}
\end{equation}
and rewrite \eqref{pk.eq1} as
\begin{equation} \label{pk.eq3}
\dot p_k =i^{-1}e^{i\ev _kt}\bket{ (u_k^+, F_1)+(u_k^-, \wbar
F_1)} + R_k.
\end{equation}
Substituting $F_1= \sum_{l,m=1}^N \la Q (2 \zeta_l \bar \zeta_m +
\zeta_l \zeta_m)$ with  $\zeta_l=z_l u_l^+  + \bar z_l
u_l^-$ and using  Convention 
\ref{convention}, we get
\begin{align*}
&i^{-1}e^{i\ev _kt}\bket{ (u_k^+, F_1)+(u_k^-, \wbar F_1)}
\\
&= \sum_{l,m=1}^N i^{-1}e^{i\ev _kt}\bket{  ( \la Qu_k^+, 2
\zeta_l \bar \zeta_m + \zeta_l \zeta_m)) + ( \la Qu_k^-,  2
\zeta_l\bar \zeta_m + \bar \zeta_l \bar\zeta_m)}
\\ &= \sum_{l,m\inomega} c_k^{lm} e^{i\ev _kt}\, z_l z_m ,
\end{align*}
for some purely imaginary constants $c_k^{lm}$ bounded by $n$. The
phase factor of a typical term $e^{i\ev _kt}\,z_l z_m $ is $\ev_k
- \ev_l - \ev_m$, which is nonzero by the observation for
\eqref{A3p}. Hence we can define
\begin{align}
p_{k,1-}&= \sum_{l,m\inomega}\frac {-ic_k^{lm}}{\ev _k -\ev _l -
\ev_m} \, e^{i\ev _kt}\,z_l z_m ,
\\
p_{k,1e}&= \sum_{l,m\inomega} \frac {ic_k^{lm}}{\ev _k - \ev _l -
\ev_m} \, e^{i(\ev _k- \ev _l -  \ev_m)t}\, (\dot p_l p_m  + p_l
\dot p_m) ,  \label{pk1e.def}
\end{align}
and we have
\[
\sum_{l,m\inomega} c_k^{lm}\, e^{i\ev _kt}\, z_l z_m =\dot
p_{k,1-} + p_{k,1e} .
\]
Because $p_{k,1e}$ are of order $n^2 z^3$, we need to extract
their leading terms. Substitute \eqref{pk.eq3} into
\eqref{pk1e.def} and collect terms. The leading terms in
$p_{k,1e}$ are cubic polynomials in $z_k$ and $p_{k,1e}$ can be rewritten as
\begin{equation}
p_{k,1e}= \sum_{l,m,n\inomega} i^{-1} e^{i\ev _kt}\,d_{k,1}^{lmj}
\, z_l z_m z_j  + g_{k, 1} ,
\end{equation}
\begin{equation} \label{gk1.def}
g_{k,1}= \sum_{l,m\inomega} \frac {ic_k^{lm}}{\ev _k - \ev _l -
\ev_m} \, e^{i(\ev _k- \ev _l -  \ev_m)t}\, (R_l p_m  + p_l R_m).
\end{equation}
Here $d_{k,1}^{lmj}$ are real constants bounded by $n^2$, and
the error terms $g_{k,1}$ are bounded by
\begin{equation}\label{gk1.est}
|g_{k, 1}| \le C \max_{k,l,m} |c_k^{lm}| |p_l| |R_m| \le C n \rho
\max_m |R_m|.
\end{equation}
We conclude
\begin{equation}\label{pk1.eq}
\dot p_k - \dot p_{k,1-}= R_k + \sum_{l,m,n\inomega} i^{-1}
e^{i\ev _kt}\, d_{k,1}^{lmj}\, z_l z_m z_j + g_{k, 1} .
\end{equation}

\medskip

\noindent{\bf Step 2} \quad Integration of terms of the form $nz \eta^{(2)}$.

We now integrate  $i^{-1}e^{i\ev _kt} (u_k^+, F_4)$, which is part of the
first term of $R_k$. Recall $F_4=2\la Q[(\zeta +\bar \zeta)
\eta^{(2)}+\zeta \bar \eta^{(2)}]$ with $\eta^{(2)}= U^{-1}
\sum_{l,m \inomega} z_l z_m \wt \eta_{lm}$.
Denote 
\begin{equation}
\eta ' =\sum_{l,m\inomega} z_l z_m \tilde \eta_{lm}, 
\end{equation}
where $ \tilde \eta_{lm}$ are defined in \eqref{eta.kl} with
$\norm{\wt \eta_{lm}}_{L^2 \loc} \les n$. Recall Lemma \ref{th:AU}
that $U^{-1}=(U_+)^* - (U_-)^*\conj $ and both $U_+$ and $U_-$
commute with $i$ and conjugation. Then $\eta ^{(2)}= U^{-1} \eta '
= (U_+)^* \eta ' - (U_-)^* \bar \eta '$ and
\begin{align}
 ( u_k^+, \, F_4 ) 
&= (2\la Q u_k^+  (\zeta+\bar \zeta),\, \eta^{(2)} ) + (2\la Q u_k^+ \bar
 \zeta ,\, \bar\eta ^{(2)}) \nonumber
\\
& =  (2\la Q  u_k^+  (\zeta+\bar \zeta),\,  [(U_+)^* \eta ' -
(U_-)^* \bar \eta ' ]) \nonumber
\\ &\quad
+(2\la Q u_k^+ \bar
 \zeta ,\, [((U_+)^* \bar \eta ' - (U_-)^*  \eta '] )\nonumber
\\
& =  (U_+[2\la Q  u_k^+  (\zeta+\bar \zeta)] - U_-[2\la Q u_k^+ \bar
 \zeta ], \ \eta')\label{3:I1}
\\ &\quad
+ (U_+ [2\la Q u_k^+ \bar \zeta ] -  U_-[2\la Q  u_k^+  (\zeta+\bar \zeta)]
 , \ \bar \eta'). \label{3:I2}
\end{align}
%
Substituting $\zeta = \sum_{j}z_j u_j^+ + \bar z_j u_j^-$
and $\eta'= \sum_{l,m \inomega} z_l z_m \wt \eta_{lm}$,
we can write
\begin{equation}
i^{-1}e^{i\ev _kt} (u_k^+, F_4)=\sum_{l,m,j\inomega} d_k^{lmj}\,
e^{i\ev _kt}\, z_l z_m z_j ,
\end{equation}
for some coefficients $d_k^{lmj}$ bounded by $n^2$.
The phase factor of a typical term $e^{i\ev _kt}\,z_l z_m z_j$ is
$\ev_k - \ev_l - \ev_m -\ev_j$. By the observation for
\eqref{A3p}, it is nonzero unless one of $|l|$, $|m|$, $|j|$ is
$k$ and the other two are the same. In this exceptional case
$e^{i\ev _kt}\,z_l z_m z_j$ is of the form $e^{i\ev _kt}\, z_k z_l
\bar z_l=|p_l|^2p_k$. For fixed $l>0$, there are six such terms if
$l\not =k$, and three terms if $l=k$. We denote the sum of their
coefficients as $D_{kl}$,
\begin{equation}\label{betakl.def}
\begin{split}
D_{kl}&=d_{k}^{kl(-l)}+d_{k}^{k(-l)l}
+d_{k}^{lk(-l)}+d_{k}^{(-l)kl}+d_{k}^{l(-l)k}+d_{k}^{(-l)lk},
\quad (l\not =k),
\\
D_{kk}&=d_{k}^{kk(-k)}+d_{k}^{k(-k)k}+d_{k}^{(-k)kk}.
\end{split}
\end{equation}
The total of these zero-phase-factor terms is $\sum_{l=1}^N
D_{kl}|p_l|^2 p_k$. The other terms can be integrated. Define
\begin{equation}\label{pk2m.def}
p_{k,2-}= \sum_{\ev _k - \ev _l -  \ev_m- \ev_j \not = 0} \frac
{-id_{k}^{lmj} } {\ev _k -\ev _l - \ev_m-\ev_j} \, e^{i\ev
_kt}\,z_l z_m z_j ,
\end{equation}
\begin{equation}\label{gk2.def}
 g_{k,2}= -\sum_{\ev _k - \ev _l -  \ev_m- \ev_j \not = 0} \frac
{-id_{k}^{lmj} } {\ev _k -\ev _l - \ev_m-\ev_j} \, e^{i(\ev _k-
\ev _l -  \ev_m-\ev_j)t}\, \frac d{dt} \bke{ p_l p_m p_j} .
\end{equation}
We have
\begin{equation}
\sum_{l,m,j\inomega} d_k^{lmj}\, e^{i\ev _kt}\, z_l z_m z_j =
\sum_{l=1}^N D_{kl} |z_l|^2 p_k + \frac d{dt} (p_{k,2-}) +
g_{k,2},
\end{equation}
and
\begin{equation}\label{gk2.est}
| g_{k,2}| \les (\max|d_{k}^{lmj}|)\, \rho^2 \max_{j}|\dot p_j|
\les n^2 \rho^2 \max_{j}|\dot p_j|.
\end{equation}

We now compute $\Re D_{kl}$. We want to collect terms of the form
$C e^{i \ev_k t} z_l \bar z_l z_k$ from \eqref{3:I1}--\eqref{3:I2}
with $\Im C \not = 0$. Hence we only need to consider those $\tilde
\eta_{lm}$ with $\Im \tilde \eta_{lm} \not = 0$, i.e., $l,m>0$.
The only terms from $\bar \eta'$ with resonance coefficients
are $\sum_{m,j>0} \wbar { z_m z_j \tilde \eta_{mj}}$, which are of
the form $\bar z_m \bar z_j$, with two bars. Hence the integral in
\eqref{3:I2} does not contain $z_l \bar z_l z_k$ and is
irrelevant.
From the integral in \eqref{3:I1}, we want to choose $z_l z_k$ from $\eta'$,
i.e., $\sum_{m,j>0} z_m z_j \tilde \eta_{mj}$,  and choose $z_l$ from
$\zeta$ or $\bar \zeta$ in
$U_+[2\la Q  u_k^+  (\zeta+\bar \zeta)] - U_-[2\la Q u_k^+ \bar
 \zeta ]$. (Recall \eqref{L2.ip} that $(f,g) = \int \bar f g dx$.)
For the second part we get $z_l \wt \Phi_{kl}$ where
\begin{equation} \label{wtPhikl.def}
\wt \Phi_{kl} =U_+ 2\la Q u_k^+  (u_l^+ + u_l^-)
- U_- 2\la Qu_k^+  u_l^- .
\end{equation}
Since $\tilde \eta_{mj}=\tilde \eta_{jm}$ for $m,j>0$,  the terms
with $z_l z_k$ in $\sum_{m,j>0} z_m z_j \tilde \eta_{mj}$ is
$(2-\delta_k^l)z_k z_l \tilde \eta_{kl}$.
Therefore, also using the definition \eqref{eta.kl} of $\wt \eta_{kl}$,
\begin{align}
\Re D_{kl} &= \Re i^{-1} \int \wt \Phi_{kl} (2-\delta_k^l)\tilde
\eta_{kl}  \, d x \nonumber
\\
&= -(2-\delta_k^l) \, \Im \bke{ \wt \Phi_{kl}, \, \frac 1{A-\ev
_k-\ev _l-0i} \PcA \Phi_{kl}}. \label{ReDkl.def}
\end{align}
Recall
\begin{equation}\label{3:order}
\begin{split}
&Q=(1+O(\epz^2)) n \phi_0+O(n^3),\quad
u_k^+=\phi_k+O(n^2),\quad u_k^-=O(n^2),\\
&U_+=1+O(n^2),\quad U_-=O(n^2).
\end{split}
\end{equation}
Hence, by \eqref{Phikl.def} and \eqref{wtPhikl.def},
\begin{equation}
\Phi_{kl} = \la Q \phi_k \phi_l +O(n^3), \qquad
\wt \Phi_{kl} = 2 \la Q \phi_k \phi_l +O(n^3),
\end{equation}
and we have
\begin{equation}\label{3-74}
\Re D_{kl} = -2 (2-\delta_k^l) (1+o(1)) n^2
\Im \bke{\phi_0 \phi_k \phi_l, \frac
1{A-\ev _k-\ev _l-0i}\PcA \phi_0 \phi_k \phi_l} + O(n^4).
\end{equation}
By Assumption A2 and Lemma \ref{th:AU} (3), we have
$-\Re D_{kl} \le 5 n^2 \gamma_0^+$
and $\Re D_{kl} \le C n^4$ for all $k,l>0$, and
$- \Re D_{kk} \ge \gamma_0 n^2$.
We conclude \eqref{3-45}.

\medskip

\noindent{\bf Step 3} \quad Integration of other terms.

We now integrate other terms in \eqref{pk1.eq} no smaller than
$n^2 z^3$. We first consider the first group of terms in  $R_k$
defined in \eqref{Rk.def}. Using
$F-F_1=F_2+F_3+F_4+F_5$ and removing $i^{-1}e^{i\ev _kt} (u_k^+,
F_4)$, we get
\[
i^{-1}e^{i\ev _kt} \bket{ (u_k^+, F_2+F_3) + (u_k^-, \wbar
F_2) } + g_{k,3},
\]
where
\begin{equation} \label{gk3.def}
g_{k,3} = i^{-1}e^{i\ev _kt} \bket{ (u_k^+, F_{5})  + (u_k^-,
\wbar F_3+\wbar F_4+\wbar F_5)}.
\end{equation}
We have, using $u_k^-=O(n^2)$ and the explicit form of $F_3+F_4$,
\begin{equation}\label{gk3.est}
|g_{k,3}| \les \norm{F_5}_{L^1 \loc} + n^2 \norm{F_3+F_4}_{L^1\loc}
\les \norm{F_5}_{L^1 \loc} + n^2 (n^2 \rho^3).
\end{equation}

We now consider the second group of terms in $R_k$, see \eqref{Rk.def},
\begin{equation} \label{Rk2nd}
i^{-1}e^{i\ev _kt} \bkt{(u_k^+, h)+(u_k^-, \wbar h)+ (u_k,R) a} F_
\theta.
\end{equation}
The only relevant terms are
$i^{-1}e^{i\ev _kt} \bkt{(u_k^+, \zeta)+(u_k^-, \wbar \zeta)} F_\theta$.
We move the other part to error term,
\begin{equation} \label{gk4.def}
g_{k,4} = i^{-1}e^{i\ev _kt}
\bkt{(u_k^+, \eta)+(u_k^-, \wbar \eta)+ (u_k,R) a} F_\theta.
\end{equation}
Note that $(u_k,R)=O(n)$ since $u_k\perp Q$ and $R=C n^{-2}Q +
O(n)$ by Lemma \ref{th:2-1}. Also, $|(u_k^+,\eta)+ (u_k^-, \wbar
\eta)| \le C n^2 \norm{\eta}_{L^2\loc}$ since $\Re \eta \perp
v_k$, $\Im \eta \perp u_k$, and the differences between
$u_k,v_k,u_k^+,u_k^-$ are bounded by $n^2$. Hence
\begin{equation}\label{gk4.est}
|g_{k,4}|\le (n^2\norm{\eta}_{L^2\loc} + n \rho^2) |F_\theta|.
\end{equation}
Using $\zeta=\sum_l z_l u_l^+ + \bar z_l u_l^-$, we get
\[
(u_k^+, \zeta)+(u_k^-, \wbar \zeta) =
\ssum_{l=1}^N \bke{ C_k^l z_l +  \wt C_k^l \bar z_l},
\]
where $C_k^l =(u_k^+, u_l^+)+ (u_k^-, u_l^-)$ and $\wt C_k^l
=(u_k^-, u_l^+)+ (u_k^-, u_l^+)$.
Because $u_k^+ = \phi_k + O(n^2)$ and  $u_k^- = O(n^2)$,
see \eqref{3:order}, we have
\[
C_k^l = \delta_k^l + O(n^2), \qquad  \wt C_k^l =O(n^2).
\]
Thus, terms no smaller than $n^2\rho^3$ in \eqref{Rk2nd} are
\begin{equation} \label{3-77}
i^{-1}e^{i\ev _kt} \bket{ z_k F_{\theta,1} + \ssum_l \bkt{ (C_k^l
- \delta_k^l) z_l + \wt C_k^l \bar z_l} F_{\theta,2} },
\end{equation}
where $F_{\theta,1}$ and $F_{\theta,2}$ are leading parts of $F_
\theta$ \eqref{Ftheta.def}, (recall $a=b+a^{(2)}$)
\begin{align}
F_{\theta,1} &\equiv -[a +\bke{ c_1R,\, \Re F_1 + F_2 } ] \cdot
\bkt{1+ (c_1R,R)b }^{-1} , \label{Ft1.def}
\\
F_{\theta,2} &\equiv -[b+(c_1R,\Re F_1)], \label{Ft2.def}
\end{align}
and \eqref{Rk2nd} is equal to $g_{k,4} +\eqref{3-77}+ g_{k,5}$ where
\begin{equation} \label{gk5.def}
g_{k,5} = i^{-1}e^{i\ev _kt} \bket{ z_k  (F_\theta- F_{\theta,1})
+  \ssum_l \bkt{ (C_k^l - \delta_k^l) z_l + \wt C_k^l \bar z_l}
(F_\theta- F_{\theta,2}) }.
\end{equation}
Since $\abs{F_\theta- F_{\theta,2}}\le \abs{F_\theta-
F_{\theta,1}}+\abs{F_{\theta,1}- F_{\theta,2}}$, we have
\[
 |g_{k,5}| \les \rho \abs{F_\theta- F_{\theta,1}} + n^2 \rho
 \abs{F_{\theta,1}- F_{\theta,2}}.
\]
Recall \eqref{Ftheta.def} $F_ \theta = -\bkt{a+ \bke{ c_1R,\, \Re F } }  \cdot
\bkt{1+ (c_1R,R)a+(c_1R,\Re h)}^{-1}$. Hence
\begin{align}
F_\theta- F_{\theta,1}&=-\bke{ c_1R,\, \Re F_3 + F_4 + F_5 }\cdot
\bkt{1+ (c_1R,R)b }^{-1} \nonumber
\\
&\quad + \frac{ [a+ \bke{ c_1R,\, \Re F } ] \cdot [
(c_1R,R)a^{(2)} +(c_1R,\Re h)]} {
 [1+(c_1R,R)a+(c_1R,\Re h)]\cdot [1+(c_1R,R)b]} .
\end{align}
Since $R=C n^{-2}Q + O(n)$ and $\Re h\perp Q$, we have $|(c_1R,\Re
h)| \les n \norm{h}_{L^1\loc}$. Hence,
\begin{align*}
\abs{F_\theta- F_{\theta,1}} &\les n^{-1} \norm{F_3+F_4 +
F_5}_{L^1 \loc} +
\bkt{|a| + n^{-1}\norm{F}_{L^1 \loc}}(\rho^2 +  n \norm{h}_{L^1\loc}) 
\\
&\les n^{-1}(n^2 \rho^3 + X) + [\rho^2 + n^{-1}(n\rho^2 +
X)](n\rho + n \norm{\eta}_{L^2\loc})
\\
&\les n \rho^3 + n^{-1} X .
\end{align*}
Here we have used Lemma \ref{th:S-2}. We also have
\begin{align}
F_{\theta,1}- F_{\theta,2}&=-[a^{(2)} + [ c_1R,\, \Re F_2)] \cdot  \bkt{1+
(c_1R,R)b }^{-1} \nonumber
\\
&\quad +  [a +\bke{ c_1R,\, \Re F_1 + F_2 } ]  \cdot [1+(c_1R,R)b]^{-1} \cdot
(c_1R,R)b.
\end{align}
Hence
\[
\abs{F_{\theta,1}- F_{\theta,2}}
\les (n^2 \rho^2 + n^{-1} \norm{F_2}_{L^1\loc})
 + (\rho^2 + n^{-1}\norm{F_1+F_2}_{L^1\loc}) n^{-2} \rho^2.
\]
Since $\norm{F_1}_{L^1\loc} \le n \rho^2$ and
$\norm{F_2}_{L^1\loc} \le  \rho^3$, we have $\abs{F_{\theta,1}-
F_{\theta,2}} \les n^2 \rho^2 + n^{-1} \rho^3$. Hence $g_{k,5}$ is
bounded by
\begin{align}
|g_{k,5}| &\les \rho \abs{F_\theta- F_{\theta,1}} + n^2 \rho
\abs{F_{\theta,1}- F_{\theta,2}} \nonumber
\\
&\les \rho (n \rho^3 + n^{-1} X) + n^2 \rho (n^2 \rho^2 + n^{-1}
\rho^3)\nonumber
\\
&\les n \rho^4 + n^4 \rho^3 + n^{-1}\rho X. \label{gk5.est}
\end{align}

Summarizing, we can rewrite eq.~\eqref{pk1.eq} as
\begin{equation}
\frac d{dt} \bke{p_k - p_{k,1-} -p_{k,2-}} = \sum_{l=1}^N D_{kl}
|z_l|^2 p_k  +  \wt R_k + \sum_{j=1}^5 g_{k,j},
\end{equation}
where
\begin{multline}
\wt R_k =  i^{-1}e^{i\ev _kt} \Bigg\{ \sum_{l,m,j\inomega}
d_{k,1}^{lmj}\, \,z_l z_m z_j +(u_k^+, F_2+F_3) + (u_k^-, \wbar
F_2)
\\
- z_k \, \frac 1{1+(c_1R,R)b}\, [a +\bke{ c_1R,\, \Re F_1 + F_2 }
]
\\
-\ssum_l \bkt{ (C_k^l - \delta_k^l) z_l + \wt C_k^l \bar z_l}
[b+(c_1R,\Re F_1)] \Bigg \}. \label{wtRk.def}
\end{multline}

We now integrate $\wt R_k $. Denote
\begin{equation} \label{B.def}
 B=(c_1R,R)b , \qquad |B|\le C n^{-2}\rho^2.
\end{equation}
Note that $\wt R_k $ is of the form
\begin{equation} \label{wtRk.form}
i e^{i\ev _kt} \bkt{ n^{-1}b^2 + bz + n^{-1}b z^2 + z^3 + n^{-1} z_k z^3},
\end{equation}
in the sense that
\begin{multline}
\wt R_k = \frac{i e^{i\ev _kt}}{1+B}  \Bigg[ f(B)n^{-1}b^2 +  \sum
_{j\inomega} f(B) b z_j + \sum _{j_1,j_2 \inomega} f(B) n^{-1}b
z_{j_1} z_{j_2}
\\
+ \sum _{j_1,j_2,j_3\inomega}
 f(B) z_{j_1} z_{j_2} z_{j_3}
+ \sum _{j_1,j_2,j_3\inomega}
 f(B) n^{-1} z_k z_{j_1} z_{j_2} z_{j_3}
\Bigg ],
\end{multline}
where $f(B)$ are polynomials in $B$ with real coefficients bounded by one.
We have omitted their dependence on the summation indexes.
The phase factors of the above summands are
\[
\ev_k, \quad
\ev_k \pm \ev_{|j_1|}, \quad
\ev_k \pm \ev_{|j_1|}\pm \ev_{|j_2|}, \quad
\ev_k \pm \ev_{|j_1|}\pm \ev_{|j_2|}\pm \ev_{|j_3|}, \quad
\pm \ev_{|j_1|}\pm \ev_{|j_2|}\pm \ev_{|j_3|},
\]
respectively.
By the observation for \eqref{A3p},
zero phase factor only occurs to $bz$ and $z^3$ terms. Those terms
with zero  phase factor are
\[
\frac{i f_{bk}(B)}{1+B} e^{i\ev _kt} b z_k,\qquad
\frac{i f_{kl}(B)}{1+B} e^{i\ev _kt} z_k z_l \bar z_l.
\]
The sum of these terms is equal to $i J_k e^{i\ev _kt} z_k= i J_k p_k$, where
\begin{equation}  \label{fk.def}
J_k(t) \equiv \frac{f_{bk}(B)}{1+B} b  + \sum_{l=1}^N \frac{
f_{kl}(B)}{1+B}  |z_l|^2.
\end{equation}
They are real functions bounded by $C\rho^2$. The other terms can
be integrated. For example, if $j\inomega$, $j \not = k$, we have
\[
\frac{i f(B)}{1+B} e^{i\ev _kt} b z_j =
\frac d{dt} \bke{\frac{i f(B)}{i(\ev _k - \ev_j)(1+B)}
e^{i(\ev _k - \ev_j)t} b p_j}  - \text{error},
\]
where
\[
\text{error} =  e^{i(\ev _k - \ev_j)t} \frac d{dt}
\bke{\frac{i f(B)}{i(\ev _k - \ev_j)(1+B)}
b p_j},
\]
and it is bounded by
\[
|\text{error}|\les |\dot b p_j| + |b\dot  p_j| + |b p_j \dot B|.
\]
Note $|\dot B|= C n^{-2} |\dot b|$.
We can integrate other terms similarly.
Summing up, we get
\begin{equation}\label{3-91}
\wt R_k = i J_k p_k  + \frac d{dt} p_{k,3-} + g_{k,6},
\end{equation}
where
$p_{k,3-}$ is of the same form \eqref{wtRk.form} as $\wt
R_k$ and, adding the estimates for the integration remainders for
all five kinds of terms in \eqref{wtRk.form}, we get
\begin{align}%
|g_{k,6}|&\les n^{-1} (|b \dot b| + b^2 |\dot B|)
 + (|\dot b|\rho + \max_j |b\dot  p_j| + |b \rho \dot B|)
 \nonumber
\\
&\quad + n^{-1} (|\dot b|\rho^2 + \max_j |b\rho \dot  p_j| + |b
\rho^2 \dot B|)\nonumber
\\
&\quad + (\rho^2 \max_j  |\dot  p_j| + \rho^3 |\dot B|) + n^{-1}
(\rho^3 \max_j  |\dot  p_j| + \rho^4 |\dot B|)\nonumber
\\
&\les  \rho^2 \max_j  |\dot  p_j| + \rho |\dot b| .\label{gk6.est}
\end{align}

\medskip

{\bf Step 4} \quad Final form. We now define
\begin{equation}  \label{qk.def}
q_k\equiv  p_k-p_{k,1-}-p_{k,2-}-p_{k,3-} .
\end{equation}
Since $p_{k,1-}\sim n z^2$, $p_{k,2-}\sim n^2 z^3$ and $p_{k,3-}
\sim n^{-1}b^2 + bz + n^{-1}b z^2 + z^3 + n^{-1} z_k z^3$, we have
\begin{equation}  
  |p_k-q_k| \le |p_{k,1}|+|p_{k,2-}| + |p_{k,3-}|\le C n \rho^2.
\end{equation}
From \eqref{wtRk.def} and \eqref{3-91} we have
\begin{align*}
\dot q_k &= \sum_{l= 1,\ldots,N} D_{kl} |p_l|^2 p_k +
i J_k p_k + \ssum_{j=1}^6 g_{k,j}
\\
&=\sum_{l= 1,\ldots,N} D_{kl} |q_l|^2 q_k + i J_k q_k + g_k,
\end{align*}
where
\begin{align}\label{gk.def}
g_k &=  g_{k,1}+\cdots +g_{k,6}+g_{k,7} ,\\
g_{k,7} &\equiv \sum_{l= 1,\ldots,N} D_{kl} \bke{|p_l|^2
p_k-|q_l|^2 q_k} + i J_k  (p_k-q_k). \label{gk7.def}
\end{align}
Since $D_{kl}=O(n^2)$ and $J_k=O(\rho^2)$, we have
\begin{equation}  \label{gk7.est}
|g_{k,7}|\les n^2 \rho^2 (n\rho^2) + \rho^2 (n \rho^2) \les  n \rho^4.
\end{equation}
Collecting estimates, we have
\begin{align*}  
|g_k| \les  \sum_{j=1}^7 |g_{k,j}| &\les n \rho \max_m |R_m| + n^2
\rho^2 \max_{j}|\dot p_j| + [\norm{F_5}_{L^1 \loc} + n^2 (n^2
\rho^3)]
\\
&\quad + ( n^2 \norm{\eta}_{L^2\loc} + n \rho^2 )\, |F_\theta| +
(n\rho^4 + n^4 \rho^3 +n^{-1}\rho X)
\\
&\quad + (\rho^2  \max_j |\dot p_j| + \rho |\dot b|) +  n \rho^4.
\end{align*}
Using $\max_m |R_m| \les \norm{F-F_1}_{L^1\loc}
+ (\rho+\norm{\eta}_{L^2\loc})|F_\theta|$ and Lemma \ref{th:S-2},
we have
\begin{align*}
|g_k| &\les  n \rho^4 + n^4 \rho^3 + n \rho \norm{F-F_1}_{L^1\loc}
+ ( n^2 \norm{\eta}_{L^2\loc} +n\rho^2) |F_\theta|
\\
&\quad   + \norm{F_5}_{L^1\loc} + n^{-1} \rho X
\\
&\les  n \rho^4 + n^4 \rho^3  + n \rho (\rho^3 + X) + ( n^2
\norm{\eta}_{L^2\loc} +n\rho^2) (\rho^2 + n^{-1}X)
\\
&\quad + (n \rho^4 + n \rho \norm{\eta^{(3)} }_{L^2\loc} + \wt X)+
n^{-1} \rho X
\\
&\les  n \rho^4 + n^4 \rho^3 + n \rho \norm{\eta^{(3)} }_{L^2\loc}
+ \rho^2  \norm{\eta}_{L^2\loc}
+ n \norm{\eta}_{L^2 \loc} ^2 + \norm{\eta^3}_{L^1 \loc}.
\end{align*}

\bigskip

{\bf Part 2. Ground state.}

We have derived the main oscillatory terms of $a(t)$ in
\eqref{a.dec}
\[
a(t) = a^{(2)}(t) + b(t) ,\qquad
 a^{(2)}=\sum_{k,l\inomega} a_{kl} z_k z_l,
\]
with $b(t)$ given by \eqref{eq:b1}. We have
\begin{equation}  \label{b.eq}
\dot b = (c_1Q, \Im [F-F_1]) + (c_1Q, \Im h) F_\theta  - A_{2,rmd},
\end{equation}
with $A_{2,rmd}$ given in \eqref{A2rmd.def},
\begin{align*} 
A_{2,rmd} & = \Im \sum_{k,l=1}^N \bket{e^{-i(\ev _k+\ev _l)s} i
a_{kl,3}\frac d{d s}(p_k p_l) + e^{-i(\ev _k-\ev _l)s} i
a_{kl,4}\frac d{d s}(p_k \bar p_l) }.
\end{align*}
As for the excited states $p_k$, we want to find a perturbation
$\beta(t)$ of $b(t)$ so that oscillatory terms no smaller than
$n^2 \rho^4$ on the right side of \eqref{b.eq} are removed.
We have observed in Lemma \ref{th:S-2} that $|\dot b| \les n \rho^3
+ n X$ and $|(c_1Q, \Im h)|\les n^3 \norm{h}_{L^1\loc}$
since $h\in M$ is almost orthogonal to $Q$.
Hence the right side of \eqref{b.eq} is of the form
\begin{align}
\dot b &= n \bket{ \NOT{z^3} + \NOT{b z}
+ \NOT{n^{-1}b z^2} + \NOT{n^{-2}b^2 z}
+ \NOT{n z \eta} + \NOT {z^2 \eta^{(2)}} + z^2 \eta^{(3)} + n \eta^2
+ (\eta^3)\loc  } \nonumber
\\
&\quad + n^3 z \bket{\NOT{z^2} + \NOT{\; b}
+ n^{-1}z(\NOT{\; z^2} + \NOT{\; b})
 + O(n^{-2}\rho^4)} + n^3 \eta F_\theta \label{b-eqform}
\\
&\quad + n^2 z \bket{\NOT{nz^2} + \NOT{z^3} + \NOT{b z} + O(\rho^4/n)
+  n z \eta + z^2 \eta + n \eta^2 + (\eta^3)\loc + \cdots}. \nonumber
\end{align}
Here $(\eta^3)\loc$ means terms with same bound as
$\norm{\eta^3}_{L^1 \loc}$. We shall calculate the normal form for
$b$ by integrating by parts those terms with orders which were
crossed out. Notice that there are resonant terms with crossed-out
orders, explicitly, terms of the form $n^2 |z_k|^2 |z_l|^2$.
These terms  cannot be integrated by parts and
will remain on the right hand side. The final normal form
equation is of the form \eqref{beta.eq}. This procedure is the
same as that for excited states and we shall not repeat it in details
but point out a few key steps.

1. There are no terms of the form $b^2$, $b^3$ or $|z_k|^2b$
in the first line of \eqref{b-eqform}.
Terms of these forms 
are eliminated by the $\Im$ operator.

2. Terms of the form $C z_k z_l b^j$ are oscillatory if $k+l \not = 0$.
These terms can be integrated. Similarly, by the observation for
\eqref{A3p}, terms of the form $C z_{m_1} z_{m_2} z_{m_3} $,
$z_m b$ and
$C z_{m_1} z_{m_2} z_{m_3} z_{m_4} z_{m_5}$ are also oscillatory with
nonzero phase factors.
These terms can be integrated.

3.  Most terms of the form $C z_{m_1} z_{m_2} z_{m_3} z_{m_4}$ are
oscillatory.  The only terms with zero phase factor,
$\ev_{m_1}+\cdots+\ev_{m_4} = 0$, are of the form
$C|z_k|^2|z_l|^2$ by the observation for
\eqref{A3p}. These terms cannot
be integrated and will remain in the final equation.
Moreover, in order to survive the $\Im$
operator, these terms from $F-F_1$ must have complex coefficients,
i.e., they must involve $\eta^{(2)}$ and are of the order $nz^2\eta^{(2)}$.

4. We need to integrate terms of the form $n^2 z_j \eta$. They are
from $(c_1Q,\Im 2 \la Q [(\zeta+\bar \zeta) \eta + \zeta \bar \eta])$
in the first line of \eqref{b-eqform}.
Since there is only one $z_j$ involved, the Green's function is approximately
$[H_0 - e_0 \pm (e_j -e_0)]^{-1}$ and is invertible in $L^2$. Hence
there is no resonance with the continuous spectrum.
This integration is carried out in details in \cite[p.193--195]{TY}.
In contrast, resonant terms are of the form $n z_k z_l \eta^{(2)}$ and
are mentioned in point 3.

In conclusion, we can find a perturbation $\beta(t)$ of $b(t)$  of
the form
\begin{align}
\beta &= b +\Re \bket{ \sum C n z_{k_1} z_{k_2} z_{k_3} +
C n z_k b + n^2 (z_k\psi_{k}, \eta) } \nonumber
\\
&\qquad + \Re \sum_{\ev_{k_1}+\cdots+\ev_{k_4}\not = 0}
C z_{k_1} z_{k_2} z_{k_3} z_{k_4}
+ \Re \sum_{\ev_{k}+\ev_{l} \not = 0} C  z_k z_l b   \label{beta.def}
\\
& \qquad +\Re \bket{ C z_{k_1} \cdots z_{k_5}/n
+  C z_{k_1} \cdots z_{k_6}/n^2 }, \nonumber
\end{align}
so that $\beta(t)$ satisfies a normal form equation,
\begin{equation}
\dot \beta = \sum_{1 \le k \le l \le N} B_{kl}|z_k|^2|z_l|^2  + g_b.
\end{equation}
Here $C$ denote complex constants bounded by one,
and $\psi_k$ denote some explicit local functions.
$B_{kl}$ are real constants bounded by $n^2$, and $g_b$ is an
error term of the form
\[
g_b \sim n z^5 + n^4 z^4 + n z^2 \eta^{(3)} +
n^3 z^2 \eta + n^2 \eta^2
+ n \eta^3 + \cdots,
\]
and we have the bounds
$|\beta - b| \les n \rho^3 + n^2 \rho \norm{\eta}_{L^2\loc}$ and
\begin{equation}
|g_b|\les n \rho^5 + n^4 \rho^4 +  n \rho^2
\norm{ \eta^{(3)}}_{L^2\loc} + n^3 \rho^2 \norm{\eta}_{L^2\loc} +
n^2 \norm{\eta}_{L^2\loc}^2 + n \norm{\eta^3}_{L^1\loc}.
\end{equation}

We now compute $B_{kl}$, the coefficients of $|z_k|^2|z_l|^2$.
The main contribution comes from
\begin{equation} \label{S-100}
\bke{c_1 Q,\Im \la \zeta^2 \wbar {\eta^{(2)}} } ,
\end{equation}
where $\la \zeta^2 \wbar {\eta^{(2)}}$ is from $F_5$.
Although there are terms from $\bke{c_1
Q,\Im 2 \la \zeta \bar \zeta \eta^{(2)} }$ and $(c_1 Q, \Im 2\la Q
[(\ell + \bar \ell)\eta^{(2)} + \ell\bar\eta^{(2)}])$ with $\ell=
a^{(2)} R+\eta^{(2)}$, their coefficients are small of order $O( n^4)$.
Since $\eta^{(2)}= \sum_{k,l\inomega} \, U^{-1} \,z_k
z_l \wt{\eta}_{kl}$ , and by \eqref{Upm.dec} $ U^{-1}= U_+^* -
U_-^* \conj$, the expression is \eqref{S-100}
is a sum of terms of the form $z_{k_1}
z_{k_2} z_{k_3} z_{k_4}$. Moreover, since $\wt{\eta}_{kl}$ has
nontrivial imaginary part only if both $k,l>0$, hence
\[
\wbar {\eta^{(2)}} = \sum_{k,l>0} \, \conj U_+^* z_k z_l
\wt{\eta}_{kl} - \sum_{k,l>0} \,   U_-^* z_k z_l \wt{\eta}_{kl} +
\text{irrelevant terms.}
\]
To get $|z_k|^2|z_l|^2$ from $\Im  \la \zeta^2 \wbar {\eta^{(2)}}
$, the relevant terms in $\zeta^2$ are $\nu_{kl} (z_k u_k^+ z_l
u_l^+)$ and $\nu_{kl}(\bar z_k u_k^- \bar z_l u_l^-)$, where
$\nu_{kl} \equiv 2-\delta_k^l$.
Therefore
\begin{align*}
\bke{c_1 Q, \, \Im  \la \zeta^2 \wbar {\eta^{(2)}}  } %
&= \sum_{k,l>0}\, \bke{ c_1 Q, \, \Im  \la \nu_{kl} (z_k u_k^+
z_l u_l^+) \, \bar z_k \bar z_l \conj U_+^*  \wt{\eta}_{kl} } %
\\
& + \sum_{k,l>0}\, \bke{ c_1 Q, \, \Im  \la\nu_{kl} (\bar z_k
u_k^- \bar z_l u_l^-) \,   z_k  z_l  (-1) U_-^* \wt{\eta}_{kl} }+
\text{(*)},
\end{align*}
where (*) denotes terms of the form $ z_{k_1} z_{k_2} z_{k_3}
z_{k_4}$ with  $\ev_{k_1}+\cdots+\ev_{k_4} \not = 0$. Irrelevant
terms with $\ev_{k_1}+\cdots+\ev_{k_4}  = 0$ are eliminated by the
$\Im$ operator. Moreover, we can disregard the second sum since
$U_- =O(n^2)$ is smaller than $U_+$.
By \eqref{3:order} and $\wt{\eta}_{kl}=O( n)$, we
have
\begin{align*} 
B_{kl}&=\bke{ U_+ \bke{ c_1 Q \la \nu_{kl} u_k^+  u_l^+}\, , \,
-\Im \wt{\eta}_{kl} } +O( n^4)
\\ 
&= c_1 \nu_{kl} \bke{U_+ \bke{\la Q \phi_k  \phi_l} \, , \,  \Im
\frac {1}{A-\ev _k-\ev _l-0i} \,  \PcA \Phi_{kl} }+O( n^4) .
\end{align*}
In view of \eqref{ReDkl.def}--\eqref{3-74}, we have
\begin{equation} \label{eq:Bkl}
B_{kl} = - \frac {c_1}2 \, \Re D_{kl}+O( n^4)  .
\end{equation}
The proof of Lemma \ref{th:S-NF} is complete.
\myendproof

\subsection{Main estimates}

Theorem \ref{th:S-1} can be proved using the following proposition and
a continuity argument. Recall
$\rho(t)\equiv [\rho_0^{-2} + N^{-1} \gamma_0  n^2t]^{-1/2}$,
$\Lambda(t)= (1+s) ^{-1/2}\rho^2(s)$, and
$D= 6 N |c_1| \gamma_0^+ /\gamma_0$.

\begin{proposition}  \label{th:S-4}
Suppose the assumptions of Theorem \ref{th:S-1} hold.
Suppose for a fixed $T> 0$ we can find the best approximation
$Q_{E(T)}$ of $\psi(T)$, i.e., $a_{E}(T)=0$ in the
decomposition \eqref{psi.dec2} of $\psi(T)$ with $E=E(T)$. Define
\begin{equation}
\begin{split}
M_T &\equiv \sup _{0\le t \le T} \max \Big \{  \rho(t)^{-1}
(\ssum_{k=1}^N |z_k(t)|^2)^{1/2} , \quad 2D^{-1}\rho^{-2}(t)|a(t)|,
\\
&\qquad
n^{-4/5}\rho(t)^{-8/5}\norm{\eta(t)}_{L^5},\quad
 [\Lambda(t) +  n^{1/3} \rho^{8/3}(t)]^{-1}
\norm{\eta^{(3)} (t)}_{L^2 \loc} ,
\\
&\qquad
2(3D)^{-1} \rho_0^{-2}|E(T) - E_0|
 \Big \}.
\end{split}
\end{equation}
Suppose, furthermore, $M_T \le 2$. Then we have $M_T\le 3/2$.
\end{proposition}

\myproof Since $M_T\le 2$, we have $|E(T) - E_0|\le 3 D \rho_0^{2}$
and, for $t \in [0,T]$,
\begin{equation} \label{S-pfasp}
\begin{split}
&(\ssum_{k=1}^N |z_k(t)|^2)^{1/2}\le 2 \rho(t) , \qquad |a(t)|\le D \rho(t)^2,
\\
&  \norm{\eta(t)}_{L^5} \le 2  n^{4/5}\rho(t)^{8/5},
\\
& \norm{\eta^{(3)} (t)}_{L^2 \loc} \le 2 \Lambda(t)
+ 2 n^{1/3} \rho^{8/3} (t) .
\end{split}
\end{equation}
Since $\eta=\eta^{(2)}+\eta^{(3)}$,
$\norm{\eta^{(2)}}_{L^2 \loc} \les n\rho^2$ and $\rho \le \epz n$,
\begin{equation}
\norm{\eta(t)}_{L^2 \loc} \le C n \rho^2 + \norm{\eta^{(3)}
(t)}_{L^2 \loc} \le C n \rho^2 + 2 \Lambda(t).
\end{equation}
Note that $\Lambda(t) \le \rho^2$ by its definition.
Therefore $\norm{\eta(t)}_{L^2 \loc} \le C \rho^2$,
$\norm{\eta(t)}_{L^2 \loc\cap L^{5}} \le C n^{2/5}\rho^{8/5}$,
and the assumptions of Lemmas \ref{th:S-2} and \ref{th:S-NF}
are satisfied with $\rho =\rho(t)$.

We have, using Lemma \ref{th:2-2}, \eqref{S-2} and \eqref{S-pfasp},
\begin{equation}
|E(T)-E_0| \le |E(T) - E(0)| + |E(0)-E_0| \le
\tfrac 98 [|a_{E(T)}(0)| + |a_{E_0}|] \le \tfrac 94 D \rho_0^2.
\end{equation}

Since $\norm{\psi(t)}_{L^2} = \norm{\psi_0}_{L^2} \ll 1$, we have
$\norm{\eta(t)}_{L^2}\ll 1$.
By H\"older inequality,
\begin{equation}\label{3-104}
\begin{split}
&\norm{\eta^3}_{L^{5/4}} \les \norm{\eta}_{L^2}^{2/3} \norm{\eta}_{L^5}^{7/3}
\les o(1) (n^{4/5}\rho(t)^{8/5})^{7/3} = o(1) n^{28/15} \rho^{56/15},
\\
&\norm{\eta^3}_{L^{1}} \les \norm{\eta}_{L^2}^{4/3} \norm{\eta}_{L^5}^{5/3}
\les o(1) (n^{4/5}\rho(t)^{8/5})^{5/3} = o(1) n^{4/3} \rho^{8/3},
\\
& \norm{\eta^3}_{L^{1}\loc} \les
\norm{\eta}_{L^2 \loc}^{4/3} \norm{\eta}_{L^5}^{5/3}
\les  \norm{\eta}_{L^2 \loc}^{4/3} n^{4/3} \rho^{8/3}.
\end{split}
\end{equation}
Recall $\wt X= \rho^2  \norm{\eta}_{L^2\loc}
+ n \norm{\eta}_{L^2 \loc} ^2 +\norm{\eta^3}_{L^{1}\loc}$ and
$X \les   n \rho  \norm{\eta}_{L^2\loc} + \wt X$.
We have
\begin{equation}\label{3-105}
\rho^2  \norm{\eta}_{L^2\loc}
+ n \norm{\eta}_{L^2 \loc} ^2
\les  \rho^2 ( n \rho^2 + \Lambda) + n ( n \rho^2 + \Lambda)^2
\les n \rho^4 +  \rho^2 \Lambda + n \Lambda^2.
\end{equation}
Since $\norm{\eta^3}_{L^{1}\loc} \ll \rho^2  \norm{\eta}_{L^2\loc}$ by
\eqref{3-104}, we have
\begin{equation} \label{X.est}
\wt X \les  n \rho^4 +  \rho^2 \Lambda + n \Lambda^2,
\qquad
X \les n^2 \rho^3 +  n \rho \Lambda + n \Lambda^2.
\end{equation}

We now estimate $\eta(t)$. Recall from \eqref{eq:eta},
\eqref{eq:eta1} that $\eta(t) = U^{-1} e^{-i \theta(t)} \tilde
\eta (t)$ and
\[
\tilde \eta(t) = U e^{t\L} \eta(0) + \int_0^t e^{-iA (t-s)} \PcA
F_\eta (s) \, d s .
\]
By Lemma \ref{th:S-2}, \eqref{S-pfasp},
\eqref{3-104}, $\norm{\eta(t)}_{L^2 \loc} \le C \rho^2$, and
$\norm{\eta(t)}_{L^2 \loc\cap L^{5}} \le C n^{2/5}\rho^{8/5}$,
\begin{align*}
\norm{F_{\eta}}_{L^{5/4}}
&\les n \rho^2 + n\rho  \norm{\eta}_{L^2\loc}
+ n \norm{\eta}_{L^2 \loc \cap L^5} ^2 + \norm{\eta^3}_{L^{5/4}} + n^2 \rho^2
\norm{\eta}_{L^{5}}
\\
&\les n \rho^2 + n\rho  (\rho^2) +
n(n^{2/5}\rho^{8/5})^2 + n^{28/15} \rho^{56/15} +  n^2 \rho^2(n^{4/5}\rho^{8/5})
\les n \rho^2 .
\end{align*}
Using \eqref{out:thS1} for $\eta(0)$, the decay estimates of
$e^{t\L}$ and $e^{-itA}$, and the boundedness in Sobolev spaces of
$U$ and $U^{-1}$ from Lemma \ref{th:AU}, we have
\begin{align*}
\norm{\eta(t)}_{L^5} &\le \norm{e^{t\L} \eta(0)}_{L^{5}} +
C\int_0^t |t-s|^{-9/10} \norm{F_\eta(s)}_{L^{5/4}} \, ds
\\
&\le n^{4/5} \rho^{8/5} + C\int_0^t |t-s|^{-9/10} n
\rho(s)^{2} \, d s \le n^{4/5} \rho(t)^{8/5} + C n^{4/5} \rho_0^{2r}
\rho^{9/5-2r},
\end{align*}
for any $r>0$.
Here we have used $\rho(s) = C n^{-1} (\Dt + s)^{-1/2}$ with $\Dt =
C n^{-2}\rho_0^{-2}$ and
\begin{equation} \label{S-99}
\int_0^t |t-s|^{-9/10} (\Dt + s)^{-1}\, d s \le \Dt ^{-r} (\Dt
+ t)^{-9/10+r},
\end{equation}
for any $r>0$. Taking $2r=1/5$,
we get $\norm{\eta(t)}_{L^5}  \le \tfrac 32 n^{4/5} \rho(t)^{8/5}$.
\donothing{
{\it Proof of \eqref{S-99}}: If $t<\Dt$, the integral is bounded by
$\Dt^{-1} \int_0^t |t-s|^{-9/10} ds = \Dt^{-1} t^{1/10} \le
\Dt^{-9/10} \sim (\Dt +t)^{-9/10}$. If $t>\Dt$, we have $t \sim \Dt
+ t$ and the integral is bounded by
\begin{align*}
&\int_0^{t/2} t^{-9/10} (\Dt+s)^{-1}ds + \int_{t/2}^t |t-s|^{-9/10}
t^{-1} ds
\\
&\qquad = C t^{-9/10} \log \frac{\Dt+t}{\Dt} + C t^{-9/10}
\\
&\qquad \sim (\Dt + t)^{-9/10} (1+\log \frac{\Dt+t}{\Dt})= \Dt
^{-r} (\Dt + t)^{-9/10+r} C(t),
\end{align*}
where $C(t) = (\frac{\Dt+t}{\Dt})^{-r} (1+\log
\frac{\Dt+t}{\Dt})$ is uniformly bounded for all $t\ge 0$.
} 

We now consider the $L^2 \loc$ estimates of $\eta$. Recall
$\eta^{(3)}=\eta^{(3)}_1+\cdots+\eta^{(3)}_4$. The estimate of
$\eta^{(3)}_1$ is by \eqref{out:thS1} and that $U^{-1} e^{-i \theta}U
= e^{-i \theta} + O(1) [i,U] =  e^{-i \theta} + O(n^2)$,
\[
\norm{\eta^{(3)}_{1}(t)}_{L^2 \loc} \le (1+o(1)) \Lambda(t).
\]
By the singular decay estimate \eqref{eq:32-2} in Lemma \ref{th:AU}
and the definition \eqref{eta.kl} of $\wt \eta_{kl}$
with $\wt \eta_{kl}=O(n)$,
we have
\[
\norm{\eta^{(3)}_{2}(t)}_{L^2 \loc} \les  n\rho_0^2 (1+t)^{-3/2}
\ll \Lambda(t),
\]
\[
\norm{\eta^{(3)}_{3}(t)}_{L^2 \loc} \les  \int_0^t
\bka{t-s}^{-3/2} n(\rho^2 |\dot \theta|+\rho \max_k|\dot p_k| )\,
ds .
\]
By the bounds of $\dot \theta$ and $\max_j|\dot p_j|$ in Lemma \ref{th:S-2},
and $\Lambda \le \rho^2$,
we have
\begin{align*}
&n(\rho^2 |\dot \theta|+\rho \max_k|\dot p_k| ) \les
n\rho^2 (\rho^2 + n^{-1} X) + n\rho ( n \rho^2 + X) \les n^2 \rho^3 + n \rho X
\\
&\les n^2 \rho^3 + n \rho (n^2 \rho^3 + n \rho \Lambda + n\Lambda ^2)
\les  n^2 \rho^3 .
\end{align*}
Hence
\[
\norm{\eta^{(3)}_{3}(t)}_{L^2 \loc} \les  \int_0^t \bka{t-s}^{-3/2}
 n^2 \rho^3 (s) \, ds
\les  n^2 \rho^3 (t).
\]
Finally, by Lemma \ref{th:S-2}, \eqref{3-104},
$\norm{\eta(t)}_{L^2 \loc\cap L^{5}} \le C n^{2/5}\rho^{8/5}$,
and $\Lambda \le \rho^2$,
\begin{align*}
\norm{F_{\eta,3}}_{L^{5/4}\cap L^1}
&\les  \rho^3 + n\rho  \norm{\eta}_{L^2\loc}
+ n \norm{\eta}_{L^2 \loc\cap L^5} ^2 + \norm{\eta^3}_{L^{5/4}\cap L^1}
+ n^2 \rho^2 \norm{\eta}_{L^{5}}
\\
&\les \rho^3 + n\rho  (n\rho^2 + \Lambda) +
n(n^{2/5}\rho^{8/5})^2 + o(1) n^{4/3} \rho^{8/3}
+ n^2 \rho^2(n^{4/5}\rho^{8/5})
\\
&\les \rho^3 + n^{4/3} \rho^{8/3} + n \rho \Lambda.
\end{align*}
Hence, bounding the integrand of $\eta^{(3)}_{4}$ by either $L^\infty$ or
$L^5$-norm, we have
\begin{align*}
\norm{\eta^{(3)}_{4}(t)}_{L^2 \loc} &\le  \int_0^t \min
\bket{|t-s|^{-3/2}, |t-s|^{-9/10}}
\norm{F_{\eta,3}(s)}_{L^{5/4}\cap L^1} \, ds
\\
&\les \int_0^t \min \bket{|t-s|^{-3/2}, |t-s|^{-9/10}}
[\rho^3 +n^{4/3} \rho^{8/3} +  n  \rho \Lambda ] (s) \, ds
\\
&\les [\rho^3 + n^{4/3} \rho^{8/3} + n  \rho \Lambda](t) .
\end{align*}
%
Summing the estimates, we conclude
\[
\norm{\eta^{(3)}(t)}_{L^2 \loc} \le \sum_{j=1}^4
\norm{\eta^{(3)}_{j}}_{L^2 \loc} \le \tfrac 54 \Lambda +C \rho^3 +
C n^{4/3} \rho^{8/3} \le \tfrac 54 (\Lambda(t) + n^{1/3}\rho^{8/3})(t) .
\]

We now estimate the error terms $g_k$ and $g_b$.
Using $\rho\le \epz n$, \eqref{S:gk.est}, \eqref{S:gb.est},
\eqref{S-pfasp}--\eqref{X.est} and $\Lambda \le \rho^2$, we have
\begin{align}
|g_k|&\les n \rho^4 + n^4 \rho^3
+ n \rho \norm{\eta^{(3)} }_{L^2\loc} + \wt X \nonumber
\\
&\les n \rho^4 + n^4 \rho^3  + n \rho(\Lambda +
n^{1/3} \rho^{8/3}) + ( n \rho^4 +  \rho^2 \Lambda + n \Lambda^2)\nonumber
\\
&\les (\epz^{2/3} + n^2) n^{2} \rho^3 + n \rho \Lambda .
\label{S:gk.est2}
\end{align}
\begin{align}
|g_b|&\les n \rho^5 + n^4 \rho^4 +  n \rho^2
\norm{ \eta^{(3)}}_{L^2\loc} + n^3 \rho^2 \norm{\eta}_{L^2\loc} +
n^2 \norm{\eta}_{L^2\loc}^2 + n \norm{\eta^3}_{L^1\loc}\nonumber
\\
&\les n \rho^5 + n^4 \rho^4  + n \rho^2 (\Lambda + n^{4/3} \rho^{8/3})
+ n^3 \rho^2(n\rho^2 + \Lambda) + n^2(n\rho^2 + \Lambda)^2\nonumber
\\
& \qquad + n (n\rho^2 + \Lambda)^{4/3} n^{4/3} \rho^{8/3}\nonumber
\\
&\les (\epz^{2/3} + n^2) n^2 \rho^4 + n \rho^2 \Lambda .
\label{S:gb.est2}
\end{align}

We now estimate $a(t)$. By \eqref{q-p.est} of Lemma \ref{th:S-NF} we
have
\[
|\beta-a| \le |\beta-b| + |b-a|
\le (C n \rho^3 + C n^2 \rho \norm{\eta}_{L^2\loc})
 + C n^2 \rho^2.
\]
Since $\norm{\eta}_{L^2\loc} \les  \rho^2$,
we get $|\beta-a| \le C n^2 \rho^2$.
Since $a(T)=0$, we have $|\beta(T)| = |\beta(T)-a(T)| \le
Cn^2\rho^2(T)$. Using \eqref{beta.eq} and \eqref{S:gb.est2} for $g_b$, we have
\begin{align*}
|\beta(t)| &\le |\beta(T)|
+ \int_t^T \sum |B_{kl}| |z_k|^2|z_l|^2 + |g_b(s)| d s
 \\
& \le C n^2\rho^2(T) +  \int_t^T (\max_{k,l} |B_{kl}|) \rho^4 + C\bkt{
(\epz^{2/3} + n^2) n^2 \rho^4 + n \rho^2 \Lambda}(s)  d s
 \\
&\le  \frac 12 D \rho^2(t) + o(1) \rho^2(t) \le \frac 58 D
\rho^2(t).
\end{align*}
Here we have used  $\rho(s) = (\rho_0^{-2} +  N^{-1}\gamma_0 n^2 s)^{-1/2}$
and $(\max _{k,l} |B_{kl}|)/(N^{-1} \gamma_0 n^2) \le D/2$.
We also have used $\Lambda(s) = n^{1/2}(1+s)^{-1/2}\rho^2(s)$,
$\rho(s) = C n^{-1} (\Dt + s)^{-1/2}$ with $\Dt = C(n\rho_0)^{-2}$,
and the following estimate: for $t\ge 0$ and $\Dt \ge 1$, $m,r>0$,
$m+r >1$,
\begin{equation}
\int_t^\infty (\Dt + s)^{-m} (1+s)^{-r} ds
\le \int_t^\infty (\Dt + s)^{-m-r} ds
\le C (\Dt + s)^{-m-r+1}.
\end{equation}
%
Hence
\[ |a(t)| \le |\beta(t)| + |a(t)-\beta(t)| \le \frac 58 D
\rho^2(t) + C n^2 \rho^2(t) \le \frac 34 D \rho^2(t).
\]

We now estimate the excited states $z_k(t)$. For their initial value,
we have
\begin{equation}
\norm{(\zeta_E - \zeta_{E_0})(0)} \le C n^{-1}(\rho_0/n)|E-E_0|
\le C n^{-2} \rho_0^3,
\end{equation}
by  \eqref{shift} of Lemma \ref{th:2-2} and $|E-E_0|\le 3D \rho_0^2$.
In particular,
$(\sum_{k=1}^N {z_{k,E}(0)}^2)^{1/2} \le \tfrac 58 \rho_0$
by \eqref{S-2}.
Let $f_k=|q_k|^2$.  By \eqref{q-p.est}, $f(0)\le \tfrac 12 \rho_0^2$.
By \eqref{qk.eq} and that $J_k(t)$ are real,
we have
\begin{equation} \label{3-111}
\dot f_k = \sum_{l=1}^N 2\Re D_{kl}\, f_l \, f_k \, + 2\Re \bar
q_k g_k,\qquad (k=1,\ldots, N).
\end{equation}
Let $f=f_1+ \cdots +f_N$. Since $|\Re D_{kl}|
\le 5 \gamma_0^+ n^2$, summing \eqref{3-111} over $k$ we get
\[
|\dot f| \le   (10 \gamma_0^+ n^2) f^2 + 2N \max_{k}|q_k
g_k|  \le C n ^2 \rho^4
+ C\rho ((\epz ^{2/3}+ n^2)n^{2} \rho^3 + n \rho \Lambda),
\]
by \eqref{S:gk.est2}. Let $t_0 = n^{-3} \ll n^{-2} \rho_0^{-2}$.
For $t\in [0,t_0]$,
$\Lambda(t) \le \rho_0^2 (1+t)^{-1/2}$ and
\[
|f(0)-f(t)| \les \int_0^t n ^2 \rho^4
+ n \rho^2 \Lambda  ds
 \les n^2 \rho_0^4 t_0 + n \rho_0^4 \sqrt{t_0} \ll \rho_0^2,
\]
since $t_0 \ll n^{-2} \rho_0^{-2}$.
We conclude $f(t) \sim f(0)$ for $t\in [0,t_0]$. For $t \ge t_0$, we have
\begin{equation} \label{3-118}
|g_k(t)|\les (\epz^{2/3} + n^2)n^{2} \rho^3 + n \rho \Lambda
\les (\epz^{2/3} + n^{1/2})n^{2} \rho^3.
\end{equation}
Since the positive part of $ \Re D_{kl}$ is bounded
by $n^4$ and $\Re D_{kk} \le -\gamma_0 n^2$, we have
\[
\dot f \le   -\sum_{k=1}^N 2\gamma_0  n^2 f_k^2
+ C n^4 f^2 + 2N \max_{k}|q_k
g_k|  \le - \frac 3{2N} \gamma_0 n^2 f^2 + 2 N \max_{k}|q_k g_k| .
\]
Using $\rho^2(t) = (\rho_0^{-2} +  N^{-1}\gamma_0 n^2 t)^{-1}$ and
\eqref{3-118}, we have
\begin{align*}
\frac d{dt} (\rho^2) &=  - N^{-1}\gamma_0 n^2 \rho^4
\ge  - \frac 3{2N} \gamma_0 n^2 (\rho^2)^2 + 2N \max_{k}|q_k g_k| .
\end{align*}
Since we have shown previously that $\rho^2(t) \ge f(t)$ for $t\in [0,t_0]$,
in particular $\rho^2(t_0) \ge f(t_0)$, we have
$\rho^2(t) \ge  f(t)$ for all $t \ge t_0$
by comparison principle.
Thus we have $\rho^2(t) \ge  f(t)$ for all $t\ge 0$.
\myendproof


We now prove Theorem \ref{th:S-1} using Proposition
\ref{th:S-4} and a continuity argument.

\textsc{Proof of Theorem \ref{th:S-1}:} By the assumption of
Theorem \ref{th:S-1}, we have $M_T \le 3/2$ for $T=0$. Denote by
$J$ the set of all $T$ such that we can find a best approximation
$Q_{E(T)}$ of $\psi(T)$ and we have $M_T \le 3/2$ with respect to $E=E(T)$.
Clearly $J$ is a closed interval containing $0$ by
the continuity of the Schr\"odinger equations.

Suppose $T' \in J$. By continuity there is a $\delta>0$ which may
depend on $\psi_0$ and $T'$ such that, for all $T\in
[T',T'+\delta]$, there is a best approximation $Q_{E(T)}$ of
$\psi(T)$, and with respect to $E=E(T)$ we have $M_T \le 2$. By
Proposition \ref{th:S-4}, we have $M_T \le 3/2$. Hence
$(0,T'+\delta) \subset J$. This shows the right end of $J$ is open
and hence $J=[0,\infty)$.

For $t<T$, we have $|E(t) - E(T)| \le \frac 54 |a_T(t)|\le
 D \rho(t)^2$ by Lemma \ref{th:2-2} and Proposition \ref{th:S-4}.
This uniform bound shows $E(t)$
has a unique limit $E_\infty$ as $t \to \infty$ and $|E(t) -
E_\infty| \le D \rho(t)^2$.
By continuity of $M_T$ in $T$ we have $M_\infty \le 3/2$.

Finally we prove the lower bound \eqref{lowerbound}.
If $(\ssum_{k=1}^N \norm{\zeta_{k}}_{L^2}^2)^{1/2} \ge \tfrac 14 \rho_0$,
we have $f(0)\ge C \rho_0^2$. We have shown previously that $f(t) \sim f(0)$
for $t \le t_0$.
Summing \eqref{3-111} over $k$ and using the error estimates, we have,
for $t\ge t_0$,
\[
\dot f \ge - 10 \gamma_0^+ n^2 f^2 + 2 N \max_k |q_k g_k|
\ge - 12 \gamma_0^+ n^2 f^2.
\]
By a similar comparison argument
we have  $f(t) \ge (f(t_0)^{-1} +  12 \gamma_0^+ n^2 (t-t_0))^{-1}$. Hence
$f(t)\ge C \rho^2(t)$ for all $t\ge 0$.
This completes the proof of Theorem \ref{th:S-1}. \myendproof

We remark that, since $B_{kl}$ are (almost) positive, $\beta(t)$ and
hence $a(t)$ are increasing as $t \to \infty$. This shows that the ground
state gains energy from excited states even in the stabilization
regime.

%
%

\section{Transition Regimes}

In this section we study the dynamics in the transition regime.
We will prove Theorem \ref{th:1-1} using Theorem \ref{th:S-1}.
In this regime the natural operator is $H_0=-\Delta+V$ and the
natural decomposition is
\[
\psi =  x_0 \phi_0 + \cdots + x_N \phi_N + \xi,\qquad
\xi \in \Hc(H_0).
\]

\subsection{Equations}

Recall $H_0=-\Delta + V$ has $N+1$ simple eigenvalues $e_k$,
$k=0,\ldots N$, with corresponding normalized eigenvectors $\phi_k$:
$H_0 \phi_k=e_k \phi_k$, $\norm{\phi_k}_2=1$.
We can decompose the solutions $\psi(t)=\psi(t,x)$ respect to
$H_0$:
\begin{equation} \label{4-1}
\psi(t)= \chi(t)+ \xi(t), \qquad \chi(t)= \sum_{k=0}^N x_k(t)
\phi_k,
\end{equation}
where $x_k(t)\in \Complex$ and $\xi(t) \in \Hc(H_0)$. Substituting
\eqref{4-1} into \eqref{Sch}, we obtain the following system for
these components:
\begin{equation} \label{4-2}
\left \{
\begin{aligned}
i \dot x_k &= e_k x_k + (\phi_k, \, G), \qquad(k=0,\cdots, N),
\\
i\pd_t \xi &= H_0 \xi + \PcH G ,\qquad G=\la\psi^2 \bar \psi ,
\end{aligned}
\right .
\end{equation}
with initial conditions $x_k(0)=x_k^0 $ and $\xi(0)=\xi_0$. From
these equations we find that each $x_k(t)$ is oscillatory with a
main oscillation factor $e^{-i e_k t}$. We say that the {\it phase
factor} of $x_k$ is $-e_k$. Define $u_k(t)$ by
\begin{equation} \label{uk.def}
x_k(t) = e^{-i e_k t} \, u_k(t).
\end{equation}
The function $u_k(t)$ has the same magnitude as $x_k(t)$ but is not as
oscillatory. In particular, $|\dot u_k|$
is smaller than $|\dot x_k|$.
We shall study the following system for $u_k$ and $\xi$, which is
equivalent to \eqref{4-2}:
\begin{equation} \label{uk.eq0}
\dot u_k(t) = e^{i e_k t} \, (\phi_k,\, G(t)), \qquad (k=0,\ldots,N),
\end{equation}
\begin{equation} \label{xi.eq}
\xi(t) = e^{-i H_0  t}\xi_0  +\int_0^t e^{-i H_0 (t-s)} \, \PcH
i^{-1} G (s)\, d s , \qquad G=\la \psi^2 \bar \psi .
\end{equation}

\subsection{Decompositions of $G$ and $\xi$}

It is useful to decompose various terms according to their orders in $n$,
so that we can identify the main terms.
We expect that $x_k=O(n)$ and $\xi=O(n^3)$ locally after an initial layer
of time. We first decompose $G=\psi^2\bar \psi$. Using \eqref{4-1} that
$\psi=\chi + \xi$ with $\chi=\ssum_{k=0}^N x_k \phi_k$, we decompose $G$ as
\begin{equation}
G=  \la \psi^2\bar \psi =\la(\chi+\xi)^2 (\bar \chi+\bar \xi)= G_3
+ G_5 + G_7,
\end{equation}
where
\begin{align}
\label{G3.def} G_3 &=  \la \chi ^2 \bar \chi,
\\
\label{G5.def} G_5 &= \la \chi ^2 \wbar \xi + 2\la |\chi |^2 \xi,
\\
\label{G7.def} G_7 &= \la  \wbar \chi \xi^2  + 2\la \chi| \xi|^2 +
\la \xi^2 \bar \xi.
\end{align}
Note that $G_3=O(n^3)$, $G_5=O(n^5)$ and $G_7=O(n^7)$.

We now identify the main term of $\xi$ using the integral
equation \eqref{xi.eq}. The main term of the integrand is $i^{-1}G_3$.
Using \eqref{uk.def} and factoring out the
main phase factors in $G_3$ we have
\begin{equation} \label{G3.def2}
G_3(t) = \sum_{l,m,j=0}^N e^{i(-e_l-e_m+e_j)t} \, u_l u_m \bar
u_j(t) \phi_{lmj} ,
\end{equation}
where
\begin{equation} \label{philmj.def}
\phi_{lmj}=\la \phi_l \phi_m \phi_j.
\end{equation}
We now integrate by parts a typical term in $i^{-1}G_3$:
\begin{align*}
& \int_0^t e^{-i H_0  (t-s)} \, \PcH   i^{-1} e^{i(-e_l-e_m+e_j)s}
\, u_l u_m \bar u_j(s) \phi_{lmj} \, d s
\\
&=\lim_{r \to 0+} i^{-1}  e^{-i H_0  t} \int_0^t  e^{is (H_0
-e_l-e_m+e_j -r i)} \, u_l u_m \bar u_j(s) \PcH \phi_{lmj} \,
d s
\\
&=\lim_{r \to 0+} i^{-1}  e^{-i H_0  t}   \frac 1{ i (H_0
-e_l-e_m+e_j -r i)}
\\
& \qquad \qquad \Bigg \{ \bkt{ e^{i H_0 s} \, e^{i(-e_l-e_m+e_j)s}
\, u_l u_m \bar u_j(s)  \PcH \phi_{lmj} }_0^t
\\
& \qquad \qquad \quad - \int_0^t  e^{i H_0  s} \,
e^{i(-e_l-e_m+e_j)s} \frac d{d s} \bke{ u_l u_m \bar u_j} \PcH
\phi_{lmj} \, d s \Bigg \}.
\end{align*}
We need to take a limit since $H_0+e_j -e_l-e_m$ may not be
invertible.
Define
\begin{equation} \label{Klmj.def}
K_{lm}^j \equiv \lim_{r \to 0+} \frac {1}{H_0 -e_l-e_m +e_j-r
i} \,\PcH  .
\end{equation}
\begin{equation} \label{xilmj.def}
\xi_{lm}^j \equiv - K_{lm}^j  \PcH \phi_{lmj} .
\end{equation}
$K_{lm}^j$ are bounded operators in ${\cal B}(L^2_r,L^2_{-r'})$
with $r,r'>1/2$ and $r+r'>2$, see \cite{JK}. Hence we
have $\xi_{lm}^j \in L^2_{-r'}$. We may have $\xi_{lm}^j \not \in
L^2$ and $\Im \xi_{lm}^j  \not = 0$ only if
\begin{equation}
  e_j -e_l-e_m < 0 ,
\end{equation}
in particular $j<l,m$.
We can now rewrite the above integral as
\begin{align*}
& x_l x_m \bar x_j(t) \, \xi_{lm}^j - x_l x_m \bar
x_j(0) \, e^{-i H_0  t} \, \xi_{lm}^j\\
&\quad - \int_0^t  e^{-i H_0 (t- s)} \, e^{i(-e_l-e_m+e_j)s} \frac
d{d s} \bke{ u_l u_m \bar u_j} \xi_{lm}^j \, d s .
\end{align*}
Note that the choice $r \to 0+$, instead of  $r \to 0-$,
ensures the local decay of $e^{-i H_0 t} \, \xi_{lm}^j = -e^{-i
H_0 t} \, K_{lm}^j \PcH \phi_{lmj}$ and of the last integral, see
Lemma \ref{th:2-3}.

Summing these terms over $l,m,j$, we can decompose $\xi(t)$ as
\begin{equation}\label{xi.dec}
\xi(t)= \xi^{(2)}(t)+\xi^{(3)}(t) = \xi^{(2)} +\bke{\xi^{(3)}_1 +
\cdots + \xi^{(3)}_5},
\end{equation}
where $\xi^{(2)}(t)$ is the main part of $\xi(t)$,
\begin{equation}\label{xi2.def}
\xi^{(2)}(t)= \sum_{l,m,j=0}^N x_l x_m \bar x_j (t)\, \xi_{lm}^j,
\end{equation}
and $\xi^{(3)}(t)=\xi^{(3)}_1 + \cdots + \xi^{(3)}_5$ is the rest,
\begin{equation}\label{xi3.dec}
\begin{split}
\xi^{(3)}_1(t)&= e^{-i H_0  t} \xi_0,
\\
\xi^{(3)}_2(t)&= -e^{-i H_0  t}\xi^{(2)}(0),
\\
\xi^{(3)}_3(t)&= - \int_0^t e^{-i H_0 (t-s)} \, \PcH \,
 \sum_{l,m,j=0}^N e^{i(-e_l-e_m+e_j)  s} \frac d{d s}
\bke{ u_l u_m \bar u_j} \, \xi_{lm}^j  \, d s,
\\
\xi^{(3)}_4(t)&=\int_0^t e^{-i H_0 (t-s)} \, \PcH  i^{-1}
\bke{G-G_3-\la \xi^2 \bar \xi} \, d s ,
\\
\xi^{(3)}_5(t)&=\int_0^t e^{-i H_0 (t-s)} \, \PcH  i^{-1}
\bke{\la \xi^2 \bar \xi} \, d s .
\end{split}
\end{equation}
We single out $\la |\xi|^2\xi$ since it is
the only non-local term in $G-G_3$.

We have the following estimates for nonlinear terms.
\begin{lemma} \label{th:4-1}
Suppose, for a fixed time $t$, for some $n\le n_0$,
\begin{equation} \label{4-18}
|x_k(t)|\le 2n, \qquad \norm{\xi(t)}_{L^2 \loc \cap L^5} \le 2n, \qquad
\norm{\xi(t)}_{L^2 } \ll 1.
\end{equation}
We have
\begin{equation} \label{4-21}
\begin{split}
\norm{G_7}_{L^1 \loc} &\les n   \norm{\xi}_{L^2\loc}^2
+ \norm{\xi}_{L^2\loc}^{4/3} \norm{\xi}_{L^5} ^{5/3},
\\
\norm{G_5+G_7}_{L^1\loc} &\les  n^2 \norm{\xi}_{L^2\loc},
\\
\norm{G}_{L^1 \loc}+\max_k |\dot u_k| &\les n^3,
\\
\norm{G_5+G_7-\la \xi^2\bar \xi}_{L^{1} \cap L^{5/4}}
&\les n^2 \norm{\xi}_{L^2\loc}.
\end{split}\end{equation}
\end{lemma}

\myproof
For the nonlocal term $\la \xi^2 \bar \xi$ we have, using
H\"older inequality,
\begin{equation}\label{4-22}
\norm{\la \xi^2 \bar \xi}_{ L^{5/4}}
\les \norm{\xi}_{L^2}^{2/3} \norm{\xi}_{L^5} ^{7/3},\qquad
\norm{\la \xi^2 \bar \xi}_{L^1}
\les \norm{\xi}_{L^2}^{4/3} \norm{\xi}_{L^5} ^{5/3}.
\end{equation}
Similarly, $\norm{\la \xi^2 \bar \xi}_{L^1 \loc}
\les \norm{\xi}_{L^2\loc}^{4/3} \norm{\xi}_{L^5} ^{5/3}$.
Using the definition \eqref{G7.def} of $G_7$, we have
\begin{align*}
\norm{G_7}_{L^1 \loc } &\les n   \norm{\xi}_{L^2\loc}^2
+ \norm{\xi}_{L^2\loc}^{4/3} \norm{\xi}_{L^5} ^{5/3}.
\end{align*}
By \eqref{G3.def} and \eqref{G5.def}, we have $\norm{G_3}\les
n^3$ and $\norm{G_5} \les n^2 \norm{\xi}_{L^2\loc}$.
By \eqref{4-18},
\[
\norm{G_5+G_7}_{L^1\loc} \les n^2 \norm{\xi}_{L^2\loc}
+ n   \norm{\xi}_{L^2\loc}^2
+ \norm{\xi}_{L^2\loc}^{4/3} \norm{\xi}_{L^5} ^{5/3}
 \les n^2 \norm{\xi}_{L^2\loc},
\]
and $\norm{G}_{L^1\loc} \le \norm{G_3}_{L^1\loc}
+ \norm{G_5+G_7}_{L^1\loc} \les n^3 + n^2 \norm{\xi}_{L^2\loc}\les n^3$.
Since $|\dot u_k|\le \norm{G}_{L^1\loc}$ by \eqref{uk.eq0},
the estimate for $\dot u_k$ follows. Finally, $G_7-\la \xi^2\bar \xi$
is of the form $n \xi^2$ and
\[
\norm{G_7-\la \xi^2\bar \xi}_{L^{1} \cap L^{5/4}} \les  n
\norm{\xi}_{L^2 \loc \cap L^5}^{2/3} \norm{\xi}_{L^2
\loc}^{4/3}\les n^2 \norm{\xi}_{L^2\loc}
\]
by \eqref{4-18}. Since $\norm{G_5}_{L^{1} \cap L^{5/4}}
\les n^2 \norm{\xi}_{L^2\loc}$, the last estimate follows.
\myendproof

\subsection{Normal forms for equations of bound states}

Recall we write  $x_k(t) = e^{- ie_k t} u_k(t)$.
In this subsection we derive the normal form for the equations of
$\dot u_k$, where terms of different phase factors
are removed. 

\begin{lemma} [Normal form] \label{th:4-2}
Suppose for some $n\le n_0$,
\begin{equation}
|x_k(t)|\le 2 n, \qquad \norm{\xi(t)}_{L^2 \loc \cap L^5} \le 2 n, \qquad
\norm{\xi(t)}_{L^2} \ll 1.
\end{equation}
There are perturbations $\mu_k(t)$ of $u_k(t)$, $k=0,1,\ldots,N$,
to be defined in \eqref{muk.def}, satisfying
\begin{equation}\label{ukmuk.diff}
|u_k(t)-\mu_k(t)|\le C n^3 ,
\end{equation}
such that
\begin{equation} \label{eq:NF}
\dot \mu_k =\sum_{l=0}^N c_{l}^{k}|\mu_{l}|^2 \mu_{k}
+\sum_{a,b=0}^N d_{ab}^k \, |\mu_a|^2 |\mu_b|^2 \mu_k + g_{k}.
\end{equation}
Here $g_k$ are error terms, to be defined in \eqref{4-gk.def},
satisfying
\begin{equation}\label{gk.est}
|g_k(t)| \les
n^7 + n^2 \norm{\xi^{(3)}}_{L^2\loc}
+ n \norm{\xi}_{L^2\loc}^2
+ \norm{\xi}_{L^2\loc}^{4/3} \norm{\xi}_{L^5} ^{5/3} .
\end{equation}
All coefficients $c_{l}^{k}$ and
$d_{ab}^k$ are bounded by one. The coefficients $c_{l}^{k}$ are
purely imaginary. The real parts of $d_{ab}^k$ are given by
\begin{equation} \label{dabk.def}
\Re d_{ab}^k =(2-\delta_a^b) \gamma_{ab}^k - 2(2-\delta_k^b)
\gamma_{kb}^a ,
\end{equation}
where, with $\phi_{abk}=\la \phi_a \phi_b \phi_k$,
\begin{equation} \label{4-26}
\gamma_{ab}^k \equiv \Im \bke{ \phi_{abk},\, \frac 1{H_0 -e_a - e_b + e_k -0i}
\, \Pc \,\phi_{abk}} \ge 0.
\end{equation}
\end{lemma}

Note $\gamma_{ab}^k$ may be positive only if $e_a - e_b + e_k <0$,
in particular $k < a,b$.
By assumption A2 we have $\gamma_{kk}^0\ge \gamma_0 >0$ for
$k>0$.
The first part of $\Re d_{ab}^k$ is positive
and the second part negative. If
$k=0$, $\gamma_{kb}^a =0$ for all $a,b$, hence $\Re d_{ab}^k \ge 0$.
If $k=N$, $\gamma_{ab}^k=0$ for all $a,b$, hence $\Re d_{ab}^k\le 0$.
For intermediate $k$ the sign of $\Re d_{ab}^k$ depends on the
relative size of these coefficients.
Let $f_k=|\mu_k |^2$, $k=0,\ldots, N$. Since $\tfrac d{dt} |\mu|^2
= 2 \Re \bar \mu \dot \mu$ and $c_{l}^{k}$ are purely imaginary,
we have
\begin{equation} \label{fk.eq}
\dot f_k = \sum_{a,b=0}^N 2 (\Re d_{ab}^k) \, f_a f_b f_k + 2 \Re
\bar \mu_k g_{k}.
\end{equation}
The example 1 in \S 1 follows by letting $N=2$, $x=\mu_0$, $y=\mu_1$
and $z=\mu_2$.

\myproof Recall \eqref{uk.eq0} that
\begin{equation} \label{uk.eq}
\dot u_k = i^{-1} e^{ie_k t} \, \bke{\phi_k, \,  G}, \qquad
(k=0,\cdots , N).
\end{equation}
We now proceed to derive the normal form in two steps.

\noindent {\bf Step 1} \quad Integration of terms of order $n^3$.

Substituting $G=G_3+G_5+G_7$ and the expression \eqref{G3.def2} for
$G_3$, we get
\begin{align}
\dot u_k(t)&=\sum_{l,m,j=0}^N  i^{-1}e^{ie_kt} x_l x_m \bar x_j(t)
(\phi_k, \, \phi_{lmj} )+R_k \label{4-16}
\\
&=\sum_{l,m,j=0}^N  i^{-1}e^{i(e_k-e_l-e_m+e_j)t} u_l u_m \bar
u_j(t) (\phi_k, \, \phi_{lmj} )+R_k, \nonumber
\\
R_k &= i^{-1}e^{ie_kt} (\phi_k, \,G_5+G_7 )(t).\nonumber
\end{align}
By assumption A3,
\begin{equation}  \label{not0}
e_k - e_l -e_m + e_j \not = 0,
\end{equation}
unless the two
sets $\bket{k,n}$ and $\bket{l,m}$ are the same. Hence
\begin{equation}  \label{4-32}
\dot u_k(t)= \sum_{l=0}^N c_{l}^{k}|u_{l}|^2 u_{k} +
\sum_{\eqref{not0} } i^{-1}e^{i(e_k-e_l-e_m+e_j)t} u_l u_m \bar
u_j(t) (\phi_k, \, \phi_{lmj} )+  R_k,
\end{equation}
where $c_{l}^{k}$ are purely imaginary constants defined by
\[
c_{l}^{k}=2i^{-1}(\phi_k, \, \phi_{llk} ), \quad (k \not =l);
\qquad c_{k}^k =i^{-1} (\phi_k, \, \phi_{kkk} ).
\]
Terms in the second group of \eqref{4-32}
are oscillatory and can be integrated. Define
\[
u_{k,1}=u_k -u_{k,1}^{-},
\]
where
\[
u_{k,1}^{-}(t) =\sum_{\eqref{not0}}
\frac{i^{-1}e^{i(e_k-e_l-e_m+e_j)t}} {i(e_k-e_l-e_m+e_j)} u_l u_m
\bar u_j (t)(\phi_k, \, \phi_{lmj} ),
\]
\[
g_{k,1}(t)=-\sum_{\eqref{not0}}
\frac{i^{-1}e^{i(e_k-e_l-e_m+e_j)t}} {i(e_k-e_l-e_m+e_j)} \frac
d{d t} \bke{ u_l u_m \bar u_j }(t) (\phi_k, \, \phi_{lmj} ).
\]
We have
\[
\dot u_{k,1}=\dot u_k - \dot u_{k,1}^{-} =\sum_{l=0}^N
c_{l}^{k}|u_{l}|^2 u_{k}  +R_k + g_{k,1}.
\]
We further define
\[
g_{k,2}=\sum_{l=0}^N c_{l}^{k}\bket{ |u_l|^2 u_k -|u_{l,1}|^2
u_{k,1}}.
\]
We have
\begin{equation} \label{uk1.eq}
\dot u_{k,1}= \sum_{l=0}^N c_{l}^{k}|u_{l,1}|^2 u_{k,1}  +
i^{-1}e^{ie_kt} (\phi_k, \,G_5+G_7) + g_{k,1} + g_{k,2}.
\end{equation}
%

\medskip

\noindent {\bf Step 2} \quad Integration of terms of order $n^5$.

We next integrate $O(n^5)$ terms. We first decompose $g_{k,1}$ and
$g_{k,2}$ since they contain $O(n^5)$ terms.
Using their explicit form we can decompose them as
\begin{align}
g_{k,1}&= e^{i e_k t} g_{k,1,5} + g_{k,1,7},
\\
g_{k,2}&= e^{i e_k t} g_{k,2,5} + g_{k,2,7},
\end{align}
where $g_{k,1,7}$ and $g_{k,2,7}$ are higher order terms with
\begin{equation} 
|g_{k,1,7}| \les n^2 \max_j|R_j|, \qquad
|g_{k,2,7}| \les n^7.
\end{equation}
$g_{k,1,5}$ and $g_{k,2,5}$ are explicit homogeneous polynomials
of degree $5$ of the form
\begin{equation}\label{4-23}
C \, x_{l_1} \, x_{l_2} \, x_{l_3}\, \bar x_{l_4} \, \bar x_{l_5},
\end{equation}
where $l_j \in \bket{ 0,\ldots,N}$ and $C$ is a {\it purely
imaginary} coefficient. The claim on $g_{k,1,5}$ can be seen by
substituting \eqref{4-16} into $\tfrac d{dt}(u_l u_m \bar u_j)$ in
the definition of $g_{k,1}$.  The claim on $g_{k,2,5}$ can be seen
from the definitions of $g_{k,2}$ and $u_{k,1}^-$.

We next decompose $i^{-1}e^{ie_kt} (\phi_k, \,G_5)$. Recall $ G_5
= \la \chi ^2 \wbar \xi + 2\la |\chi |^2 \xi$ with $\chi = \sum
_{l=0}^N x_l \phi_l$. Using the decomposition $\xi = \xi^{(2)}
+\xi^{(3)}$ with $\xi^{(2)}=\sum_{l,m,j=0}^N x_l x_m \bar x_j \, \xi_{lm}^j$,
we can decompose
\[
G_5 = G_{5,1} + G_{5,2} + G_{5,3},
\]
where
\begin{align*}
G_{5,1} &=\la \chi ^2 \sum_{l,m,j=0}^N \bar x_l \bar x_m x_j \,
(-i\Im \xi_{lm}^j) + 2\la |\chi |^2 \sum_{l,m,j=0}^N x_l x_m \bar
x_j \, (i \Im \xi_{lm}^j),
\\
G_{5,2} &=\la \chi ^2 \sum_{l,m,j=0}^N \bar x_l \bar  x_m  x_j \,
(\Re \xi_{lm}^j) + 2\la |\chi |^2 \sum_{l,m,j=0}^N x_l x_m \bar
x_j \, (\Re \xi_{lm}^j),
\\
G_{5,3} &= \la \chi ^2 \wbar \xi^{(3)} + 2\la |\chi |^2 \xi^{(3)}.
\end{align*}
Recall that $\Im \xi_{lm}^j \not = 0$ only if $e_j -e_l -e_m <0$.
The term $G_{5,3}$ will be shown to be smaller than $G_{5,1}$ and
$G_{5,2}$. Both $G_{5,1}$ and $G_{5,2}$ are homogeneous
polynomials of degree 5 of the form \eqref{4-23}. However, the
coefficients $C$ of  $G_{5,2}$ are real-valued $L^2\loc$ functions, while
those of $G_{5,1}$ are purely imaginary.

We can now rewrite \eqref{uk1.eq} as
\begin{align*}
\dot  u_{k,1}
&=\sum_{l=0}^N c_{l}^{k}|u_{l,1}|^2 u_{k,1} + \wt R_k + \bke{
i^{-1}e^{ie_kt} (\phi_k, \, G_{5,3} +
G_7)  + g_{k,1,7} + g_{k,2,7} },
\end{align*}
where $\wt R_k$ consist of  terms of order $O(n^5)$,
\begin{equation} \label{4-36}
\wt R_k = e^{ie_kt} \bket{i^{-1} (\phi_k, \,G_{5,1}+G_{5,2}) + g_{k,1,5} +
g_{k,2,5}}.
\end{equation}
Inside the bracket are polynomials in $x_l$ and $\bar x_l$,
$l=0,\ldots,N$. The only terms with real coefficients are those
from $G_{5,1}$. As in Step 1, the above expression can be
separated into two groups: those with zero phase factor and those
with non-zero phase factors. The phase factor of a typical term in
the above expression is of the form
\[
e_k - e_{l_1} - e_{l_2} - e_{l_3} +e_{l_4} + e_{l_5},
\]
where $l_j \in \bket{0,\ldots, N}$. To get zero, we must have the
form $e_k
-e_k - e_a - e_b+ e_a+ e_b $ for some $a,b$ by assumption A3. Thus
a typical term in the first group is of the form $C e^{i e_k t}x_k
x_a x_b \bar x_a \bar x_b = C|u_a|^2 |u_b|^2 u_k$.

We can integrate by parts those terms with non-zero phase factors
as in Step 1. The other terms remain. Hence we can rewrite $\wt R_k$
as
\[
\wt R_k = \sum_{a,b=0}^N d_{ab}^k  \, |u_a|^2 |u_b|^2 u_k + \frac d{dt}
(u_{k,2}^- ) + g_{k,3},
\]
for some order one constants $d_{ab}^k$, some explicit homogeneous polynomials
$u_{k,2}^-$ of degree $5$, and remainder terms $g_{k,3}$ with
\begin{equation} \label{4-39}
|g_{k,3}| \les n^4 \max_j |\dot u_j| .
\end{equation}
We will identify $\Re d_{ab}^k$ in a moment. Define
\begin{equation} \label{muk.def}
\mu_{k} \equiv u_{k,1} -u_{k,2}^- = u_k - u_{k,1}^- -u_{k,2}^-.
\end{equation}
We have
\begin{align*}
\dot \mu_{k} & =\dot  u_{k,1} -\dot u_{k,2}^-
\\
&=\sum_{l=0}^N c_{l}^{k}|u_{l,1}|^2 u_{k,1}  +\sum_{a,b=0}^N
d_{ab}^k \, |u_a|^2 |u_b|^2 u_k + g_{k,3}
\\
&\qquad + \bke{ i^{-1}e^{ie_kt}
(\phi_k, \, G_{5,3}+G_7 ) +g_{k,1,7} + g_{k,2,7} }
\\
&=\sum_{l=0}^N c_{l}^{k}|\mu_{l}|^2 \mu_{k}  +\sum_{a,b=0}^N
d_{ab}^k \, |\mu_a|^2 |\mu_b|^2 \mu_k + g_{k},
\end{align*}
where
\begin{equation} \label{4-gk.def} g_k =  i^{-1}e^{ie_kt}
(\phi_k, \, G_{5,3}+G_7 ) +g_{k,1,7} + g_{k,2,7} + g_{k,3} +
g_{k,4},
\end{equation}
\[
g_{k,4} =\sum_{l=0}^N c_{l}^{k}\bket{|u_{l,1}|^2 u_{k,1}
-|\mu_{l}|^2 \mu_{k} }+\sum_{a,b=0}^N d_{ab}^k \, \bket{|u_a|^2
|u_b|^2 u_k - |\mu_a|^2 |\mu_b|^2 \mu_k}.
\]
We have $|g_{k,5}| \les n^7$.
Collecting estimates, using
the explicit form of $G_{5,3}$, $|R_k| \les \norm{G_5+G_7}_{L^1\loc}$,
and Lemma \ref{th:4-1}, we get
\begin{align*}
|g_k| &\les \norm{G_{5,3}}_{L^1\loc} + \norm{G_{7}}_{L^1\loc}
+ n^2 \max_j|R_j| + n^7 + n^4 \max_j |\dot u_j| + n^7
\\
&\les n^7 + n^2 \norm{\xi^{(3)}}_{L^2\loc}
+ n^4 \norm{\xi}_{L^2\loc} + n \norm{\xi}_{L^2\loc}^2
+\norm{\xi^3}_{L^1\loc}.
\end{align*}
Using $\norm{\xi}_{L^2\loc} \les n^3 +  \norm{\xi^{(3)}}_{L^2\loc}$,
we can remove $n^4 \norm{\xi}_{L^2\loc}$ and get \eqref{gk.est}.

We have noted that $c_l^k$ are purely imaginary. We now
compute the real parts of $d_{ab}^k$. From the previous
discussions on $g_{k,1,5}$, $g_{k,2,5}$, $G_{5,1}$ and $G_{5,2}$,
we know that the only contribution to $\Re d_{ab}^k$ is from
$i^{-1}e^{ie_kt} (\phi_k, \,G_{5,1})$. We have
\begin{align*}
i^{-1}e^{ie_kt} (\phi_k, \,G_{5,1})&= \sum_{a,b,l,m,j} - e^{ie_kt}
\, x_a x_b \bar x_l \bar x_m x_j \; (\phi_k ,\, \la \phi_a \phi_b
\Im \xi_{lm}^j )
\\
& +\sum_{a,b,l,m,j} 2 e^{ie_kt} \, x_a \bar x_b x_l x_m \bar x_j
\; (\phi_k ,\, \la \phi_a \phi_b \Im \xi_{lm}^j ),
\end{align*}
and we are interested in those terms of the form $C
e^{ie_kt}|u_a|^2 |u_b|^2 u_k$. As mentioned, $\Im \xi_{lm}^j \not
= 0$ only if $e_j -e_l - e_m<0$, in particular $j\not =l,m$ (not
sufficient though). Therefore, to get products in the first
summation we need $j=k$ and the two sets $\bket{l,m}$ and
$\bket{a,b}$ are the same; in the second summation we need $j=a$
and the two sets $\bket{l,m}$ and $\bket{k,b}$ are the same. Hence
$\sum_{a,b=0}^N (\Re d_{ab}^k) \, |u_a|^2 |u_b|^2 u_k$ is equal to
\begin{align*}
& \sum_{a,b=0}^N - (2-\delta_a^b) \, |x_a|^2 |x_b|^2 u_k \;
(\phi_k ,\, \la \phi_a \phi_b \Im \xi_{ab}^k )
\\
& +\sum_{a,b=0}^N 2 (2-\delta_k^b)\, |x_a|^2 |x_b|^2 u_k \;
(\phi_k ,\, \la \phi_a \phi_b \Im \xi_{kb}^a ).
\end{align*}
Note we have the factors $(2-\delta_a^b)$ and $(2-\delta_k^b)$
since we have two choices for assigning $l$ and $m$ if $a \not =
b$ (or $k\not = b$). Recall $ \la \phi_a \phi_b \phi_k=
\phi_{abk}$, $\xi_{lm}^j = - K_{lm}^j  \PcH \phi_{lmj}$, and
$\gamma_{lm}^j= ( \phi_{lmj} , \Im K_{lm}^j  \PcH \phi_{lmj} )$.
We conclude
\begin{align*}
\Re d_{ab}^k &= - (2-\delta_a^b)(\phi_{abk}, \, \Im \xi_{ab}^k ) +
2 (2-\delta_k^b)(\phi_{kba} ,\,  \Im \xi_{kb}^a ) \nonumber
\\
&= (2-\delta_a^b) \gamma_{ab}^k - 2(2-\delta_k^b) \gamma_{kb}^a .
\end{align*}
\myendproof

\subsection{Main estimates}

In this subsection we prove estimates in the transition regime
for the solution $\psi(t)$.
The following proposition is the main result of this section. It proves
the same result of Theorem \ref{th:1-1} under
weaker assumptions \eqref{4-41}--\eqref{out:t0} below. Note that
the assumption $\norm{\xi_0}_Y \le n/2$ in \eqref{psiz.asp} of
Theorem \ref{th:1-1} implies, using Lemma \ref{th:2-3},
\[
\norm{e^{-itH_0}\xi_0}_{L^5} \les n (1+t)^{-9/10},\qquad
\norm{e^{-itH_0}\xi_0}_{L^2 \loc} \les n (1+t)^{-3/2}.
\]
Hence it implies \eqref{out:t0}. Eq.~\eqref{out:t0} is motivated
by \eqref{out:t2} and \cite{TY4}, and may be more convenient for
future application when we study the asymptotic profiles of all
small solutions. See \cite{TY4} for two bound states case. Also
note that, by \eqref{4-41} below, we have $\norm{\psi(t)}_{L^2} =
\norm{\psi_0}_{L^2} \ll 1$ and hence
$\max_k|x_k(t)|,\norm{\xi(t)}_{L^2} \ll 1$. However, we do not
assume any bound of $\norm{\psi_0}_{L^2}$ in terms of $n$. This is
essential for future application.

Notice that we define $t_2$ by \eqref{4-45} to be the time when
the size of the excited states decays to the order $n^{1+\sigma}$.
This is more than sufficient since we only need it to be smaller
than $\epz n$ in order to apply Theorem \ref{th:S-1}.
%

\begin{proposition}  \label{th:4-3}
Assume the assumptions A0--A3 given in \S 1. There is a small
constant $n_0>0$ such that the following hold. Suppose
$\psi(t)=\psi(t,x)$ is a solution of \eqref{Sch} with the initial
data $\psi(0)=\psi_0 = x_0^0\phi_0 + \cdots + x_N^0\phi_N + \xi_0$
satisfying, for some $n \le n_0$, $0<\sigma < 1/10$,
\begin{equation}  \label{4-41}
\begin{split}
&\tfrac 34 \, n^2 \le |x_0^0|^2 + \cdots + |x_N^0|^2 \le n^2,
\qquad \norm{\psi_0}_{H^1} \ll 1,
\\
&|x_0^0| \ge n^{3- \epsilon}, \qquad
\bke{\ssum_{k=1}^N |x_k^0|^2}^{1/2} \ge 2 |x_0^0|^{1 + \sigma},
\end{split}
\end{equation}
and for some $t_* \in [1, n^{-4-2\sigma}]$, for all $t\ge 0$,
\begin{equation}\label{out:t0}.
\begin{split}
\norm{  e^{-i t H_0} \xi_0}_{L^5} & \le \Lambda_5(t) \equiv
n (1+t)^{-9/10} + C  n^3 t_* (t_*+t) ^{-9/10} ,
\\
\norm{ e^{-i t H_0}  \xi_0}_{L^2 \loc} & \le  \Lambda(t) \equiv
n (1+t)^{-3/2} +  C  n^3 \frac{t_*}{t_*+ t } \ (1+t) ^{-1/2} .
\end{split}
\end{equation}
Let $t_0 = n^{-2}$. There exist $t_2 \ge t_1\ge t_0$ such that
\begin{equation}
\label{4-43}
t_1\le C_4 n^{-4} \log \frac 1 n , \qquad
C_4^{-1}n^{-4} \le t_2 \le C_4 n^{-4-2\sigma} ,
\end{equation}
for some constant $C_4>1$ independent of $n$, and we have,
for $0\le t\le t_2$,
\begin{equation} \label{4-44}
\begin{split}
|x_0(t)|&\ge \tfrac 34 \,\sup_{0\le s\le t} \, |x_0(s)| ,
\\
|x_k(t)|&\le \tfrac 54 n, \qquad (k=0,\dots,N),
\\
\norm{\xi(t)}_{L^5} &\le C_5\,  n^3 t^{1/10}  + \frac 54 \, \Lambda_5(t),
\\
\norm{\xi(t)}_{L^2 \loc} &\le  \, C_5\,  n^{3}  + \frac 54\, \Lambda(t),
\\
\norm{\xi^{(3)}(t)}_{L^2 \loc} &\le C_5\,  n^{5},
\qquad \text{for }t\ge t_0,
\end{split}
\end{equation}
where $C_5$ is an explicit constant.
We also have, at $t=t_1$ and $t=t_2$,
\begin{equation} \label{4-45}
|x_0(t_1)|^2,  |x_0(t_2)|^2 \ge \frac 34 \, 2^{-2-N}  n^2 , \quad
\bke{\ssum_{k=1}^N |x_k(t_2)|^2}^{1/2} \le |x_0(t_2)|^{1 + \sigma}.
\end{equation}

Moreover, we have the following out-going estimates for $\xi(t_2)$: For
$\tau \ge 0$,
\begin{equation}\label{out:t2}
\begin{split}
\norm{  e^{-i\tau H_0} \xi(t_2)}_{L^5} & \le C  n^3 (t_*+t_2) (t_2+\tau
) ^{-9/10} ,
\\
\norm{ e^{-i\tau H_0}  \xi(t_2)}_{L^2 \loc} & \le  C  n^3
\frac{t_*+t_2}{t_2+\tau } \ (1+\tau) ^{-1/2} .
\end{split}
\end{equation}
\end{proposition}

\myproof
We will prove estimates \eqref{4-43}--\eqref{4-45} using
a continuity argument. Hence we can assume the
following weaker estimates:  For $0\le t\le t_2$:
\begin{equation}\label{A:pf}
\begin{split}
|x_0(t)|&\ge \tfrac 12 \,\sup_{0\le s\le t} \, |x_0(s)| ,
\\
|x_k(t)|&\le \tfrac 32 n, \qquad (k=0,\dots,N),
\\
\norm{\xi(t)}_{L^5} &\le 2 C_5 n^3 t^{1/10} + 2 \Lambda_5 (t),
\\
\norm{\xi(t)}_{L^2 \loc} &\le  \, 2 C_5\,  n^{3} + 2\Lambda (t),
\\
\norm{\xi^{(3)}(t)}_{L^2 \loc} &\le 2 C_5\,  n^{5},
\qquad \text{for }t\ge t_0.
\end{split}
\end{equation}
If we can prove \eqref{4-43}--\eqref{4-45} assuming \eqref{A:pf},
then \eqref{4-43}--\eqref{4-45} hold true for all $t \in [0,t_2]$
since \eqref{A:pf} is never violated.
By definition we have $\Lambda_5(t),\Lambda(t)\le n$.
By \eqref{A:pf} we have
$\norm{\xi(t)}_{L^2\loc \cap L^5} \le 2 n$. Hence
the assumptions of Lemmas \ref{th:4-1}
and Lemma \ref{th:4-2} are satisfied. Since
$\norm{\psi(t)}_{L^2} = \norm{\psi_0}_{L^2} \ll 1$, we have
$\norm{\xi(t)}_{L^2}\ll 1$.
Also, since  $t \le t_2\le C_4 n^{-4-2\sigma}$,
\begin{equation} \label{4-L5est}
\norm{\xi}_{L^5} \les n^3 t_2^{1/10} +
n (1+t)^{-9/10} + C  n^3 t_*^{1/10}
\les  n^{13/5 - \sigma/5}+ n(1+t)^{-9/10},
\end{equation}
\begin{equation} \label{4-L2locest}
\norm{\xi}_{L^2\loc} \les n^3 + n (1+t)^{-3/2} + C  n^3 (1+t)^{-1/2}
\les n^3 + n (1+t)^{-3/2}.
\end{equation}

We now estimate $\xi(t)$. Recall \eqref{xi.eq},
\begin{equation*} 
\xi(t) = e^{-it H_0}\xi_0 + 
\int_0^{t} e^{-i(t -s)H_0} \PcH i^{-1}G(s) \, d s  .
\end{equation*}
Note $\norm{G}_{L^{5/4}} \les \norm{G_3}_{L^{5/4}}
+ \norm{G-G_3-\la \xi^2 \bar \xi}_{ L^{5/4}} + \norm{\la \xi^2 \bar \xi}_{ L^{5/4}}$ and  $\norm{G_3}_{L^{5/4}} \le C n^3$.
Using  Lemma \ref{th:4-1}, \eqref{4-L5est} and \eqref{4-L2locest},
\begin{equation}\label{4-51}
\norm{G-G_3-\la \xi^2 \bar \xi}_{L^1\cap L^{5/4}} \les
n^2 \norm{\xi}_{L^2\loc} \les n^2 (n^3 + n(1+s)^{-3/2}),
\end{equation}
\begin{equation}
\norm{\la \xi^2 \bar \xi}_{ L^{5/4}}
\les \norm{\xi}_{L^2}^{2/3} \norm{\xi}_{L^5} ^{7/3} \le
o(1) [n^{13/5 - \sigma/5}+ n(1+s)^{-9/10}]^{7/3}.
\end{equation}
Hence $\norm{G(s)}_{L^{5/4}} \les n^3 +  n^{7/3} \bka{s}^{-21/10}$
and
\begin{align*}
\norm{  \xi(t) }_{L^5} &\le
\norm{e^{-it H_0}\xi_0}_{L^5}  +
C \int _0^{t} |t-s|^{-9/10}
\norm{G(s)}_{L^{5/4}} \, d s
\\
&\le \Lambda_5(t) + C\int _{0}^{t} |t-s|^{-9/10}
 [n^3 +  n^{7/3} \bka{s}^{-\frac{21}{10}}]d s
\le \tfrac 54 \Lambda_5(t) + C n^3 t^{1/10}.
\end{align*}
Hence $\norm{\xi(t)}_{L^5}$ is estimated.
For the $L^2 \loc$ norm, we use the decomposition \eqref{xi.dec} that
$\xi(t) = \xi^{(2)} + \xi^{(3)}$ with $\xi^{(3)}=\sum_{j=1}^5 \xi^{(3)}_j$.
We have $\norm{\xi^{(2)}}_{L^2 \loc}\le C n^3$ by its explicit form.
By \eqref{out:t0} for $\xi_0$ we have
\[
\norm{  \xi^{(3)}_1(t) }_{L^2 \loc}  \le \Lambda(t).
\]
By the singular decay estimate \eqref{eq:22-1B} in Lemma \ref{th:2-3}
and the definition \eqref{xilmj.def} of $\xi_{lm}^j$,
we have
\[
\norm{  \xi^{(3)}_2(t) }_{L^2 \loc}  \le C n^3 (1+t)^{-3/2}
\le C n^2 \Lambda(t).
\]
Using \eqref{eq:22-1B}, \eqref{xilmj.def} again and the bound of
$\max_j |\dot u_j|$ in
Lemma \ref{th:4-1}, we have
\begin{align*}
\norm{  \xi^{(3)}_3(t) }_{L^2 \loc}  &\le C \int _{0}^{t} \bka{t-s}^{-3/2}
n^2 \max_j |\dot u_j| ds \le C \int _{0}^{t} \bka{t-s}^{-3/2}
n^5 ds  \le C n^5.
\end{align*}
For $\xi^{(3)}_4(t)$,
bounding its integrand by either $L^\infty$ or
$L^5$-norm and using \eqref{4-51}, we have
\begin{align*}
\norm{  \xi^{(3)}_4(t) }_{L^2 \loc}  &\les \int _{0}^{t} \min \bket{
|t- s |^{-3/2} , \, |t- s |^{-9/10} }
\norm{(G-G_3-\la \xi^2 \bar \xi)(s)}_{L^1\cap L^{5/4}} \, d  s
\\
&\les \int _{0}^{t} \min \bket{ |t- s |^{-3/2} , \, |t- s
|^{-9/10} } n^2 (n^3 + n(1+s)^{-3/2})\, d s
\\
&\les  n^{5} + n^2 \Lambda(t).
\end{align*}
Finally we give two estimates for $\xi^{(3)}_5$.
By H\"older inequality and \eqref{4-L5est},
\begin{align}
\norm{\xi^3}_{L^1 \cap L^{5/4}} &\le \norm{\xi}_{L^2 \cap L^{5}}^{4/3}
\norm{\xi}_{L^5}^{5/3} \le o(1)
(n^{13/5- \sigma/5}+ n(1+s)^{-9/10})^{5/3}. \nonumber
\\
\label{4-49}
& \les n^{13/3 - \sigma/3} + n^{5/3} (1+s)^{-3/2} .
\end{align}
The same estimate for $\xi^{(3)}_4$ shows
\[
\norm{  \xi^{(3)}_5(t) }_{L^2 \loc} \les n^{13/3 - \sigma/3}
+  n^{2/3} \Lambda(t).
\]
Summing the estimates we conclude
\[
\norm{\xi}_{L^2 \loc} \le \norm{\xi^{(2)}}_{L^2 \loc} +
\sum_{j=1}^5 \norm{\xi^{(3)}_j}_{L^2 \loc} \le C_5 n^3 + \tfrac 54 \Lambda(t).
\]
We now estimate $\xi^{(3)}_5$ in another way. We have
\begin{align*}
\norm{\xi^2 \bar \xi}_{L^{6/5}\cap L^{5/4}} &\le
\norm{\xi}_{L^2\cap L^5}^{7/9}
\norm{\xi}_{L^5}^{20/9} \le o(1) (n^{13/5 - \sigma/5} + n(1+s)^{-9/10})^{20/9}
\\
&\le n^{52/9 - 4\sigma/9} + n^{20/9} (1+s)^{-2}.
\end{align*}
Bounding the integrand of  $\xi^{(3)}_5(t)$ by either $L^6$ or
$L^5$ norm we have
\begin{align*}
\norm{  \xi^{(3)}_5(t) }_{L^2 \loc}  &\les \int _{0}^{t} \min \bket{
|t- s |^{-1} , \, |t- s |^{-9/10} }
\norm{\la \xi^2 \bar \xi(s)}_{L^{6/5}\cap L^{5/4}} \, d  s
\\
&\les \int _{0}^{t} \min \bket{ |t- s |^{-1} , \, |t- s
|^{-9/10} } \bke{n^{52/9 - 4\sigma/9} + n^{20/9} (1+s)^{-2} }\, d s
\\
&\les n^{52/9 - 4\sigma/9} \log (1+t_2)  + n^{20/9} (1+t)^{-1} \log (1+t_2).
\end{align*}
Summing the estimates and using $t_2 \le C_4 t^{-4-2\sigma}$, we get
\[
\norm{\xi^{(3)}}_{L^2 \loc} \le
\sum_{j=1}^5 \norm{\xi^{(3)}_j}_{L^2 \loc} \le Cn^5
+ \tfrac 54 \Lambda(t)+ n^{20/9} (1+t)^{-1}\log (1+t_2).
\]
Clearly we have $\norm{\xi^{(3)}(t)}_{L^2 \loc} \le C n^{5}$ for
$t>t_0 = n^{-2}$.

\medskip

We now estimate the error
term $g_k(t)$ in the equation \eqref{eq:NF} of $u_k$. Recall \eqref{gk.est},
\[
|g_k(t)| \les n^7 + n^2\norm{\xi^{(3)} }_{L^2\loc}
+ n\norm{\xi }_{L^2\loc} ^2
+ \norm{\xi }_{L^2\loc}^{4/3} \norm{\xi }_{L^5}^{5/3}.
\]
Using \eqref{A:pf}--\eqref{4-L2locest},
$\Lambda_5(t),\Lambda(t) \le n$, and the computation in \eqref{4-49},
\begin{align*}
|g_k(t)| &\les n^7 + n^2\norm{\xi^{(3)} }_{L^2\loc}
+  n(n^3 + \Lambda)^2
+ (n^3 + \Lambda)^{4/3}(n^{13/3 - \sigma/3} +  n^{2/3} \Lambda(s))
\\
&\les  n^7 + n^2\norm{\xi^{(3)} }_{L^2\loc}
+ n^{14/3 -\sigma/3} \Lambda + n\Lambda^2.
\end{align*}
For $t\le t_0=n^{-2}$ we use the first estimate of
$\norm{\xi^{(3)}_5 }_{L^2\loc}$ and we have $\norm{\xi^{(3)}
}_{L^2\loc} \le \sum_{j=1}^5 \norm{\xi^{(3)}_j }_{L^2\loc} \les
n^{13/3 - \sigma/3} +  \Lambda$. Hence
\begin{equation} \label{4-51A}
|g_k(t)| \les n^{6+1/3 - \sigma/3} + n^2 \Lambda, \qquad (t\in [0,t_0]).
\end{equation}
For $t\ge t_0$ we have $\norm{\xi^{(3)}(t)}_{L^2 \loc} \le C n^{5}$
and $\Lambda(t)\les n^2$.
Thus
\begin{equation} \label{4-51B}
|g_k(t)| \les n^7, \qquad (t\in [t_0,t_2]).
\end{equation}

We now estimate bound states. As illustrated in Example 1 of \S 1, some
bound states may grow at intermediate time.
We will use the following monotonicity properties.
Recall that $f_k=|\mu_k |^2$, $k=0,\ldots, N$, satisfy \eqref{fk.eq}.
%
%
Motivated by \eqref{model2}, we consider the following functions:
\begin{equation}
f(t) = f_0(t), \qquad g(t)=\sum_{k=1}^N f_k, \qquad
g_+(t)=\sum_{k=1}^N 2^{-k} f_k.
\end{equation}
By \eqref{4-41}, \eqref{out:t0} and \eqref{ukmuk.diff} we have
\begin{equation} \label{fg0.est}
f(0) \ge  2^{-1} n^{6-2\epsilon} ,\quad 2 n^2 \ge f(0)+ g(0) \ge
2^{-1} n^2 , \quad f(0)+ g_+(0) \ge 2^{-1-N}n^2,
\end{equation}
if $n$ is sufficiently small. Although $g$ and $g_+$ are
comparable, $(f+g)(t)$ is almost monotone decreasing while
$(f+g_+)(t)$ is almost monotone increasing in the following sense:
\begin{equation}\label{4-52}
\begin{split}
\frac d{d t} ( f+g)(t) &\le 2(N+1) \max_k |\bar \mu_k g_k| ,
\\
\frac d{d t} (f+ g_+)(t) &\ge - 2(N+1) \max_k |\bar \mu_k g_k|.
\end{split}
\end{equation}
We now prove \eqref{4-52}. By \eqref{dabk.def} and \eqref{fk.eq},
\begin{align}
&\frac d{d t} (f+g) - \sum_{k=0}^N 2 \Re  \bar \mu_k g_{k} =
\sum_{a,b,k=0}^N 2 (\Re d_{ab}^k) \, f_a f_b f_k \nonumber
\\
&= \sum_{a,b,k=0}^N 2\bkt{(2-\delta_a^b) \gamma_{ab}^k -
2(2-\delta_k^b) \gamma_{kb}^a }f_a f_b f_k. \label{4-48}
\end{align}
Switching $a$ and $k$ in the terms with minus sign, we see that
the above sum is non-positive. Similarly,
\begin{align*}
&\frac d{d t} (f+g_+) - \sum_{k=0}^N 2^{1-k} \Re  \bar \mu_k g_{k}
= \sum_{a,b,k=0}^N 2^{1-k} (\Re d_{ab}^k) \, f_a f_b f_k
\\
&= \sum_{a,b,k=0}^N 2^{1-k} \bkt{(2-\delta_a^b) \gamma_{ab}^k -
2(2-\delta_k^b) \gamma_{kb}^a } f_a f_b f_k
\\
&= \sum_{a,b,k=0}^N \bkt{2^{1-k} (2-\delta_a^b) \gamma_{ab}^k -
2^{2-a}(2-\delta_a^b) \gamma_{ab}^k } f_a f_b f_k .
\end{align*}
In the last line we have switched $a$ and $k$ for terms with minus
sign. If $k \ge a$ then $\gamma_{ab}^k =0$. Hence $2^{1-k}\ge
2^{2-a}$ for nonzero terms and the above sum is non-negative. We
have shown \eqref{4-52}.
We now estimate the bound states in three steps.

\noindent{\bf Step 1.} Initial layer.

In this period the dispersive part disperses away so much that it
becomes negligible locally. The time it takes is less than $t_0 =
n^{-2}$, which is not long enough to change the magnitudes of
the bound states. Explicitly, by \eqref{fk.eq}, \eqref{A:pf}
 and \eqref{4-51A} we have
\[
|\dot f_k(s)| \le C n^4 \al^2 + \al \bke{
n^{6+1/3 - \sigma/3} + n^2 \Lambda(s) },
\]
for $k=0,1,\ldots,N$, where $\al = \sqrt {f_k(0)}$.
Thus, for $t \in [0,t_0]$,
\begin{align*}
\abs{f_k(t) - f_k(0)} &\les \int_0^t   n^4 \al^2 + \al \bke{
n^{6+1/3 - \sigma/3} + n^2 \Lambda(s) } ds
\\
&\les n^4 \al^2 t_0 + \al  n^{6+1/3 - \sigma/3} t_0 +
\al n^2 (n + n^3 \sqrt{t_0})
\\
&\les n^2 \al^2 + n^3 \al
\les n^{\epsilon} \al^2 + n^{6-\epsilon}.
\end{align*}
Since $f_k(0) \ge n^{6-2\epsilon}$, we get
$\abs{f_k(t) - f_k(0)} \ll f_k(0)$.

\medskip

\noindent{\bf Step 2.} Transition regime (i).

After time $t=t_0$ most mass of the dispersive wave is far away
and has no effect on the local dynamics. Two possible situations
can occur: We either have $f(t_0) \ge 2^{-2 -N} n^2 $ or the
opposite. In the first case we define $t_1=t_0$ and jump to next
step. We now focus on the second case.
In the second case the ground state begins to grow exponentially
until it is of order $ n$. The time it takes is of order
$n^{-4}\log n^{-1}$. Define
\begin{equation}
t_1\equiv\inf_{t \ge t_0} \bket{t:f_0(t) \ge 2^{-2 -N} n^2 } .
\end{equation}
By assumption $t_1>t_0$.
We want to show that
\begin{equation} \label{4-58}
t_0\le t_1\le  t_1' \equiv t_0 + 2^{8+2N} N { \gamma_0}^{-1}\ n^{-4}
\log \frac 1 n \ .
\end{equation}
Suppose \eqref{4-58} fails, that is, $f_0(t) < 2^{-2 -N} n^2 $
for all $t\le t_1'$.
We have by \eqref{fg0.est}, \eqref{4-51B} and \eqref{4-52},
for all $t\in [t_0, t_1']$,
\begin{align*}
g_+(t) &\ge (f_0 + g_+)(t_0) - f_0(t) - \int _{t_0}^t
\max_k |\mu_k| |g_k| ds
\\
&\ge 2^{-1-N}n^2 - 2^{-2 -N} n^2 - C n n^7 \ge
2^{-3 -N} n^2.
\end{align*}
In particular, the coefficient of the linear term in \eqref{fk.eq} of
$\dot f_0$ has a lower bound,
\begin{align}
\sum_{a,b=0}^N 2 (\Re d_{ab}^0) f_a f_b &= \sum_{a,b=0}^N 2
(2-\delta_a^b) \gamma_{ab}^0 f_a f_b \ge \sum_{a=1}^N 2 \gamma_0
f_a^2 \ge 2  \gamma_0  N^{-1} g^2 \label{4-55}
\\ &\ge 2  \gamma_0  N^{-1} g_+^2
 \ge 2^{-5 -2N}\gamma_0  N^{-1} n^4. \nonumber
\end{align}
Since $\sqrt{f_0(t)} =f_0(t) /|\mu_0(t)| \le 3 f_0(t)/|\mu_0(0)|
\le 4 f_0(t) n^{-3+ \epsilon}$ by the first line of \eqref{A:pf},
we have for $t\ge t_0$,
\begin{align*}
\dot f_0(t) &\ge  2^{-5 -2N}\gamma_0  N^{-1} n^4 f_0 - \sqrt{f_0} C n^7
\ge  2^{-6 -2N}\gamma_0  N^{-1} n^4 f_0 ,
\end{align*}
if $ n$ is sufficiently small. Hence
\begin{equation}\label{4-56}
f_0(t) \ge f_0(t_0) \exp \bket{ 2^{-6 -2N}\gamma_0 N^{-1} n^4
(t-t_0)} ,
\end{equation}
for $t\in [t_0, t_1']$. However, using the definition of $t_1'$
and
\[
f_0(t_0) \ge  \tfrac 34 |x_0(t_0)|^2 \ge  \tfrac 3{16} |x_0(0)|^2
\ge \tfrac 3{16} n^{6-2\epsilon},
\]
\eqref{4-56} implies $f_0(t_1') > 2^{-2 -N} n^2$, which is a
contradiction. This contradiction shows the existence of $t_1\in
[t_0, t_1']$ so that $f_0(t_1) = 2^{-2-N} n^2$. We also have
\eqref{4-44} and \eqref{4-56} for all $t\le t_1$.

\medskip

\noindent{\bf Step 3.} Transition regime (ii).

For $t\ge t_1$, $f_0(t)$ is large enough to cause $g(t)$ to decay
in a rate we can control. Define
\begin{equation}
t_2 \equiv \inf_{t \ge t_1} \bket{t:g(t ) \le n^{2 + 2\sigma} }.
\end{equation}
We want to show that
\begin{equation} \label{4-62}
t_1\le t_2\le t_2'\equiv t_1 + n^{-4-2\sigma}.
\end{equation}

Suppose the contrary, then $g(t) \ge n^{2 + 2\sigma}$ for all
$t\le t_2'$.
By \eqref{A:pf}, for $t>t_1$ we have $\sqrt{f_0(t)} \le f_0(t)
(\tfrac 12 f_0(t_1))^{-1/2} \le 2^{(3+N)/2}n f_0(t)$. Recall
from \eqref{4-55} that $\sum_{a,b=0}^N 2 (\Re d_{ab}^0) f_a f_b \ge 2
\gamma_0 N^{-1} g^2$. Hence for $t \in [t_1 , t_2']$,
\[
\dot f_0(t) \ge  2  \gamma_0  N^{-1} g^2  - \sqrt{f_0} C n^7 > 0.
\]
Thus $\dot
f_0 >0$ and $f_0(t) \ge 2^{-2-N} n^2$ for $t \in [t_1 , t_2']$. In
particular \eqref{4-44} holds for $t>t_1$ and \eqref{4-45} holds for
$x_0(t_2)$ should $t_2$ exist.
By eq.\eqref{fk.eq} for $\dot f_0$,
\begin{align*}
\dot g(t) &\le - \dot f_0(t) + {\textstyle \sum_{k=0}^N} 2 \Re
\bar \mu_k g_{k}
\\
& = - \bket{{\textstyle \sum_{a,b=0}^N 2 (\Re d_{ab}^0) f_a f_b }} f_0 +
{\textstyle \sum_{k=1}^N} 2 \Re
\bar \mu_k g_{k}
\\
&\le - 2  \gamma_0 N^{-1} g^2 f_0 + {\textstyle \sum_{k=1}^N } 2
\Re \bar \mu_k g_{k}
\\
&\le - 2^{-1-N} n^2 \gamma_0 N^{-1} g^2 +C n n^7
\\
&< - 2^{-2-N}\gamma_0 N^{-1}  n^2  g^2 ,
\end{align*}
if $ n$ is sufficiently small. In the second
line we have cancelled $2 \Re \bar \mu_0 g_0$. In the third line
we used \eqref{4-55}. Thus, by comparison principle,
\[
g(t) < [g(t_1)^{-1} + 2^{-2-N} \gamma_0 N^{-1} n^2 (t-t_1)]^{-1}
, \quad (t_1 < t \le t_2').
\]
for $t \in (t_1 , t_2']$. However, by the definition of $t_2'$
this implies $g(t_2') < n^{2+2\sigma}$, which is a
contradiction. This contradiction shows the existence of $t_2$
satisfying \eqref{4-62}.
Similarly, since
$\dot g(t) > - C n^2 g(t)^2$ for $t_1 \le t \le t_2$,
we have a lower bound
$g(t) \ge [g(t_1)^{-1} + C n^2 (t-t_2)]^{-1}$ by comparison
principle. This implies $t_2-t_1 \ge C n^{-4}$.
We have proven \eqref{4-43}--\eqref{4-45}
in Proposition \ref{th:4-3} using assumption
\eqref{A:pf}. Since \eqref{A:pf} holds for $t=0$,  it
holds for all $t \le t_2$ by continuity. Hence \eqref{4-43}--\eqref{4-45}
are proven.

\medskip

Finally we prove \eqref{out:t2} for $e^{-i\tau H_0}
\xi(t_2)$. Let $t=t_2+ \tau $. We have
\begin{equation*} 
e^{-i \tau H_0}\xi(t_2) = e^{-it H_0}\xi_0 + J(t), \qquad J(t)
\equiv \int_0^{t_2} e^{-i(t -s)H_0} \PcH G(s) \, d s  .
\end{equation*}
The estimate of $e^{-i t H_0}\xi_0$ is by
\eqref{out:t0}, since $\Lambda_5(t) \le C n^3 (t_*+t_2) (1+t)^{-9/10}$
and $\Lambda(t) \le C n^3 (t_*+t_2) (1+t)^{-1} (1+\tau)^{-1/2}$ if $t_2\ge
n^{-2}$. To estimate $ J(t)$,
we use the following integral inequalities: For $t\ge T \ge 1$,
\begin{equation}\label{cal-1}
\int _{0}^{T} |t-s|^{-9/10} \, d s \le C T t^{-9/10}.
\end{equation}
\begin{equation}  \label{cal-2} \int_{0}^{T}
\min \bket{(t-s)^{-3/2}, \ (t-s)^{-9/10} }  \, d s \le C T t^{-1}
\bka{t-T}^{-1/2}.
\end{equation}
See \cite[Lemma 2.6]{TY4}. Since $\norm{G(s)}_{L^1\cap L^{5/4}} \le C n^3$
for $s \in [0,t_2]$, we have
\begin{align*}
\norm{  J(t) }_{L^5} &\le C \int _0^{t_2} |t-s|^{-9/10}
\norm{G_\xi(s)}_{L^{5/4}} \, d s
 \le C\int _{0}^{t_2} |t-s|^{-9/10} n^3 d s
\\
&\le C n^3 t_2 (1+ t)^{-9/10}.
\end{align*}
\begin{align*}
\norm{  J(t) }_{L^2 \loc}  &\le C \int _{0}^{t_2} \min \bket{
|t- s |^{-3/2} , \, |t- s |^{-9/10} } \norm{G(s)}_{L^1\cap
L^{5/4}} \, d  s
\\
&\le C\int _{0}^{t_2} \min \bket{ |t- s |^{-3/2} , \, |t- s
|^{-9/10} } n^3 d s
\\
&\le C n^3  {t_2}t^{-1} (1+t-t_2)^{-1/2}.
\end{align*}
The proof of Proposition \ref{th:4-3} is complete.
\myendproof

\textsc{Proof of Theorem \ref{th:1-1}} \quad To prove Theorem
\ref{th:1-1} using Theorem \ref{th:S-1}, we want to use
$\psi(t_2)$ as initial data and decompose $\psi(t_2)$
according to \eqref{psi.dec2}, i.e., with respect to $\L$, and
check the assumptions \eqref{S-2}--\eqref{out:thS1}.
If the last condition in \eqref{4-41} fails, we simply set $t_2=0$.
Let  $\rho_0 = (|x_1(t_2)|^2 + \cdots +|x_N(t_2)|^2 )^{1/2}$.
We have
\begin{equation}
|x_0(t_2)| \sim n, \qquad \rho_0 = (1+o(1))n^{1+\sigma}, \qquad
\norm{\xi(t_2)}_{L^2\loc} \le C n^3.
\end{equation}
We may redefine $n=|x_0(t_2)|$.
Let $Q_{E_0}$ be the unique nonlinear ground state so that
$Q_{E_0} = n\phi_0 + k$ with $k \perp \phi_0$, $k=O(n^3)$. Also choose
$e^{i \theta_0} =  n^{-1} x_0(t_2)$.
We have $\norm{\psi(t_2) - Q_{E_0}e^{i \theta_0}}_{L^2\loc} \le
\norm{k}_{L^2\loc} + \norm{\psi(t_2) - x_0(t_2)\phi_0}_{L^2\loc} \les
n^3 + \rho_0 + n^3 \les \rho_0 $, which is much smaller than $n$.
By Lemma \ref{th:2-2}, for each $E$ close to $E_0$, we can decompose
$\psi(t_2)$ as
$\psi(t_2) = \bkt{Q_E + aR_E + \zeta + \eta }e^{i \theta }$ using the
decomposition \eqref{psi.dec2} with respect to $\L_E$. We have
\begin{equation} \label{4-65}
e^{-i\theta }\psi(t_2)  - Q_{E } = aR + \zeta + \eta =
 k  + Z + \xi ,
\end{equation}
where $Z = e^{-i\theta } [x_1(t_2) \phi_1+ \cdots +  x_N(t_2)]$
and $\xi = e^{-i\theta } \xi(t_2)$.
With $E=E_0$ in \eqref{4-65}, we have
$a_{E_0} = (c_1Q, \Re k  + Z + \xi) = c_1(n\phi_0 + k, \Re k + Z + \xi)$.
Since $\phi_0 \perp (k + Z + \xi)$, we have
\[
|a_{E_0}| \le  n^3 (n^3+\rho_0 + n^3) \les C n^3 \rho_0,
\]
which is much smaller than $\rho_0^2$.
Denote $P_\eigen = \ssum_{k=1}^NP_{\eigen_k}$.
We also have
\[
\norm{\zeta -P_\eigen Z} = \norm{P_\eigen (k + \xi)} \les n^3
+ n^2 \norm{\xi}_{L^2\loc} \les n^3 ,
\]
and hence $\norm{\zeta} = \rho_0 + O(n^3)$. We have verified
\eqref{S-2} with $E=E_0$. For all $E$ with $|E-E_0| \le D \rho_0^2$,
write
 $\eta_1 = \PcL (k  + Z)$ and $\eta_2 = \PcL \xi = \PcL
e^{-i\theta } \xi(t_2)$. We have
\[
e^{s\L}\eta = e^{s\L}\eta_1 + e^{s\L}\eta_2.
\]
Since $\eta_1$ is a local function bounded by $\norm{k}+ n^2\norm{Z}
\les n^3$, we get
\begin{equation*}
\begin{split}
\norm{e^{s\L}\eta_1}_{L^{5}} &\le C n^3 (1+s)^{-9/10}
\ll n^{4/5} \rho(s) ^{8/5},
\\
\norm{e^{s\L}\eta_1}_{L^{2}\loc} &\le C n^3 (1+s)^{-3/2}\ll
n^{1/2} (1+s)^{-1/2} \rho^2(s).
\end{split}
\end{equation*}
Recall $\rho(s)=[\rho_0^{-2} + N^{-1}\gamma_0 n^2 s]^{-1/2}$ with
$\rho_0 \sim n^{1+\sigma}$.
Write $\L$ as $\L = -i (H_0-E) + W$, where $W h = 2 \la Q^2 h +
\la Q^2 \bar h$ is a local operator bounded by $n^2$. We have
\[
e^{t\L} \eta_2 = \PcL e^{-it(H_0-E)} e^{-i\theta } \xi(t_2) +
\int_0^t e^{(t-s)\L} W e^{-is(H_0-E)} e^{-i\theta} \xi(t_2)\, ds.
\]
Hence
\[
\norm{e^{t\L}\eta_2}_{L^{5}} \le \norm{e^{-itH_0}\xi(t_2)}_{L^5} +
\int_0^t |t-s|^{-9/10} n^2 \norm{e^{-isH_0}\xi(t_2)}_{L^2 \loc} \,
ds,
\]
\begin{align*}
\norm{e^{t\L}\eta_2}_{L^2\loc} &\le
\norm{e^{-itH_0}\xi(t_2)}_{L^2\loc}
\\
&\quad + \int_0^t \min \bket{ |t-s|^{-3/2},\, |t-s|^{-9/10} } n^2
\norm{e^{-isH_0}\xi(t_2)}_{L^2 \loc} \, ds.
\end{align*}
Using the out-going estimates \eqref{out:t2} for $e^{-isH_0}\xi(t_2)$,
we conclude the
estimates \eqref{out:thS1} for $\eta_E$ for all $E$ close to $E_0$
with $|E-E_0| \le D\rho_0^2$.
\myendproof

\subsection*{Acknowledgments}

The author would like to thank H.-T. Yau for his very helpful
comments and discussions. The author was partially supported by
NSF grant DMS-9729992.


\begin{thebibliography}{XX}




\bibitem{BP1} V.S.~Buslaev and G.S.~Perel'man,
Scattering for the nonlinear Schr\"odinger equations, states close to
a soliton, St. Petersburg Math J. {\bf 4} (1993) 1111--1142.

\bibitem{BP2} V.S.~Buslaev and G.S.~Perel'man,
On the stability of solitary waves for nonlinear Schr\"odinger
equations. Nonlinear evolution equations, 75--98, Amer. Math. Soc.
Transl. Ser. 2, 164, Amer. Math. Soc., Providence, RI, 1995.

\bibitem{BS} V.S.~Buslaev and C. Sulem,
On the asymptotic  stability of solitary waves of nonlinear
Schr\"odinger equations, preprint.

\bibitem{C1} S.~Cuccagna, Stabilization of solutions to nonlinear
Schr\"odinger equations. {\it Comm. Pure Appl. Math.} {\bf 54}
(2001), no. 9, 1110--1145.

\bibitem{C2} S.~Cuccagna, On asymptotic stability of ground states
of NLS, preprint.







\bibitem{G0} M.~Grillakis:
Linearized instability for nonlinear Schr\"odinger and
Klein-Gordon equations.  {\it Comm. Pure Appl. Math.} {\bf 41}
(1988), no. 6, 747--774.


\bibitem{G} M.~Grillakis: Analysis of the linearization around a
critical point of an infinite dimensional Hamiltonian system, {\it
Comm. Pure Appl. Math.} {\bf 43} (1990), 299--333.



\bibitem{JK} A. Jensen and T.~Kato,
Spectral properties of Schr\"odinger operators and time-decay of
the wave functions, {\it Duke Math. J.} {\bf 46} no. 3, (1979),
583--611.

\bibitem{JSS}
J.-L.~Journe, A.~Soffer and C.D.~Sogge, Decay estimates for
Schr\"odinger operators. \textit{Comm. Pure Appl. Math.} {\bf 44}
(1991), no. 5, 573--604.


\bibitem{PW} C.-A.~Pillet and C.E.~Wayne,
Invariant Manifolds for a class of dispersive, Hamiltonian,
partial differential equations,
{\it J. Diff. Equations} {\bf 141}, (1997), 310--326.


\bibitem{Ra} Rauch, J. Local decay of scattering solutions to
Schr\"odinger's equation. {\it Comm. Math. Phys}. {\bf 61} (1978),
no. 2, 149--168.





\bibitem{RW}
H.A.~Rose and M.I.~Weinstein, On the bound states of the nonlinear
Schr\"odinger equation with a linear potential, {\it Physica D}
{\bf 30} (1988), 207--218.


\bibitem{SS} J. Shatah and  W. Strauss, Instability of
nonlinear bound states. Comm. Math. Phys. 100 (1985), no. 2,
173--190.



\bibitem{SS2} J. Shatah and  W. Strauss, Spectral condition for
instability. Nonlinear PDE's, dynamics and continuum physics
(South Hadley, MA, 1998), 189--198, Contemp. Math., 255, Amer.
Math. Soc., Providence, RI, 2000.


\bibitem{Sigal} I. M. Sigal, Non-Linear wave
and Schr\"odinger equations I, instability of periodic and
quasiperiodic solutions, {\it Comm. Math. Phys.}  {\bf 153} (1993)
297-320.




\bibitem{SW1}
A.~Soffer and M.I.~Weinstein,
Multichannel nonlinear scattering theory for nonintegrable equations I, II,
{\it Comm. Math. Phys.} {\bf 133} (1990), 119--146;
{\it J. Diff. Equations} {\bf 98}, (1992), 376--390.

\bibitem{SW2}
A.~Soffer and M.I.~Weinstein,
Resonances, radiation damping and instability
in Hamiltonian nonlinear wave equations,
{\it Invent.~math.} {\bf 136}, (1999), 9--74.

\bibitem{SW3}
A.~Soffer and M.I.~Weinstein, Selection of the ground state for
nonlinear Schr\"odinger equations, preprint.

\bibitem{TY}
T.-P. Tsai and H.-T. Yau, Asymptotic dynamics of nonlinear
Schr\"odinger equations, resonance dominated and dispersion
dominated solutions, {\it Comm. Pure Appl. Math.} {\bf 55} (2002)
153--216.


\bibitem{TY2} T.-P. Tsai and H.-T. Yau,
Relaxation of excited states in nonlinear Schr\"odinger equations,
{\it IMRN}, to appear.


\bibitem{TY3} T.-P. Tsai and H.-T. Yau,
Stable directions for excited states of nonlinear Schr\"odinger
equations, {\it Comm. P.D.E.}, to appear.

\bibitem{TY4} T.-P. Tsai and H.-T. Yau,
Classification of asymptotic profiles for
nonlinear Schr\"odinger equations with small initial data, preprint.



\bibitem{W2}
M.I.~Weinstein, Lyapunov stability of ground states of nonlinear
dispersive evolution equations, \textit{Comm. Pure Appl. Math.} {\bf
39}, (1986), 51--68.

\bibitem{Y}
K.Yajima, The $W^{k,p}$ continuity of wave operators for
Schr\"odinger operators, {\it J.Math. Soc. Japan} {\bf 47}, no. 3,
(1995), 551--581.

\end{thebibliography}
\end{document}